\documentclass{ar-1col}
\usepackage{xspace}
\usepackage{natbib}
\usepackage{biblio}

\usepackage[total={30.5pc,47pc},centering]{geometry}

\begin{document}
\jname{Annu. Rev. Astron. Astrophys.}
\jyear{2016}
\jvol{54}
\fstpage{529}
\endpage{596}
\setcounter{page}{529}
\doi{10.1146/annurev-astro-081915-023441}

\def\Sigtot {$\Sigma_{\rm tot}$}
\def\Rhotot {$\rho_{\rm tot}$}
\def\Etot {$\epsilon_{\rm tot}$}

\newcommand{\FIRE}{\textsc{fire}\xspace}
\newcommand{\spaceshuttle}{\textsc{space shuttle}\xspace}
\newcommand{\spacelab}{\textsc{spacelab}\xspace}
\newcommand{\VIKING}{\textsc{viking}\xspace}
\newcommand{\DENIS}{\textsc{denis}\xspace}
\newcommand{\CLUES}{\textsc{clues}\xspace}
\newcommand{\WMAP}{\textsc{wmap}\xspace}
\newcommand{\cobe}{\textsc{cobe}\xspace}
\newcommand{\COBE}{\textsc{cobe}\xspace}
\newcommand{\dirbe}{\textsc{dirbe}\xspace}
\newcommand{\Hipparcos}{\textsc{hipparcos}\xspace}
\newcommand{\hipparcos}{\textsc{hipparcos}\xspace}
\newcommand{\HST}{\textsc{Hubble Space Telescope}\xspace}
\newcommand{\GALAH}{\textsc{galah}\xspace}
\newcommand{\RAVE}{\textsc{rave}\xspace}
\newcommand{\LEGUE}{\textsc{legue}\xspace}
\newcommand{\APOGEE}{\textsc{apogee}\xspace}
\newcommand{\GCS}{\textsc{gcs}\xspace}
\newcommand{\GES}{\textsc{ges}\xspace}
\newcommand{\WFIRST}{\textsc{wfirst}\xspace}
\newcommand{\JWST}{\textsc{jwst}\xspace}
\newcommand{\GAMA}{\textsc{gama}\xspace}
\newcommand{\APOGEETWO}{\textsc{apogee-2}\xspace}
\newcommand{\FOURMOST}{\textsc{4most}\xspace}
\newcommand{\TESS}{\textsc{tess}\xspace}
\newcommand{\POSS}{\textsc{poss}\xspace}
\newcommand{\PLATO}{\textsc{plato}\xspace}
\newcommand{\LSST}{\textsc{lsst}\xspace}
\newcommand{\WEAVE}{\textsc{weave}\xspace}

\newcommand{\GMT}{\textsc{gmt}\xspace}
\newcommand{\GCLEF}{\textsc{g-clef}\xspace}

\newcommand{\PIONEER}{\textsc{pioneer 10}\xspace}
\newcommand{\vlbi}{\textsc{vlbi}\xspace}
\newcommand{\nmagic}{\textsc{nmagic}\xspace}
\newcommand{\brava}{\textsc{brava}\xspace}
\newcommand{\argos}{\textsc{argos}\xspace}
\newcommand{\gibs}{\textsc{gibs}\xspace}
\newcommand{\twomass}{\textsc{2mass}\xspace}
\newcommand{\TWOMASS}{\textsc{2mass}\xspace}
\newcommand{\ukidss}{\textsc{ukidss}\xspace}
\newcommand{\glimpse}{\textsc{glimpse}\xspace}
\newcommand{\GLIMPSE}{\textsc{glimpse}\xspace}
\newcommand{\ogle}{\textsc{ogle}\xspace}

\newcommand{\vvv}{\textsc{vvv}\xspace}
\newcommand{\VVV}{\textsc{vvv}\xspace}
\newcommand{\SWEEPS}{\textsc{sweeps}\xspace}
\newcommand{\spitzer}{\textsc{spitzer}\xspace}

\newcommand{\sdss}{\textsc{sdss}\xspace}
\newcommand{\SDSS}{\textsc{sdss}\xspace}

\newcommand{\DES}{\textsc{des}\xspace}
\newcommand{\PFS}{\textsc{pfs}\xspace}
\newcommand{\MOONS}{\textsc{moons}\xspace}
\newcommand{\VLT}{\textsc{vlt}\xspace}
\newcommand{\KTWO}{\textsc{k2}\xspace}
\newcommand{\COROT}{\textsc{corot}\xspace}

\newcommand{\segue}{\textsc{segue}\xspace}
\newcommand{\SEGUE}{\textsc{segue}\xspace}

\newcommand{\gaia}{\textsc{gaia}\xspace}
\newcommand{\Gaia}{\textsc{ESA gaia}\xspace}


\newcommand{\SEWM}{\rm{SE}\xspace}
\newcommand{\UEWM}{\rm{UE}\xspace}
\newcommand{\UUEWM}{\rm{UUE}\xspace}
\newcommand{\imf}{\rm{IMF}\xspace}
\newcommand{\hmsfr}{\rm{HMSFR}\xspace}
\newcommand{\rcg}{\rm{RCG}\xspace}
\newcommand{\pl}{\rm{PL}\xspace}
\newcommand{\hb}{\rm{HB}\xspace}
\newcommand{\los}{\rm{LOS}\xspace}
\newcommand{\propm}{\rm{PM}\xspace}
\newcommand{\rv}{\rm{RV}}
\newcommand{\smbh}{\rm{SMBH}\xspace}
\newcommand{\nsc}{\rm{NSC}\xspace}
\newcommand{\nsd}{\rm{NSD}\xspace}
\newcommand{\gb}{\rm{GB}\xspace}
\newcommand{\rms}{\textsc{rms}\xspace}
\newcommand{\vlos}{\ifmmode{\textrm{v}_{\rm los}} \else {$\textrm{v}_{\rm los}$}\fi}

\newcommand{\narreq}{\!=\!}
\newcommand{\nsim}{\!\sim\!}
\newcommand{\ncol}{\!:\!}
\newcommand{\nsimeq}{\!\simeq\!}

\mathchardef\mhyphen="2D

\newcommand{\dg}{^\circ}
\newcommand{\mum}{\,\mu\rm{m}\xspace}
\newcommand{\mm}{\,\rm{mm}\xspace}
\renewcommand{\vec}[1]{\boldsymbol{#1}}
\newcommand{\AU}{\,\rm{AU}\xspace}
\newcommand{\pc}{\,\rm{pc}\xspace}
\newcommand{\kmskpc}{\,\rm{km\,s^{-1}\,kpc^{-1}}\xspace}
\newcommand{\arcsec}{\,\rm{arcsec}\xspace}
\newcommand{\masyr}{\,\rm{mas\,yr^{-1}}\xspace}
\newcommand{\micas}{\,\mu\rm{as}\xspace}
\newcommand{\SM}{10^{10}\rm{M}_{\odot}\xspace}
\newcommand{\bb}{\rm{b\mhyphen b}}
\newcommand{\R}{\mathcal{R}}
\newcommand{\Ms}{M_{\rm{s}}}
\newcommand{\MDM}{M_{\rm{DM}}}
\newcommand{\Gyr}{\,\rm{Gyr}}
\newcommand{\F}{\mathcal{F}\xspace}
\newcommand{\vel}{\textrm{v}}

\def\ML {$\Upsilon^\star$} 
\def\fb{$f_{\rm bary}$}
\def\yr {yr$^{-1}$}

\def\Rzero {\ifmmode{R_0} \else {$R_0$}\fi} 
\def\zzero {\ifmmode{z_0} \else {$z_0$}\fi} 
\def\Vcirc {\ifmmode{\Theta_0} \else {$\Theta_0$}\fi}  
\def\Wcirc {\ifmmode{\Omega_0} \else {$\Omega_0$}\fi}  
\def\Omcirc {\ifmmode{\Omega_0} \else {$\Omega_0$}\fi}  
\def\Vgsol  {\ifmmode{V_{g,\odot}} \else {$V_{\g,\odot}$}\fi}           
\def\Omgsol {\ifmmode{\Omega_{g,\odot}} \else {$\Omega_{g,\odot}$}\fi}  

\def\frho {\ifmmode{f_\rho} \else {$f_\rho$}\fi} 
\def\fsig {\ifmmode{f_\Sigma} \else {$f_\Sigma$}\fi} 
\def\ColRatio {$f_\rho$} 
\def\Usol {\ifmmode{U_\odot} \else {$U_\odot$}\fi} 
\def\Vsol {\ifmmode{V_\odot} \else {$V_\odot$}\fi} 
\def\Wsol {\ifmmode{W_\odot} \else {$W_\odot$}\fi} 
\def\Vsolar {\ifmmode{{\bf v}_\odot} \else {${\bf v}_\odot$}\fi} 
\def\ULSR {\ifmmode{U_{\mathrm{LSR}}} \else {$U_{\mathrm{LSR}}$}\fi} 
\def\VLSR {\ifmmode{V_{\mathrm{LSR}}} \else {$V_{\mathrm{LSR}}$}\fi} 
\def\Vstr {\ifmmode{{\bf v}_{\rm str}} \else {${\bf v}_{\rm str}$}\fi}  
\def\Vmean {$\bar{v}_\phi$} 
\def\Mvir {\ifmmode{M_{\rm vir}} \else {$M_{\rm vir}$}\fi} 
\def\Rvir {\ifmmode{r_{\rm vir}} \else {$r_{\rm vir}$}\fi} 
\def\Msfr {$\dot{m}_{\star}$} 
\def\Mvir {\ifmmode{M_{\rm vir}} \else {$M_{\rm vir}$}\fi} 
\def\Mtot {\ifmmode{M_{\rm tot}} \else {$M_{\rm tot}$}\fi} 
\def\Mbarytot {$M_{\rm bary,tot}$} 
\def\Mhot {$M_{\rm hot}$} 
\def\Mstar {\ifmmode {M_\star} \else {$M_\star$}\fi} 
\def\Lstar {\ifmmode {L_\star} \else {$L_\star$}\fi} 
\def\Mhalo {$M_{\rm h}$} 

\def\Mbary {$M_{\rm bary}$} 
\def\fbary{$f_{\rm bary}$}

\def\Mcold {$M_{\rm bary}$} 
\def\fcold {$f_{\rm bary}$}

\def\Mbul   {$M_{\rm b}$}                    
\def\Mbdyn  {$ M_{\rm b}^{\rm dyn}$}           
\def\Mbstar {$ M_{\rm b}^{*}$}                
\def\Mclfr  {$M_{\mathrm{clb}}/M_{\rm b}^{*}$}  
\def\BPBphi { \ifmmode{\phi_{\rm bp}} \else {$\phi_{\rm bp}$}\fi}    
\def\BPBba  { \ifmmode{(b/a)_{\rm bp}} \else {$(b/a)_{\rm bp}$}\fi}  
\def\BPBca  { \ifmmode{(c/a)_{\rm bp}} \else {$(c/a)_{\rm bp}$}\fi}  
\def\BPBhz  { \ifmmode{h_{\rm bp}} \else {$h_{\rm bp}$}\fi}  
\def\BPBxX  { \ifmmode{x_X} \else {$x_X$}\fi}                      
\def\Rbul   {$r_{\rm b}$} 
\def\Mlbth  {$M_{\rm tlb}$}               
\def\Mlbst  {$M_{\rm slb}$}               
\def\LBphi  { \ifmmode{\phi_{\rm lb}} \else {$\phi_{\rm lb}$}\fi}   
\def\RLB    { \ifmmode{R_{\rm lb}} \else {$R_{\rm lb}$}\fi}         
\def\LBhth  { \ifmmode{h_{\rm tlb}} \else {$h_{\rm tlb}$}\fi}       
\def\LBhst  { \ifmmode{h_{\rm slb}} \else {$h_{\rm slb}$}\fi}       
\def\Omegab { \ifmmode{\Omega_{b}} \else {$\Omega_{b}$}\fi}        
\def\RCR    { \ifmmode{R_{\mathrm{CR}}} \else {$R_{\mathrm{CR}}$}\fi}
\def\MBH    {$M_\bullet$}
\def\rinfl  {$r_{\mathrm{infl}}$}
\def\MNSC   {$M_{\rm NSC}$}                                                      
\def\MNSD   {$M_{\rm NSD}$}                                                      
\def\RNSC   { \ifmmode{r_{\mathrm{NSC}}} \else {$r_{\mathrm{NSC}}$}\fi}           
\def\NSCca  { \ifmmode{(c/a)_{\mathrm{NSC}}} \else {$(c/a)_{\mathrm{NSC}}$}\fi}   
\def\RNSD   { \ifmmode{r_{\mathrm{NSD}}} \else {$r_{\mathrm{NSD}}$}\fi}           
\def\hNSD   { \ifmmode{h_{\mathrm{NSD}}} \else {$h_{\mathrm{NSD}}$}\fi}           
\def\Mshalo { \ifmmode{M_{\rm s}} \else {$M_{\rm s}$}\fi}          
\def\Mssub  { \ifmmode{M_{\rm sub}} \else {$M_{\rm sub}$}\fi}      
\def\Rshalo { \ifmmode{r_{\rm s}} \else {$r_{\rm s}$}\fi}          
\def\ainsh  { \ifmmode{\alpha_{\mathrm{in}}} \else {$\alpha_{\mathrm{in}}$}\fi}   
\def\aoutsh { \ifmmode{\alpha_{\mathrm{out}}} \else {$\alpha_{\mathrm{out}}$}\fi} 
\def\qinsh  { \ifmmode{q_{\mathrm{in}}} \else {$q_{\mathrm{in}}$}\fi}             
\def\qoutsh { \ifmmode{q_{\mathrm{out}}} \else {$q_{\mathrm{out}}$}\fi}           
\def\vshalo { \ifmmode{{\overline{v}}^{\,\rm s}_\phi} \else {${\overline{v}}^{\,\rm s}_\phi$}\fi}   
\def\Rhalo {$r_{\rm h}$} 

\def\Mthin { \ifmmode{M^{\rm t}} \else {$M^{\rm t}$}\fi} 
\def\Mthick { \ifmmode{M^{\rm T}} \else {$M^{\rm T}$}\fi} 
\def\Rthin { \ifmmode{R^{\rm t}} \else {$R^{\rm t}$}\fi} 
\def\Rthick { \ifmmode{R^{\rm T}} \else {$R^{\rm T}$}\fi} 
\def\Rdisk {\ifmmode{R_{\rm d}} \else {$R_{\rm d}$}\fi} 
\def\RdiskM { \ifmmode{R_{\rm d}} \else {$R_{\rm d}$}\fi} 

\def\Rmaxthin {$R^{\rm t}_{\rm max}$} 
\def\Rmaxthick {$R^{\rm T}_{\rm max}$} 
\def\Rdiskmass {\ifmmode {R_{\rm M}} \else {$R_{\rm M}$}\fi} 
\def\Zacc {\ifmmode{K_{\rm z}} \else {$K_{\rm z}$}\fi} 
\def\Zthin {\ifmmode{z^{\rm t}} \else {$z^{\rm t}$}\fi} 
\def\Zthick {\ifmmode{z^{\rm T}} \else {$z^{\rm T}$}\fi} 
\def\Nthin {$\rho^{\rm t}$} 
\def\Nthick {$\rho^{\rm T}$} 
\def\sigRthin {$\sigma_R^{\rm t}$} 
\def\sigPthin {$\sigma_\phi^{\rm t}$} 
\def\sigZthin {$\sigma_z^{\rm t}$} 
\def\sigRthick {$\sigma_R^{\rm T}$} 
\def\sigPthick {$\sigma_\phi^{\rm T}$} 
\def\sigZthick {$\sigma_z^{\rm T}$} 
\def\RsigthinR {$R^{\rm t}_{\sigma,R}$} 
\def\RsigthickR {$R^{\rm T}_{\sigma,R}$} 
\def\RsigthinZ {$R^{\rm t}_{\sigma,z}$} 
\def\RsigthickZ {$R^{\rm T}_{\sigma,z}$} 
\def\Zsigthin {$z^{\rm t}_\sigma$} 
\def\Zsigthick {$z^{\rm T}_\sigma$} 
\def\betaR {$\beta_R$} 
\def\betaP {$\beta_\phi$} 
\def\betaZ {$\beta_z$} 
\def\alphaR {$\alpha_R$}  
\def\alphaZ {$\alpha_z$}  
\def\DF{${\scriptstyle\rm DF}$}
\def\EDF{${\scriptstyle\rm EDF}$}
\def\GALAXIA{${\scriptstyle\rm GALAXIA}$}
\def\kms{\ifmmode {\>{\rm\ km\ s}^{-1}}\else {\ km s$^{-1}$}\fi}
\def\kkms{\ifmmode {{\rm km\ s}^{-1}}\else {km s$^{-1}$}\fi}
\def\perkpc {\ifmmode {{\rm kpc}^{-1}}\else {\ kpc$^{-1}$}\fi}
\def\cc {\ cm$^{-3}$}
\def\ppcsq {\ pc$^{-2}$}
\def\ppccb {\ pc$^{-3}$}
\def\msol{\ifmmode {\>M_\odot}\else {$M_\odot$}\fi}
\def\mvir{\ifmmode {\>M_{\rm vir}}\else {$M_{\rm vir}$}\fi}
\def\rvir{\ifmmode {\>r_{\rm vir}}\else {$r_{\rm vir}$}\fi}
\def\msun{\ifmmode {\>{\rm M}_\odot}\else {M$_\odot$}\fi}
\def\Lsun{\ifmmode {\>{\rm L}_\odot}\else {L$_\odot$}\fi}
\def\cmsq{\ifmmode {\>{\rm\ cm}^2}\else {cm$^2$}\fi}
\def\psqcm{\ifmmode {\>{\rm cm}^{-2}}\else {cm$^{-2}$}\fi}
\def\psqpc{\ifmmode {\>{\rm pc}^{-2}}\else {pc$^{-2}$}\fi}
\def\pcsq{\ifmmode {\>{\rm\ pc}^2}\else {pc$^2$}\fi}
\def\Tkev{\ifmmode{T_{\rm kev}}\else {$T_{\rm keV}$}\fi}
\def\hubunits{\ifmmode {\>{\rm km\ s^{-1}\ Mpc^{-1}}}\else {km\: s$^{-1}$ Mpc$^{-1}$}\fi}
\def\gta{\;\lower 0.5ex\hbox{$\buildrel > \over \sim\ $}}
\def\lta{\;\lower 0.5ex\hbox{$\buildrel < \over \sim\ $}}
\def\phiIV{\ifmmode{\varphi_4}\else {$\varphi_4$}\fi}
\def\phiI{\ifmmode{\varphi_i}\else {$\varphi_i$}\fi}
\def\be{\begin{equation}}
\def\ee{\end{equation}}
\def\bea{\begin{eqnarray}}
\def\eea{\end{eqnarray}}
\def\beas{\begin{eqnarray*}}
\def\eeas{\end{eqnarray*}}
\def\gtrapprox{\;\lower 0.5ex\hbox{$\buildrel >\over \sim\ $}}
\def\lessapprox{\;\lower 0.5ex\hbox{$\buildrel < \over \sim\ $}}
\def\deg   {$^\circ$}
\def\Ftwo  {$F_{-21}$}
\def\Pcos  {$\Phi^0$}
\def\Jtwo  {$J_{-21}$}
\def\Fcos  {$F_{-21}^0$}
\def\Jcos  {$J_{-21}^0$}
\def\Em    {${\cal E}_m$}
\def\ALL   {A_{\scriptscriptstyle LL}}
\def\JLL   {J_{\scriptscriptstyle LL}}
\def\nuLL  {\nu_{\scriptscriptstyle LL}}
\def\sigLL {\sigma_{\scriptscriptstyle LL}}
\def\tauLL {\ifmmode{\tau_{\scriptscriptstyle LL}}\else {$\tau_{\scriptscriptstyle LL}$}\fi}
\def\nuOB  {\nu_{\scriptscriptstyle {\rm OB}}}
\def\aB    {\alpha_{\scriptscriptstyle B}}
\def\nH    {n_{\scriptscriptstyle H}}
\def\Em{\ifmmode{{\rm E}_m}\else {{\rm E}$_m$}\fi}
\def\NH{\ifmmode{{\rm N}_{\scriptscriptstyle\rm H}}\else {{\rm N}$_{\scriptscriptstyle\rm H}$}\fi}
\def\Ha    {H$\alpha$}
\def\Hb    {H$\beta$}
\def\OVII{O${\scriptstyle\rm VII}$}
\def\OVIII{O${\scriptstyle\rm VIII}$}
\def\HI    {H${\scriptstyle\rm I}$}
\def\HII   {H${\scriptstyle\rm II}$}
\def\CO  {CO}
\def\eg    {{\it e.g.}}
\def\ie    {{\it i.e.}}
\def\cf    {{\it cf. }}
\def\qv    {{\it q.v. }}
\def\etal  {\ et al.}
\def\Em{\ifmmode{{\cal E}_m}\else {{\cal E}$_m$}\fi}
\def\Dm{\ifmmode{{\cal D}_m}\else {{\cal D}$_m$}\fi}
\def\fesc{\ifmmode{\hat{f}_{\rm esc}}\else {$\hat{f}_{\rm esc}$}\fi}
\def\fescs{\ifmmode{f_{\rm esc}}\else {$f_{\rm esc}$}\fi}
\def\rsolar{\ifmmode{r_\odot}\else {$r_\odot$}\fi}
\def\emunit{\ifmmode{{\rm cm}^{-6}{\rm\ pc}}\else {
cm$^{-6}$ pc}\fi}
\def\intensity{\ifmmode{{\rm erg\ cm}^{-2}{\rm\ s}^{-1}
      {\rm\ Hz}^{-1}{\rm\ sr}^{-1}}
      \else {erg cm$^{-2}$ s$^{-1}$ Hz$^{-1}$ sr$^{-1}$}\fi}
\def\flux{\ifmmode{{\rm erg\ cm}^{-2}{\rm\ s}^{-1}}\else {erg
cm$^{-2}$ s$^{-1}$}\fi}
\def\fluxdensity{\ifmmode{{\rm erg\ cm^{-2}\ s^{-1}\ Hz^{-1}}}\else {erg
cm$^{-2}$ s$^{-1}$ Hz$^{-1}$}\fi}
\def\phoflux{\ifmmode{{\rm phot\ cm}^{-2}{\rm\ s}^{-1}}\else {phot
cm$^{-2}$ s$^{-1}$}\fi}
\def\phorate{\ifmmode{{\rm phot\ s}^{-1}}\else {phot s$^{-1}$}\fi}



\def\xvir{\ifmmode {\> x_{vir}}\else {$x_{vir}$}\fi}
\def\tratio{\ifmmode {\> \tau}\else {$\tau$}\fi}
\def\K{\ifmmode {\> {\rm K}}\else {K}\fi}
\def\kpc{\ifmmode {\> {\rm kpc}}\else {kpc}\fi}
\def\pcc{\ifmmode {\> {\rm cm}^{-3} }\else {${\rm cm}^{-3}$}\fi}
\def\keV{\ifmmode {\> {\rm keV} }\else {keV}\fi}
\def\Zsun{\ifmmode {\>Z_\odot}\else {${\rm Z}_\odot$}\fi}
\def\Msun{\ifmmode {\>M_\odot}\else {${\rm M}_\odot$}\fi}
%

\def\iaus{{\it IAU Symp}}
\def\asp{{\it ASP Conf}}
\def\assl{{\it Astr Space Sci Lib}}

\markboth{Bland-Hawthorn \& Gerhard}{The Galaxy in Context}

\title{The Galaxy in Context: Structural, Kinematic \& Integrated Properties}

\author{Joss Bland-Hawthorn$^1$, Ortwin Gerhard$^2$
\affil{$^1$Sydney Institute for Astronomy, School of Physics A28, 
University of Sydney, NSW 2006, Australia; 
email: jbh@physics.usyd.edu.au}
\affil{$^2$Max Planck Institute for extraterrestrial Physics, PO Box 1312, Giessenbachstr.,
   85741 Garching, Germany; 
email: gerhard@mpe.mpg.de}
}

\begin{abstract}
Our Galaxy, the Milky Way, is a benchmark for understanding disk
galaxies. It is the only galaxy whose formation history can be
studied using the full distribution of stars from faint dwarfs to
supergiants. The oldest components provide us with unique insight
into how galaxies form and evolve over billions of years.  The Galaxy
is a luminous ($L_\star$) barred spiral with a central box/peanut bulge, a
dominant disk, and a diffuse stellar halo. Based on global
properties, it falls in the sparsely populated ``green valley" region of
the galaxy colour-magnitude diagram.  Here we review the key
integrated, structural and kinematic parameters of the Galaxy, and
point to uncertainties as well as directions for future progress. 
Galactic studies will continue to play a fundamental role far into the
future because there are measurements that can only be made in the
near field and much of contemporary astrophysics depends on such
observations.
\end{abstract}


\maketitle

\section{PROLOGUE}

\begin{figure*}
\centering
\parbox{6cm}{
\includegraphics[width=6cm]{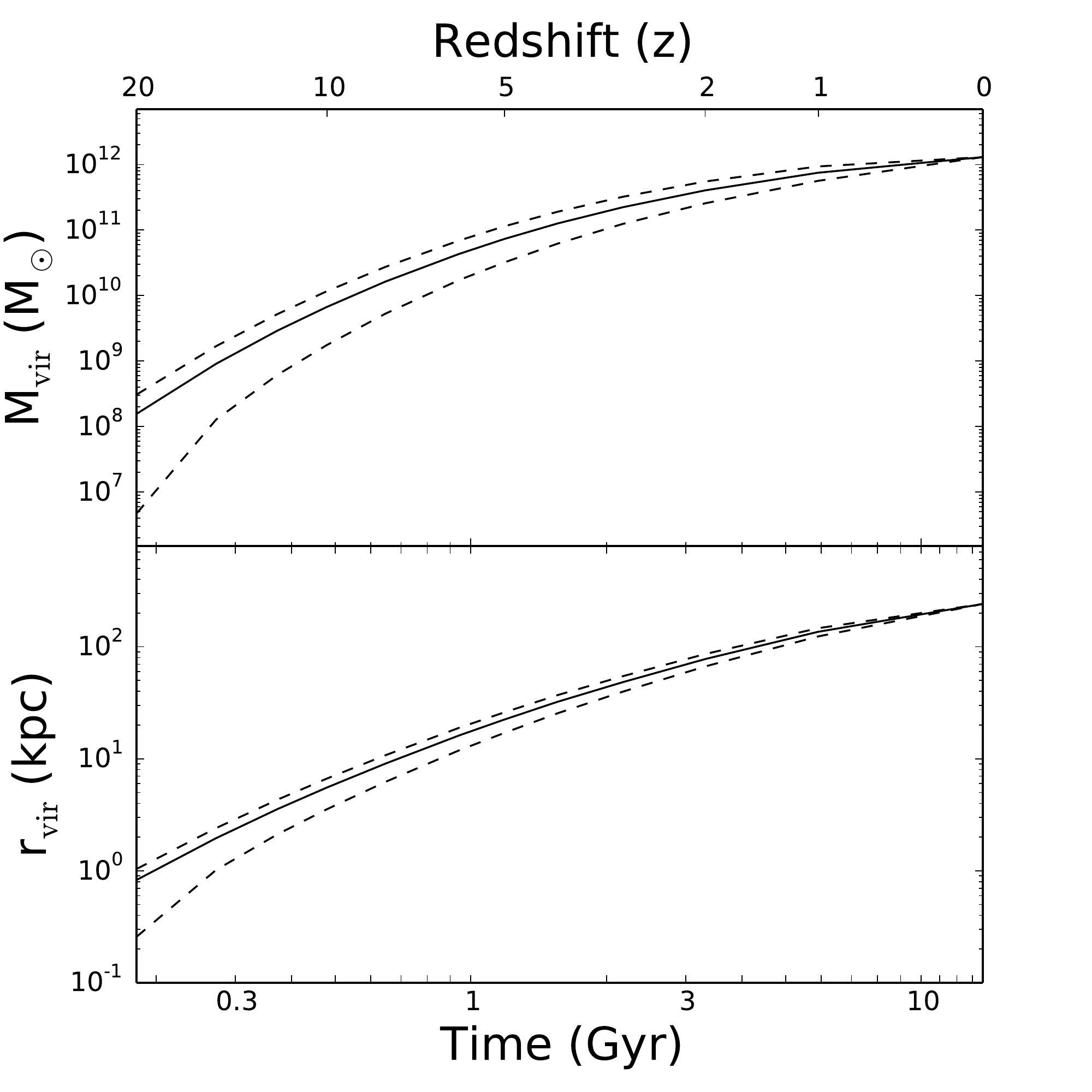}
}
\qquad
\begin{minipage}{6cm}
\includegraphics[width=6cm]{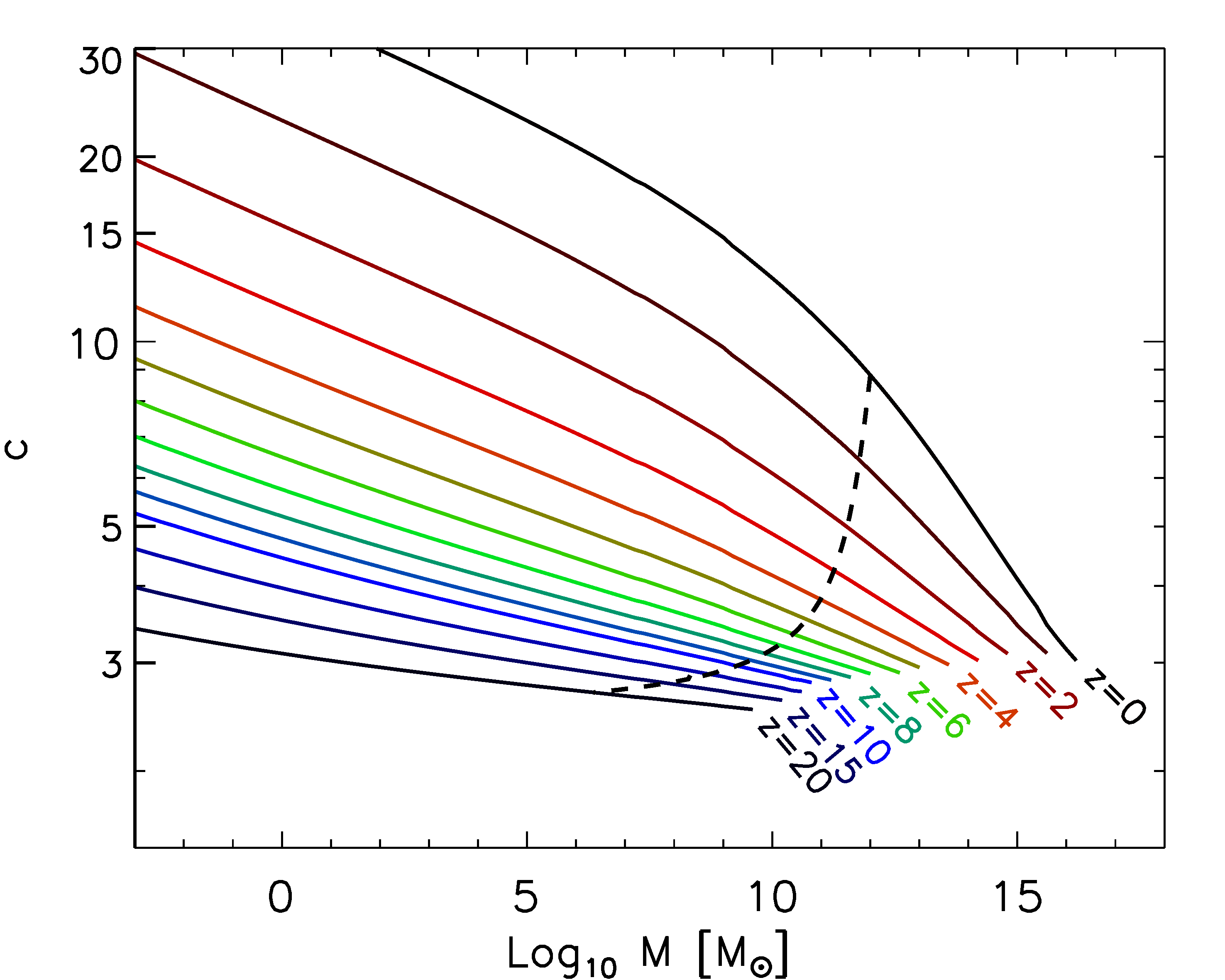}
\end{minipage}
\caption{Left: The estimated growth of the Galaxy's virial mass (\Mvir) and radius (\Rvir) from 
$z=20$ to the present day, $z=0$. Throughout this
review, we use $h=0.7$, $\Omega_{\rm M}=0.3$ and $\Omega_\Lambda=0.7$.
The virial radius at any epoch is the approximate extent 
over which the growing Galaxy has stabilized. This plot was
derived from 5000 runs of the tree merger code of \citet{Parkinson2008}: the
dashed lines encompass 67\% of the predicted halos at each epoch ($1\sigma$ uncertainty).
Right: The discrete curves show how the NFW concentration parameter $c$
depends on halo mass \Mvir\ as a function of cosmic epoch indicated by the different line
colours. If Galactic dark matter is correctly described by an NFW halo, the dashed
line shows how $c$ evolves with mass (and therefore cosmic time). The central regions form
early on and most of the accreted matter settles to the outer halo
\citep[adapted from][]{Correa2015}.
}
\label{f:CDM}
\end{figure*}

Galactic studies are a fundamental cornerstone of contemporary astrophysics.
Nowadays, we speak of near-field cosmology where detailed studies of the
Galaxy underlie our understanding of universal processes over cosmic time
\citep{Freeman2002}.
Within the context of the cold dark matter paradigm,  
the Galaxy built up over billions of years through a process of hierarchical accretion
(see Fig.~\ref{f:CDM}). Our Galaxy has recognisable components
that are likely to have emerged at different stages of the formation process.
In particular, the early part of the bulge may have collapsed first seeding the early stages of
a massive black hole, followed by a distinct phase that gave rise to the thick
disk. The inner halo may have formed about the same time while the outer halo has
built up later through the progressive accretion of shells of material over
cosmic time \citep{Prada2006}. The dominant thin disk reflects
a different form of accretion over the same long time frame \citep{Brook2012}.

\begin{figure}
\includegraphics[width=3.5in]{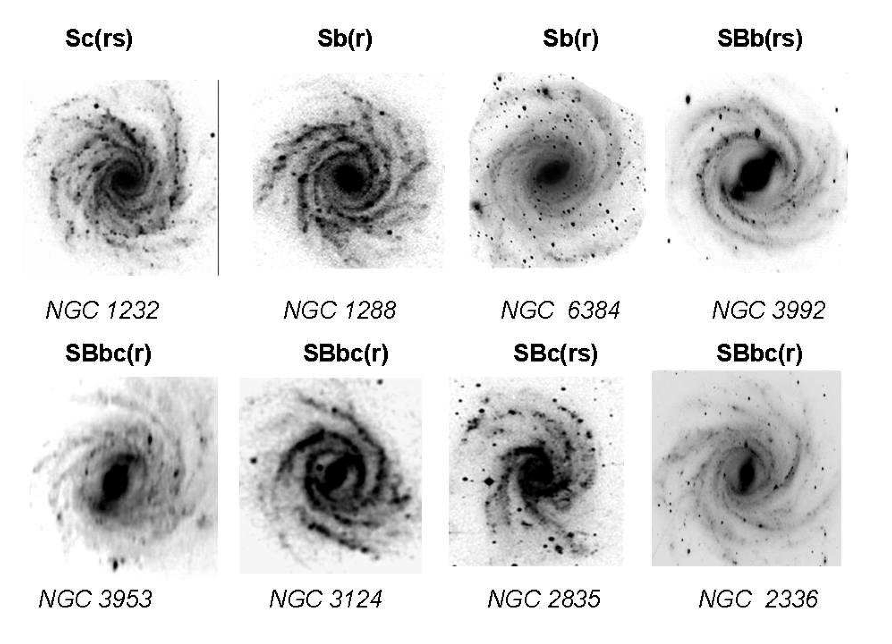}
\caption{Milky Way analogues $-$ a selection of galaxies that broadly resemble the
Galaxy \citep{Efremov2011}. All images have been rectified to a flat projection.
Classifications within the de Vaucouleurs morphological classification scheme are also shown.}
\label{fig2}
\end{figure}

We live in an age when vast surveys have been carried out across the entire sky in many 
wavebands. At optical and infrared wavelengths, billions of stars have
been catalogued with accurate photometric magnitudes and colours \citep{Skrutskie2006,Saito2012,Ivezic2012}. 
But only a fraction of these stars have high quality spectral classifications, radial velocities and 
distances, with an even smaller fraction having useful elemental abundance determinations.
The difficulty of measuring a star's age continues to hamper progress in Galactic studies, but
this stumbling block will be partly offset by the \Gaia\ astrometric survey already under way.
By the end of the decade, this mission will have measured accurate distances and velocity 
vectors for many millions of stars arising from all major components of the Galaxy \citep{deBruijne2015}.
In light of this impending data set, we review our present understanding of the main
dynamical and structural parameters that describe our home, the Milky Way.

\begin{table}
\caption{\label{t:params}
Description of Galactic parameters$^a$.}
\begin{center}
\begin{tabular}{|l|l|} 
\hline
\Rzero                         & Distance of Sun from the Galactic Centre (\S3.2) \\ 
\zzero                         & Distance of Sun from the Galactic Plane (\S3.3) \\ 
$\Vcirc, \Wcirc$               & Circular speed, angular velocity at Sun with respect to Galactic Centre (\S3.4, \S6.4.2) \\ 
$\Usol, \Vsol, \Wsol$          & $U$, $V$, $W$ component of solar motion with respect to LSR (\S5.3.3) \\
\Vsolar                        & Solar motion with respect to LSR (\S5.3.3) \\
\VLSR                          & Possible LSR streaming motion with respect to $\Vcirc$ (\S5.3.3, \S6.4.2) \\
$A$, $B$                       & Oort's constants (\S6.4.2) \\
\hline 
\Rvir                          & Galactic virial radius (\S6.3) \\
\Mvir, $M_{\rm vir,timing}$     & Galactic virial mass, virial timing mass (\S6.3) \\
\Mstar, \Msfr	               & Galactic stellar mass, global star formation rate (\S2.2, \S6.4) \\
\Mbary, \fbary$^b$             & Galactic baryon mass, baryon fraction (\S6.2, \S6.4.3) \\
\hline
\Mbdyn, \Mbstar, \Mclfr        & Bulge dynamical mass and stellar mass, classical bulge fraction (\S4.2.4) \\
$\sigma^b_x,\sigma^b_y,\sigma^b_z,\sigma^b_{\rm rms}$ & Half-mass bulge velocity dispersions in $(x,y,z)$ and rms (\S4.2.3) \\
\BPBphi, \BPBba                & b/p-bulge orientation and axis ratio from top (\S4.2.1) \\
\BPBhz, \BPBca                 & Central vertical scale-height and edge-on axis ratio of b/p-bulge (\S4.2.1) \\
\BPBxX                         & Radius of maximum X (\S4.2.1) \\
\hline
\Mlbth, \Mlbst                 & Stellar masses of thin and superthin long bar (\S4.3) \\
\LBphi, \RLB                   & Long bar orientation and half-length (\S4.3) \\
\LBhth, \LBhst                 & Vertical scale heights of thin and superthin long bar (\S4.3) \\
\Omegab, \RCR                  & Bar pattern speed and corotation radius (\S4.4) \\
\hline
\MBH, \rinfl                   & Mass and dynamical influence radius of supermassive black hole (\S3.4) \\
\MNSC, \MNSD                   & Masses of nuclear star cluster and nuclear stellar disk (\S4.1) \\
\RNSC, \NSCca                  & Nuclear star cluster half-mass radius and axis ratio (\S4.1) \\
\RNSD, \hNSD                   & Nuclear stellar disk break radius and vertical scale-height (\S4.1) \\
\hline 
\Mhot	                       & Coronal (hot) halo mass (\S6.2) \\
\Mshalo, \Mssub                & Stellar halo mass and substructure mass (\S6.1.2) \\
\ainsh, \aoutsh, \Rshalo       & Stellar halo inner, outer density slope, break radius (\S6.1.1) \\
\qinsh, \qoutsh                & Inner and outer mean flattening (\S6.1.1) \\ 
$\sigma_r^{\rm s},\;\sigma_\theta^{\rm s},\;\sigma_\phi^{\rm s}$ 
                               & Stellar halo velocity dispersions in $r$, $\theta$, $\phi$ near the Sun (\S6.1.3) \\
\vshalo                        & Local halo rotation velocity (\S6.1.3) \\
\hline
\Mthin, \Mthick               & Thin, thick disk stellar masses (\S5.1.3, \S5.2.2) \\
\Rthin, \Rthick                & Thin, thick disk exponential scalelength in $R$ (\S5.1.3, \S5.2.2) \\
\Zthin, \Zthick                 & Thin, thick disk exponential scaleheight in $z$ (\S5.1.3) \\
\frho, \fsig			     & Thick disk fraction in local density, in integrated column density (\S5.1.3) \\
\sigRthin, \sigZthin        & Old thin disk velocity dispersion in $R$, $z$ at 10 Gyr (\S5.4) \\ 
\sigRthick, \sigZthick     & Thick disk velocity dispersion in $R$, $z$ (\S5.4) \\ 
\Sigtot, \Rhotot, \Etot       & Local mass surface density, mass density, dark matter energy density (\S5.4.2) \\
\hline
\end{tabular}
\end{center}
\begin{tabnote}
$^a$Our convention is to use $R$ for a projected radius in two dimensions (e.g. disk) and $r$ for a radius in three dimensions (e.g. halo). 
$^b$\fbary\ is the ratio of baryonic mass to total mass integrated to a radius where both quantities have
been determined (e.g. \Rvir).
\end{tabnote}
\end{table}

We run into a problem familiar to cartographers. How is one 
to describe the complexity of the Galaxy? No two galaxies are identical; even the best
morphological analogues with the Milky Way have important differences (Fig. 2).
Historically, astronomers have resorted to 
defining discrete components with the aim of measuring their structural parameters
(see Table~\ref{t:params}). We continue to see value in this approach and proceed
to define what we mean by each subsystem, and the best estimates that can be
made at the present time.  In reality, these `components' exhibit strong overlap by 
virtue of sharing the same evolving Galactic potential \citep[e.g.][]{Guedes2013}
and the likelihood that stars migrate far from their formation sites 
\citep{Sellwood2002,Minchev2010}. 

Even with complete data (density field, distribution function, chemistry), it is
unlikely that any one component can be entirely separated from another. 
In particular, how are we to 
separate the bar/bulge from the inner disk and inner halo? A distinct possibility
is that most of our small bulge (compared to M31) has formed through a disk instability
associated with bar formation, rather than during a dissipational early collapse phase.
The same challenge exists in separating the
thin disk from the thick disk. Some have argued for a gradual transition but there
is now good evidence that a major part of the thick disk is chemically 
distinct from the dominant thin disk, suggesting a different origin.
In this context,
\citet{Binney2013} has argued that the Galaxy's stellar populations are better 
described by phase-space distribution functions (DF) that are self-consistent with the
underlying gravitational potential (\S 5). 

Our goal here is to identify the useful structural and kinematic parameters that aid comparison with other galaxies and place our Galaxy in context. These 
``measurables'' are also important for comparing with numerical simulations of synthetic galaxies. 
A simulator runs a disk simulation and looks to compare the evolutionary phase where 
the bar/bulge instability manifests itself.
In principle, only a statistical `goodness of fit' is needed without resorting to any parametrisation 
\citep{Sharma2011}. But in practice, the comparison is likely to involve 
global properties like rotation curves, scale lengths, total mass (gas, stars, dark matter),
stellar abundances and the star formation history. Here we focus on establishing what are
the best estimates for the Galaxy's global properties and the parameters that describe
its traditional components.

\cite{Kerr1986} revisited the 1964 IAU standards for the Sun's distance
and circular velocity ($\Rzero = 10$ kpc; $\Vcirc = 250$\kms) relative to the Galactic Centre 
and proposed a downward revision ($\Rzero = 8.5$ kpc; $\Vcirc = 220$\kms). 
Both values can now be revised further to reflect the
new observational methods at our disposal three decades on. Our
goal in this review is to provide an at-a-glance summary of the key structural
and kinematic parameters to aid the increasing focus on Galactic studies. 
For reasons that become clear in later sections, we cannot yet provide
summary values that are all internally consistent in a plausible
dynamical description of the Galaxy; this is an important aim for the
next few years. For more scientific context, we recommend reviews over the past
decade that consider major components of the Galaxy: 
\cite{Helmi2008}, \cite{Ivezic2012}, \cite{Rix2013} and \cite{Rich2013}.

This is an era of extraordinary interest and investment in Galactic studies exemplified
above all by the \gaia\ astrometric mission and many other space-based and ground-based surveys.
In the next section (\S 2), we provide a context for these studies.
The Galaxy is then described in terms 
of traditional components: Galactic Centre (\S 3), Inner Galaxy (\S 4), 
Disk Components (\S 5) and Halo (\S 6). Finally, we discuss the likely developments
in the near term and attempt to provide some pointers to the future (\S 7).

\section{THE GALAXY IN CONTEXT}

We glimpse the Galaxy at a moment in time when globally averaged star formation 
rates (SFR) are in decline and nuclear activity is low. In key respects, the Milky Way is 
typical of large galaxies today in low density
environments \citep{Kormendy2010}, especially with a view to global parameters 
(e.g. current SFR \Msfr, baryon fraction \fbary) given its total stellar mass \citep[cf.][]{deRossi2009},
as we discuss below.
But in other respects, it is relatively unusual, caught in transition between the
`red sequence' of galaxies and the `blue cloud' \citep{Mutch2011}. Moreover, 
unlike M31, our Galaxy has not experienced a major merger for the past 10 Gyr
indicating a remarkably quiescent accretion history for a luminous galaxy \citep{Stewart2008}. 
Most \Lstar\ galaxies lie near the turnover of the galaxy luminosity
function where star formation quenching starts to become effective \citep{Benson2003}. 
But this may not be the last word for the Galaxy: 
the Magellanic gas stream is evidence for very substantial ($\sim 10^9$\ \Msun)
and ongoing gas accretion in the present day \citep{Putman1998,Fox2014}.
The Galaxy stands out in another respect: it is uncommon for an $L_\star$ galaxy 
to be orbited by two luminous dwarf galaxies that are both forming stars \citep{Robotham2012}.
None the less, the Galaxy will always be the most
important benchmark for galaxy evolution because it provides
information that few other galaxies can offer $-$ the fully resolved constituents that make 
up an $L_\star$ galaxy in the present epoch.

\subsection{Environment \& Evolution}
The Galaxy is one of the two dominant members of the Local Group, a low-mass system 
constituting a loosely bound collection of spirals and dwarf galaxies. The Local Group has an 
internal velocity dispersion of about 60\kms\ \citep{vandenBergh1999} and is located in a low-density 
filament in the far outer reaches of the
Virgo supercluster of galaxies \citep{Tully2014}. Galaxy groups with one or two dominant spirals 
are relatively common, but close analogues of the Local Group are rare. The 
presence of an infalling binary pair $-$
the Small and Large Magellanic Clouds (SMC, LMC) $-$ around an $L_\star$ galaxy is only seen 
in a few percent of cases in the Galaxy and Mass Assembly (\GAMA) survey \citep{Driver2011}. 
This frequency drops to less than 1\% if we add the qualification that the 
massive binary pair are actively forming stars.

With a view to past and future evolution, it is instructive to look at numerical simulations of the 
Local Group.
The Constrained Local Universe Simulations $-$ the \CLUES\ project (www.clues-project.org) $-$ 
are optimised for a study of the formation of the Local Group within a cosmological
context \citep{Forero-Romero2013,Yepes2014}. The accretion history of
the Local Group is relatively quiet, consistent with its cold internal dynamics. The largest
simulations with the most advanced prescriptions for feedback \citep[e.g. \FIRE,][]{Hopkins2014} 
are providing new insight on why only a small fraction of the dark minihalos
in orbit about the Galaxy are visible as dwarf galaxies \citep{Wetzel2016}.
But we are still a long way from a detailed understanding of how the dark matter and baryons
work together to produce present day galaxies.

The future orbital evolution and merging of the Local Group has been 
considered by several groups \citep{Cox2008,vanderMarel2012a,Peebles2013}.
These models are being successively 
refined as  proper motions of stars in M31 become available.
We learn that the Galaxy and M31 will reach pericentre passage in about 4 Gyr and 
finally merge in $\sim$6 Gyr. The models serve to remind us that the Galaxy is undergoing at 
least three strong interactions (LMC, SMC, Sgr) and 
therefore cannot be in strict dynamical equilibrium.
This evolution can be accommodated
in terms of structural and kinematic parameters (adiabatic invariants)
that are slowly evolving at present \citep{Binney2013}. 


\begin{table}
\caption{ \label{t:licquia1} Global magnitudes, colour indices and mass-to-light ratios
for the Galaxy.}
\begin{center}
\vspace{0.3cm}
\begin{tabular}{|llllll|} 
\hline
Absolute magnitude$^a$, $M_{\Delta\lambda}$                  & $u$ &  $g$ &  $r$  &  $i$  & $z$ \\               
                                                                                     & -19.87 & -21.00 & -21.64 & -21.87 & -22.15 \\ 
                                                                                     &  $U$ &  $B$ &  $V$ & $R$ & $I$ \\
                                                                                     & -20.67 & -20.70 & -21.37 & -21.90 & -22.47 \\ 
                                                                                    \hline
Colour index$^b$, ($M_{\Delta\lambda_1}$$-$$M_{\Delta\lambda_2}$) & {\it u$-$r} & {\it u$-$g} & {\it g$-$r} & {\it r$-$i} & {\it i$-$z} \\
                                                                                        & 1.96 & 1.29 & 0.65 & 0.28 & 0.28 \\
                                                                                       & {\it U$-$V} & {\it U$-$B} & {\it B$-$V} & {\it V$-$R}  & {\it R$-$I}  \\
                                                                                       & 0.86 & 0.14 & 0.73 & 0.54 & 0.58 \\
                                                                                     \hline
Mass-to-light ratio, $\Upsilon^\star$                             &  $u$ &  $g$ &  $r$  &  $i$  & $z$ \\
                                                                                     & 1.61 & 1.77 & 1.50 & 1.34 & 1.05 \\ 
                                                                                     &  $U$ &  $B$ &  $V$ & $R$ & $I$ \\
                                                                                     & 1.66 & 1.73 & 1.70 & 1.45 & 1.18 \\ 
                                                                                   \hline
\end{tabular}
\end{center}
\vspace{0.2cm}
\begin{tabnote} 
Magnitudes and colours derived for the Galaxy from Milky Way analogues drawn from the \SDSS\
survey using the Kroupa initial mass function.
All values assume $\Rzero = 8.2$ kpc and $\Rdisk = 2.6$ kpc for the Galaxy.
$^a$The \SDSS\ and Johnson 
photometry are calibrated (typical errors $\sim$ 0.1 mag) with respect to the AB and Vega magnitude systems, respectively;
$^b$Different calibration schemes are needed for the \SDSS\ total 
magnitudes and the unbiassed galaxy colours which leads to
inconsistencies between magnitude differences and colour indices \citep[courtesy of][]{Licquia2015}.
\end{tabnote}
\end{table}

\begin{figure}
\centering
\parbox{6cm}{
\includegraphics[width=6cm]{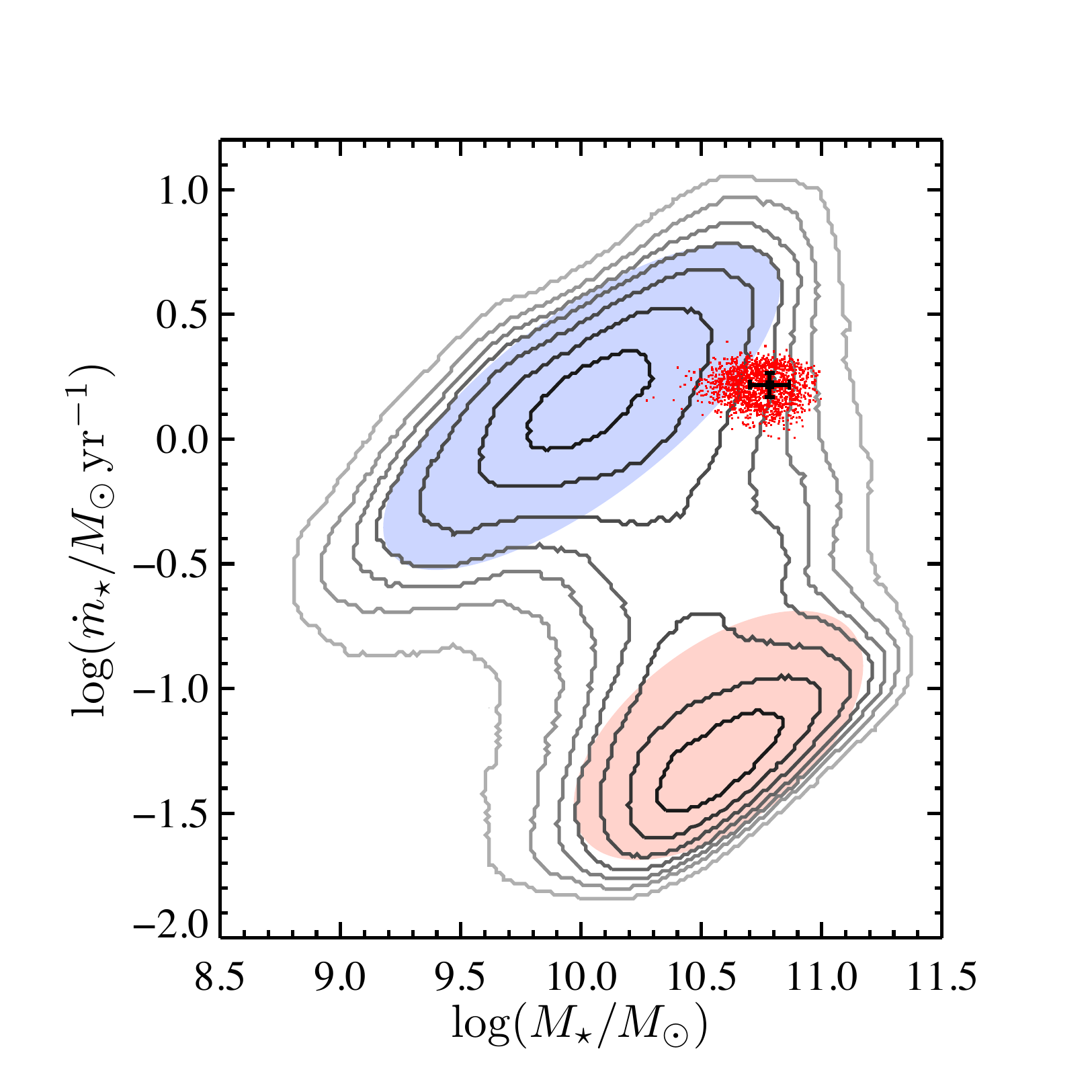}
}
\qquad
\begin{minipage}{6cm}
\includegraphics[width=6cm]{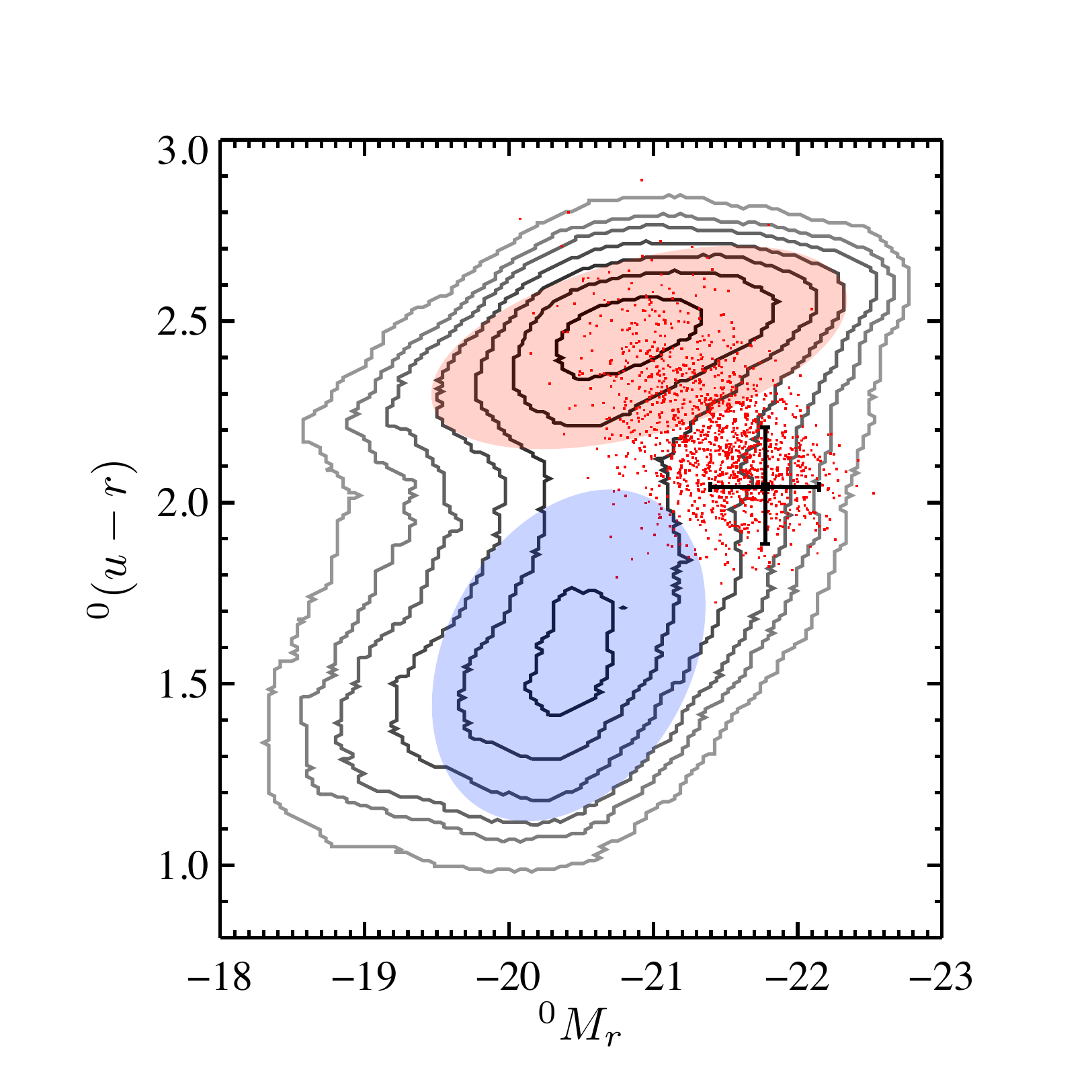}
\end{minipage}
\caption{Left: Sample of 3500 Milky Way analogues (red dots) drawn from the \SDSS survey 
with the same stellar mass \Mstar\ and star formation rate \Msfr, given the
measurement uncertainties. The Galaxy indicated by a cross resides in the `green valley' 
just below the `blue cloud' (see also \cite{Mutch2011}).
Right: When mapped to absolute magnitude-colour space, 
the Milky Way analogues are dispersed over a wider region extending into the red sequence. 
The error bar is noticeably offset from the centre of the distribution due to corrections for inclination reddening and Eddington bias.  The increased scatter compared to the left panel is a consequence of 
the broad range of $g-r$ colours that can correspond
to a specific star formation rate. The magnitude and color superscript ``0" indicates that the results are 
corrected to $z=0$. In both figures, the contours indicate the density of galaxies
in the projected plane \citep[courtesy of][]{Licquia2015}.
}
\label{f:licquia}
\end{figure}

\subsection{Galaxy classification \& integrated properties}

For most of the last century, galaxy studies made use of morphological classification
to separate them into different classes. But high-quality, multi-band photometric 
and spectroscopic surveys have provided us with a different perspective
\citep{Blanton2009}. The bulge
to disk ratio, e.g. from Sersic index fitting to photometric images, continues to play an important 
role in contemporary studies \citep{Driver2011}. 
Many observed properties correlate with the total stellar mass
\Mstar\ and the global star formation rate \Msfr. Galaxies fall into a `red sequence'
where star formation has been largely quenched, and a `blue cloud' (`main sequence')
where objects are actively forming stars, with an intervening `green valley.'  \cite{Mutch2011}
first established that the Galaxy today appears to fall in the green valley, much like
M31 interestingly. Their method is based on the Copernican principle that the Galaxy is
unlikely to be extraordinary given global estimates for \Mstar\ and \Msfr. In large
galaxy samples, these quantities
are expected to correlate closely with photometric properties like absolute magnitude
and color index with a scatter of about 0.2 dex. 
In Fig.~\ref{f:licquia}, we show the results of the most recent analysis of this kind 
\citep{Licquia2015}.

\cite{Flynn2006} recapitulate the long history in deriving Galaxy global properties. 
\cite{Chomiuk2011} present an exhaustive study across many wavebands
and techniques to arrive at a global star formation rate \Msfr\ for the Galaxy. They attempt to 
unify the choice of initial mass function and stellar population synthesis model across methods, 
and find \Msfr\ $\approx$ 1.9 \Msun\ \yr, within range of the widely quoted value
of $1-3$ \Msun\ \yr\ by \cite{McKee1997}. \cite{Licquia2015a} 
revisit this work and conclude that \Msfr\ $=$ $1.65\pm 0.19$ \Msun\ \yr\ for an adopted
Kroupa IMF. Likewise, there
are many estimates of the Galaxy's total stellar mass, most from direct integration of starlight and
an estimate of the mass-to-light ratio, \ML (see below), from which they estimate
\Mstar\ $=$ $6\pm 1\times 10^{10}$ \Msun\  (\S6) for a Galactocentric 
distance of \Rzero\ $\approx$ 8.3 kpc. In \S\ref{s:rotcur}, we estimate a total stellar mass of 
\Mstar\ $=$ $5\pm 1\times 10^{10}$ \Msun\ (and a revised \Rzero) from combining estimates of 
the mass of the
bulge and the disk from dynamical model fitting to stellar surveys
and to the Galactic rotation curve.

To transform these quantities into the magnitude system, one method is to select a large set of 
disk galaxies from the \SDSS photometric survey with measured bolometric 
properties over
a spread in inclination and internal extinction. \cite{Licquia2015} select possible analogues that 
match \Mstar\ and \Msfr\ in the Galaxy given the uncertainties. 
In Fig.~\ref{f:licquia}$a$, we present the
total absolute magnitudes and unbiassed galaxy colors for the analogues
using the \SDSS $ugriz$ bands; most appear to fall in the green valley, in agreement with \cite{Mutch2011}.
Given these data,
the likely values for the Galaxy are presented in Table~\ref{t:licquia1} without the uncertainties. 
Note the lack of consistency between the quoted \SDSS magnitudes and colors because they 
are derived using different calibration methods as recommended for \SDSS DR8 
\citep{Aihara2011},
but the differences are mostly below the statistical errors quoted by \cite{Licquia2015}. 
In Fig.~\ref{f:licquia}$b$,
the analogues show a lot of scatter when transformed to color-magnitude space where the green valley is
less well defined. While the Galaxy's location has moved, it
still resides in the green valley.

The same inconsistency is seen in the Johnson magnitudes (Table~\ref{t:licquia1}) 
which are derived via color transformations from the \SDSS magnitudes \citep{Blanton2007}. The
 {\tt cmodel} and {\tt model} magnitudes in $ugriz$ bands are converted to an equivalent set of $UBVRI$ 
 magnitudes and color indices respectively (see http://www.sdss.org/dr12/algorithms/magnitudes/\#mag\_model).
 This results in small differences between the color derived 
 from single-band absolute magnitudes as opposed to the color inferred from the best \SDSS color measurement, but these differences mostly fall below the uncertainties.
 
For either the \SDSS or the Johnson magnitudes, there are no earlier works that cover the five bands;
most studies concentrate on $B$ and $V$ \citep{deVaucouleurs1978,Bahcall1980}.
\cite{deVaucouleurs1983}, updating his earlier work,
derived $M_B=-20.2\pm 0.15$ assuming a distance of $\Rzero = 8.5\pm 0.5$ kpc
and a color term $B-V=0.53\pm 0.04$, similar to values derived by \cite{Bahcall1980}: $M_B=-20.1$,
$M_V=-20.5$ and $B-V=0.45$. \cite{vanderKruit1986} took the novel approach of using observations from
the {\sl Pioneer} probes en route to Jupiter and beyond to measure optical light from the Galaxy and found
$M_B=-20.3\pm 0.2$ and $B-V=0.83\pm 0.15$.  These values are mostly dimmer and bluer than the 
modern values in Table~\ref{t:licquia1}. The latter find strong support with the dynamically determined
$I$-band values ($M_I=-22.4$, $\Upsilon^\star_I=1.4$) from \cite{Piffl2014a}, and ($M_I=-22.5$, $\Upsilon^\star_I=1.3$) from \cite{Bovy2013}, and are in close agreement with \cite{Flynn2006}. 

It is clear from Fig.~\ref{f:licquia} that the Milky Way is a very luminous, reddish galaxy, somewhat at odds with its traditional classification as an Sb$-$Sbc galaxy. This has raised questions
in the past about where it falls on the Tully-Fisher relation \citep[e.g.,][]{Malhotra1996,Hammer2012}. 
But this high luminosity 
is consistent with its high circular velocity ($\sim 240$\kms) discussed in \S\ref{s:rotcur}. In Table~\ref{t:licquia1},
the Galaxy is slightly fainter in $ugriz$ absolute magnitude and slightly less massive in both stellar and baryonic mass than the average for its rotation speed. When considering both the current uncertainties in Milky Way properties and the scatter of galaxies about the relation, the Galaxy is consistent with the Tully-Fisher relation to better than $1\sigma$ uncertainty. The Milky Way appears to be a kinematically typical spiral galaxy
for its intrinsic luminosity.

\begin{marginnote}
\entry{$M_{\Delta\lambda}$}{Galaxy absolute magnitude; Table~\ref{t:licquia1}}
\entry{$M_{\Delta\lambda_1}-M_{\Delta\lambda_2}$}{Galaxy color index; Table~\ref{t:licquia1}}
\entry{$\Upsilon^\star_{\Delta\lambda}$}{Galaxy mass-to-light ratio; Table~\ref{t:licquia1}}
\entry{\Msfr}{$1.65\pm 0.19$ \Msun\ \yr; Galaxy total star formation rate}
\end{marginnote}

\section{Galactic Center} 
\label{s:GC}

\subsection{Location}
\label{s:location}

The Galactic Center, as we understand it today, was first identified through the discovery of Sgr A by radio astronomers
\citep{Piddington1951}. Based on its unique radio emission properties and its precise coincidence with the dynamical center
of the rotating inner \HI\ disk \citep{Oort1960}, the IAU officially adopted Sgr A as the center of the Galaxy, making
its position the zero of longitude and latitude in a new system of Galactic coordinates \citep{Blaauw1959}. Later
\citet{Balick1974} discovered the unresolved source Sgr A$^*$, now known to be at the location of the Milky Way's
supermassive black hole, at $(l_c,b_c)\narreq(-0.056\dg, -0.046\dg)$ \citep{Reid2004}.

Thirty years ago, \citet{Kerr1986} gave this working definition of the Galactic Center: {\sl Currently it is
    assumed that the Galactic Centre coincides sufficiently well with the Galaxy's barycentre that a distinction
  between the point of greatest star density (or any other central singularity) and the barycentre (centre of mass) is
  not necessary. It is also assumed that to sufficient accuracy for the internal dynamics of the Galaxy, the
  Galactic Centre defines an inertial coordinate system. These assumptions could prove to be untrue if for instance the
  centre of the distribution of the mysterious mass in the heavy halo were displaced from the mass centre of the visible
  Galaxy.}

Indeed, in a hierarchical universe the first assumption is almost certainly violated beyond tens of kpc: we will see in
\S~\ref{s:halo} that the Galaxy continues to accrete satellite galaxies carrying both stars and dark matter;
furthermore, the Milky Way's dark matter halo interacts both with infalling dark matter and with other halos in the
Local Group. However, as discussed below, the inner Milky Way appears to have ``settled down'' to a well-defined and
well-centered midplane; thus we may assume that the region of greatest star density coincides with the barycenter of the
mass within the Solar Circle.

The second part of the definition may ultimately also come into question.  Numerical simulations reveal that dark matter
halos tumble at the level of a few radians per Hubble time.  The baryonic components are largely bound
to the dark matter but may slosh around within them. Furthermore, the spin axis of the Galactic Plane is likely to
precess with respect to a celestial coordinate frame defined by distant quasars or radio sources. Over the lifetime of
the \gaia mission, this precession ($\sim 30\;\mu$as yr$^{-1}$) should be detectable \citep{Perryman2014}.

\begin{marginnote}
\entry{Galactic Center}{Location of radio source Sgr A$^*$}
\entry{$(l_c,b_c)$}{$(-0.056\dg, -0.046\dg)$, Galactic coordinates of Sgr A$^*$}
\entry{SMBH}{Milky Way's supermassive black hole}
\end{marginnote}

\subsection{Distance}
\label{s:distance}

The distance of the Sun to the Galactic Center, $R_0$, is one of the fundamental scaling parameters for the Galaxy.  All
distances determined from angular sizes or from radial velocities and a rotation model are proportional to $R_0$. Also
the sizes, luminosities, and masses of objects such as molecular clouds scale with $R_0$, as do most estimates of global
Galactic luminosity and mass. Because $R_0$ is one of the key parameters, we consider its measured value in some detail
here. Previous reviews on this subject can be found in \citet{Kerr1986, Reid1993, Genzel2010, Gillessen2013}, and a
recent compilation of results is given in \citet{Malkin2013}.

Similar to their discussion, we divide methods of determining $R_0$ into direct (primary), model-based, and 
secondary. Direct methods compare an angular dimension or velocity near the Galactic Center with a physical length
scale or radial velocity (RV), with minimal modelling assumptions and without having to use additional calibrations.
Model-based methods determine $R_0$ as one of the model parameters through a global fit to a set of data.  Secondary
methods finally use standard candle tracers whose distances are based on secondary calibrations such as
period-stellar luminosity relations, and whose distributions are known or assumed to be symmetric with respect to the
Galactic Center. In the following, we briefly review the different methods.  Table~\ref{tableR0} gives the list of
independent recent determinations of $R_0$ which we use for obtaining an overall best estimate below.

\subsubsection{Direct estimates}
\label{s:direct}
As discussed more fully in \S\S\ref{s:blackhole}, \ref{s:rotcur}, the \smbh is at rest at the dynamical center of the
Milky Way within the uncertainties \citep{Reid2004, Reid2008}. Thus $R_0$ can be determined by measuring the distance to
the \smbh's radiative counterpart, Sgr A$^*$. At $\sim\!8\kpc$ distance, the expected parallax of Sgr A$^*$ is $\sim\! 100
\micas$. This would be resolvable with Very Long Baseline Interferometry (\vlbi), but unfortunately the image of Sgr
A$^*$ is broadened by interstellar scattering \citep[e.g.,][]{Bower2004}. However, \citet{Reid2009SgrB2} measured {\rm
  trigonometric parallaxes of H$_2$O masers in Sgr B2}, a molecular cloud complex which they estimated is located $\sim\!
130\pc$ in front of Sgr A$^*$.

A second direct estimate of the distance to the \smbh comes from monitoring proper motions (\propm) and line-of-sight
(LOS) velocities for {\rm stellar orbits near Sgr A$^*$} \citep{Eisenhauer2003, Ghez2008, Gillessen2009, Morris2012}, in
particular the star S2 which by now has been followed for a complete orbit around the dynamical center. The main
systematic uncertainties are source confusion, tying the \smbh to the astrometric reference frame, and the potential
model; relativistic orbit corrections lead to an increase of $R_0$ by $\sim\!0.1\kpc$ \citep{Genzel2010}.  See also
\citet{Gillessen2009a} who combined the existing two major data sets in a joint analysis.

The {\rm statistical parallax of the nuclear star cluster (\nsc)} obtained by comparing stellar \propm\ and LOS has been
used as a third direct estimate of $R_0$ \citep{Genzel2000,Trippe2008,Do2013}. This method has now become accurate
enough that projection and finite field-of-view effects in combination with orbital anisotropies need to be modelled
\citep{Chatzopoulos2015}. 

\subsubsection{Model-based estimates}
\label{s:modelbased}
\vlbi astrometry has provided accurate {\rm parallaxes and proper motions} for over $100$ OH, SiO, and Methanol masers in
High Mass Star Formation Regions (\hmsfr) in the Galactic disk \citep{Honma2007, Honma2012, Reid2009a, Reid2014a,
  Sato2010}. Most of these sources are located in the Galactic spiral arms. By fitting a spiral arm model together with
a model for Galactic circular rotation to the positions and velocities of the \hmsfr, precise estimates of $R_0$ can be
obtained together with rotation curve and other parameters \citep[for a different
analysis, see][]{Bajkova2015}. Because distances are determined geometrically, no assumptions on tracer luminosities or
extinction are necessary. The main remaining systematic uncertainties thus are the assumption of axisymmetric rotation,
the detailed parametrisation of the rotation curve model, and the treatment of outliers.

Another long-standing approach is based on analyzing the {\rm velocity field near the Solar Circle}, using \propm\ and
LOS velocities of various tracers. These methods assume an axisymmetric velocity field and, in some cases, include the
perturbing effects of a spiral arm model. In the traditional form, the data are used to solve for the Oort constants A,
B (\S\ref{s:rotcur}) from the PMs and for $2AR_0$ from the RVs, to finally estimate $R_0$ \citep{Mihalas1981,
  Zhu2013}. Another variant is to use young stars or star formation regions assumed to follow circular orbits precisely
\citep{Sofue2011, Bobylev2013}.  An analysis less sensitive to the uncertain perturbations from spiral arms and other
substructure is that by \citet{Schonrich2012}, who uses a large number of stars with PMs and RVs from the \segue survey
\citep{Yanny2009}, to determine $R_0$ by combining the rotation signals in the radial and azimuthal velocities for
stars within several kpc from the Sun.

A promising new method is based on the detailed {\rm dynamical modelling of halo streams}. While stream modelling has
mostly been used to obtain estimates for the mass and shape of the dark matter halo (see \S~\ref{s:darkhalo}),
\citet{Kupper2015} showed that with accurate modelling $R_0$ can also be well-constrained as part of a multi-parameter
fit to detailed density and LOS velocity measurements along the stream.

\citet{Vanhollebeke2009} compared predictions from a {\rm stellar population model for the Galactic bulge} and
intervening disk to the observed star counts. They used a density model based on NIR data from \citet{Binney+97} and varied the
star formation history and metallicity distribution of bulge stars. The value of $R_0$ from their best fitting models
depends heavily on the magnitudes of red clump stars, showing a close connection to secondary methods.


\begin{table}
\tabcolsep5pt
\caption{Recent measurements of distance $R_0$ to Galactic Center}
\label{tableR0}
\begin{center}
\begin{tabular}{@{}l|c|c|c|c|c@{}}
\hline
{\rm Label} & {\rm Reference} & {\rm Method} & {\rm Loc} & {\rm T } & {\rm $R_0$ [kpc]} \\
\hline
Rd+09 &  Reid et al.            2009  &  Trig. Parallax of Sgr B           &   GC  &   d   &  $ 7.90 \pm 0.75 $ \\
Mo+12 &  Morris et al.          2012  &  Orbit of S0-2 around Sgr A*       &   GC  &   d   &  $ 7.70 \pm 0.40 $ \\
Gi+09 &  Gillessen et al.       2009  &  Stellar orbits around Sgr A*      &   GC  &   d   &  $ 8.33 \pm 0.35 $ \\
Ch+15 &  Chatzopoulos et al.    2015  &  NSC statistical parallax          &   GC  &   d   &  $ 8.27 \pm 0.13 $ \\
Do+13 &  Do et al.              2013  &  NSC statistical parallax          &   GC  &   d   &  $ 8.92 \pm 0.56 $ \\
BB15  &  Bajkova \& Bobylev     2015  &  Trig. Parallaxes of HMSFRs        &  DSN  &   m   &  $ 8.03 \pm 0.32 $ \\
Rd+14 &  Reid et al.            2014  &  Trig. Parallaxes of HMSFRs        &  DSN  &   m   &  $ 8.34 \pm 0.19 $ \\
Ho+12 &  Honma et al.           2012  &  Trig. Parallaxes of HMSFRs        &  DSN  &   m   &  $ 8.05 \pm 0.45 $ \\
ZS13  &  Zhu \& Shen            2013  &  Near-R0 rotation yg tracers       &  DSN  &   m   &  $ 8.08 \pm 0.62 $ \\
Bo13  &  Bobylev                2013  &  Near-R0 rotation SFRs+Cephs       &  DSN  &   m   &  $ 7.45 \pm 0.66 $ \\
Sch12 &  Sch\"onrich            2012  &  Near-R0 rotation SEGUE stars      &  DSN  &   m   &  $ 8.27 \pm 0.41 $ \\
Ku+15 &  K\"upper et al.        2015  &  Tidal tails of Pal-5              &   IH  &   m   &  $ 8.30 \pm 0.35 $ \\
VH+09 &  Vanhollebeke et al.    2009  &  Bulge stellar popul. model        &   B   &   m   &  $ 8.70 \pm 0.50 $ \\
Pi+15 &  Pietrukowicz et al.    2015  &  Bulge RR Lyrae stars              &   B   &   s   &  $ 8.27 \pm 0.40 $ \\
De+13 &  Dekany et al.          2013  &  Bulge RR Lyrae stars              &   B   &   s   &  $ 8.33 \pm 0.15 $ \\
Da09  &  Dambis                 2009  &  Disk/Halo RR Lyrae stars          &  DSN  &   s   &  $ 7.58 \pm 0.57 $ \\
Ma+13 &  Matsunaga et al.       2013  &  Nuclear bulge T-II Cepheids       &   B   &   s   &  $ 7.50 \pm 0.60 $ \\
Ma+11 &  Matsunaga et al.       2011  &  Nuclear bulge Cepheids            &   B   &   s   &  $ 7.90 \pm 0.36 $ \\
Gr+08 &  Groenewegen et al.     2008  &  Bulge Cepheids                    &   B   &   s   &  $ 7.98 \pm 0.51 $ \\
Ma+09 &  Matsunaga et al.       2009  &  Bulge Mirae                       &   B   &   s   &  $ 8.24 \pm 0.43 $ \\
GrB05 &  Groenewegen \& Bl.     2005  &  Bulge Mirae                       &   B   &   s   &  $ 8.60 \pm 0.81 $ \\
FA14  &  Francis \& Anderson    2014  &  Bulge red clump giants            &   B   &   s   &  $ 7.50 \pm 0.30 $ \\
Ca+13 &  Cao et al.             2013  &  Bulge red clump giants            &   B   &   s   &  $ 8.20 \pm 0.20 $ \\
Fr+11 &  Fritz et al.           2011  &  NSC red clump giants              &   GC  &   s   &  $ 7.94 \pm 0.76 $ \\
FA14  &  Francis \& Anderson    2014  &  All globular clusters             &  BIH  &   s   &  $ 7.40 \pm 0.28 $ \\
Bi+06 &  Bica et al.            2006  &  Halo globular clusters            &   IH  &   s   &  $ 7.10 \pm 0.54 $ \\
\hline
\end{tabular}
\end{center}
\begin{tabnote}
  Recent determinations of the distance to the Galactic Center, $R_0$, used for weighted averages and in
  Fig.~\ref{f:R0}.  Columns give label for Fig.~\ref{f:R0}, reference, location (Galactic Center, bulge, disk and
  solarneighbourhood, inner halo), type of measurement (direct, model-based, secondary), $R_0$.  The listed
  uncertainties include statistical and systematic errors, added in quadrature when both are published.  Where a
  systematic error is not quoted by the authors or included in their total published error, we estimated it from results
  obtained by them under different assumptions, when possible (GrB05, Fr+11, BB15, Rd+14, Bo13, Bi+06). For Ca+13 who
  did not give an error, we estimated one from the metallicity dependence of the \rcg calibration of
  \citet{Nataf2013}. In the remaining cases (ZS13, Sch12, Kue+15, Da09) we took the systematic error to be equal to the
  statistical error.
\end{tabnote}
\end{table}

\subsubsection{Secondary estimates}
\label{s:secondary}
There is a long history of secondary $R_0$ measurements based on the distance distributions of RR Lyrae stars, Cepheids,
Mira stars, red clump giants (\rcg), and globular clusters \citep{Reid1993}. Individual distances are derived from
external calibrations of period-luminosity (PL) relations for the variable stars, and from horizontal branch (HB) magnitudes
for \rcg and globular clusters. The main systematic errors in these methods come from uncertainties in the
calibrations, but also from the extinction corrections and reddening law, and from how the distance distribution in the
survey volume is related to the center of the Galaxy.

{\rm RR Lyrae}, pulsationally unstable HB stars with characteristic absolute magnitudes $M_I\narreq0.2$,
$M_K\narreq-0.5$ \citep{Catelan2004}, are the primary distance tracers for old, metal-poor populations.  Large numbers
of RR Lyrae stars identified by the \ogle \citep{Udalski2008} and \vvv surveys \citep{Minniti2010} were recently used to
map out the old metal-poor population in the bulge \citep[][respectively]{Pietrukowicz2015,Dekany2013}.  Individual
distances are more accurate by a factor $\sim\!\!2$ in the NIR than in optical data, due to higher precision and reduced
metallicity dependence in the PL-relation, and less sensitivity to reddening. With the large bulge samples the centroid
of the distribution can be determined more accurately than in earlier work
\citep[e.g.,][]{Groenewegen2008,Majaess2010}. On larger Galactic scales, \citet{Dambis2009} calibrated thick disk and
halo RR Lyrae populations separately by statistical parallax to then estimate $R_0$.

{\rm Type II Cepheids} trace old, typically metal-poor populations; they are brighter than RR Lyrae stars but much less
numerous. \citet{Groenewegen2008} used NIR data for 39 population II Cepheids in the bulge from the \ogle survey; a
related work is by \citet{Majaess2010}.  \citet{Matsunaga2011, Matsunaga2013} analysed 3 {\rm classical Cepheids} and 16
type II Cepheids from a NIR survey of the inner $\sim\!30\pc$ in the nuclear bulge.

\citet{Groenewegen2005} obtained a PL relation for 2691 {\rm Mira long period variables} in the \ogle bulge fields while
\citet{Matsunaga2009} studied 100 Miras in the nuclear bulge. For these red giants the extinction corrections are
smaller, but calibrations from the Large Magellanic Cloud (LMC) and globular clusters were needed to estimate $R_0$ from
these data.

Since the work of \citet{Paczynski1998cc}, {\rm red clump giants} (\rcg) have been recognized as important distance
probes in the Galaxy (see Girardi, this volume). \rcg are He-core burning stars with a narrow range of luminosities,
especially in medium-to-old age populations such as in the Galactic bulge. Typical absolute magnitudes are
$M_I\narreq-0.5$, $M_K\narreq-1.6$, with a dispersion $\sim\!0.1$-$0.2$ mag, and systematic effects due to age and
metallicity variations are relatively small and fairly well understood. Surveys towards the inner Galaxy are frequently
done in the NIR to minimize extinction \citep{Babusiaux2005, Nishiyama2006} or in the $I$-band \citep{Paczynski1998cc,
  Nataf2013}. The $K$-band studies tend to give slightly shorter distances. \citet{Fritz2011} used \rcg in several NIR
bands to determine a distance to the central \nsc.

The centroid of the distance distribution of {\rm globular clusters}, the basis for the famous early work by
\citet{Shapley1918}, continues to be used for estimating $R_0$ \citep{Bica2006, Francis2014}. These studies are based on
the catalogue of \citet[][and earlier]{Harris2010}, where individual distances are estimated from HB magnitudes
and reddening. The distance distribution of the clusters is somewhat asymmetric, and $R_0$ values found are on the low
side of the distribution in Table~\ref{tableR0}. Systematic effects could be due to missing clusters behind the Galactic
Center, or to errors in the HB magnitudes, e.g., from extinction uncertainties or stellar confusion in crowded cluster
fields \citep{Genzel2010}. However, the method provides a rare opportunity to estimate $R_0$ from Galactic halo tracers.

\begin{figure}[t]
\begin{center}
\includegraphics[width=12cm]{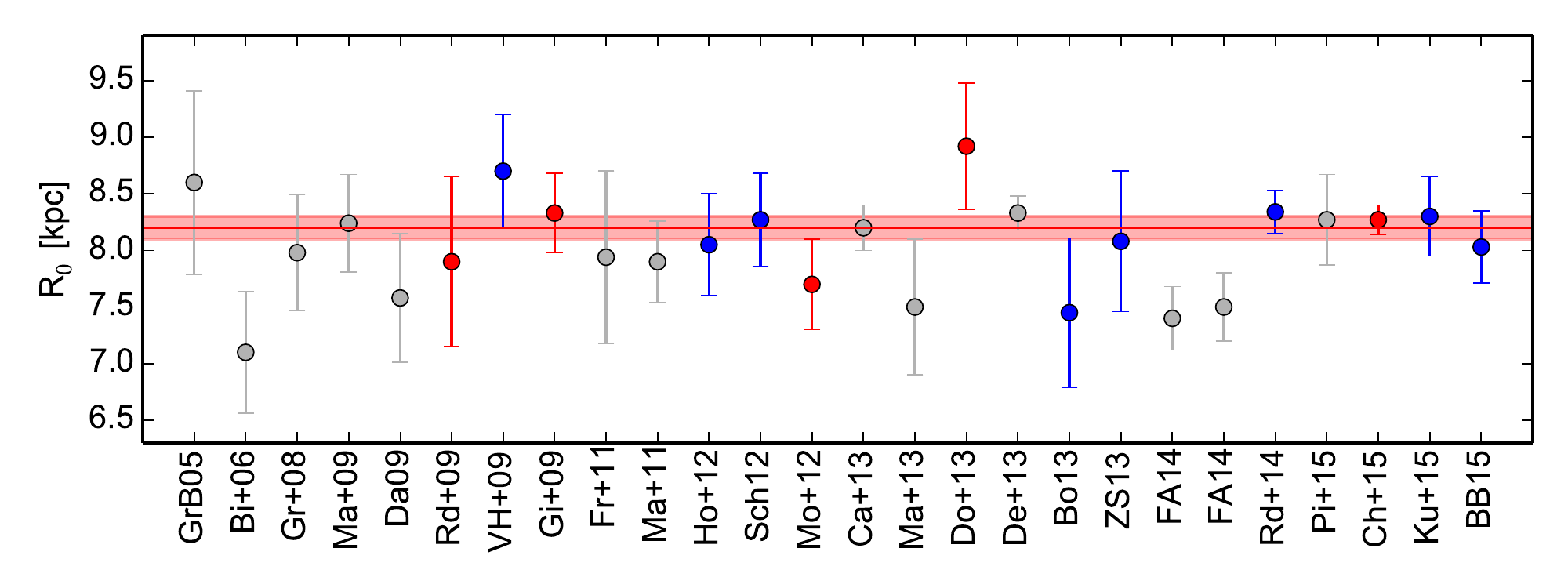}
\includegraphics[width=12cm]{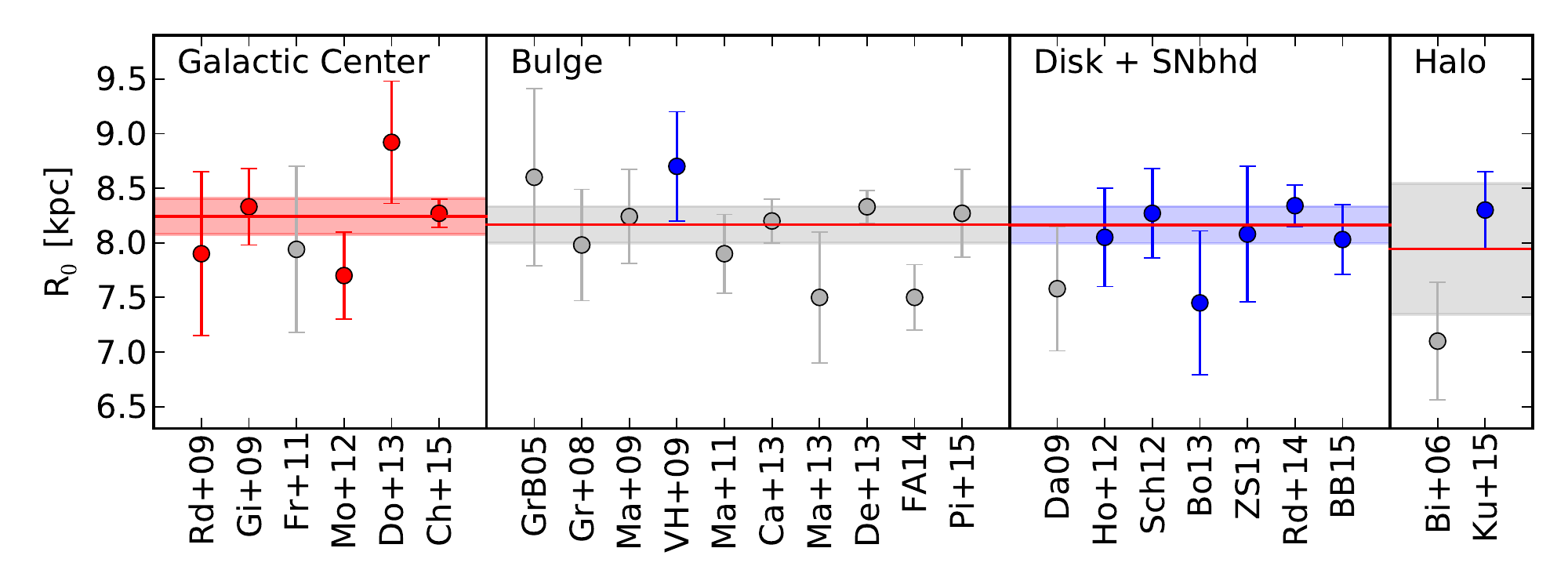}
\end{center}
\caption{Recent measurements of $R_0$ from Table~\ref{tableR0}, using different methods. Red, blue, and grey points
  denote direct, model-based, and secondary estimates. Top: time sequence for all, with our adopted best estimate,
  $R_0\narreq8.2\pm0.1\kpc$.  Bottom: separate time-sequences for determinations in the Galactic Center, bulge, disk and
  solar neighbourhood, and inner halo (not using the FA14 globular cluster value which includes the inner metal-rich
  clusters). The horizontal lines show weighted mean values for the respective components, and colored bands show
  $1\sigma$ \UUEWM-errors. See text for details.}
\label{f:R0}
\end{figure}

\subsubsection{Overall best estimate and discussion}
\label{s:bestR0}
Table~\ref{tableR0} gives the list of recent determinations which we use for obtaining an overall best estimate for
$R_0$.  For each method we kept at most three determinations to prevent overweighting often-employed techniques with
similar systematic uncertainties; for example, there are many determinations of $R_0$ using \rcg as standard candles.  We
omitted determinations which were later updated by the same group of authors based on improved data, but kept
independent reanalyses of published data by other authors. We also did not include $R_0$ estimates which used priors
based on measured values already taken into account \citep[e.g.][]{McMillan2011}.

Fig.~\ref{f:R0} shows the overall distribution of these $26$ measurements with time, and the separate distributions for
tracers in the Galactic Center, bulge, disk, and halo, respectively. To obtain a best estimate for $R_0$, we consider
weighted means for the total and various subsamples.  For any $N_m$ measurements, we compute (i) the standard error of
the weighted mean, \SEWM; (ii) the unbiassed standard error of the weighted mean (\UEWM, square root of $1/N_m$ times
the unbiassed weighted sample variance); (iii) following \citet{Reid1993}, we consider possible correlations between
some of the measurements.  These can arise, e.g., because HB or PL calibrations use very similar theoretical models or
calibrators, or are based on a common LMC distance; because two data sets, although independent, can only be obtained
for the same small number of stars; or because several measurements are all based on the assumption that the velocity
field near the Solar Circle is axisymmetric.  Clearly, some of these correlations have stronger influence than
others. For a conservative error evaluation, we retain $N_{uc}\narreq11$ independent measurements for the total sample,
and $N_{uc}\narreq(4,4,3,2)$ for the separate (GC, B, DSN, IH) tracer samples. We note that the sets of methods kept for
the different regions are largely independent.  In each case, we determine a \UUEWM uncorrelated sample error by
replacing $N_m$ in \UEWM\ by $N_{uc}$.

Using the notation $R_0 \pm \UUEWM [\UEWM, \SEWM]$, we find $R_0\narreq8.14 \pm 0.10\ [0.07, 0.06] \kpc$ for the total
sample; $R_0\narreq8.17 \pm 0.12\ [0.09, 0.07] \kpc$ for measurements 2013 or later; $R_0\narreq8.29 \pm 0.03 [0.03,
0.08] \kpc$ for the 4 measurements with errors $\le0.2\kpc$ (Ch+15, Rd+14, De+13, Ca+13); $R_0\narreq8.22 \pm 0.07\
[0.07, 0.07] \kpc $ for the 11 uncorrelated best determinations; $R_0\narreq8.21 \pm 0.08\ [0.05, 0.06] \kpc$ excluding
the $2$ values outside their $(2\sigma)$ of the overall weighted mean; and $R_0\narreq8.19 \pm 0.08 [0.06, 0.06] \kpc$
for all values excluding halo.  All data samples give very consistent results; the fact that \SEWM\ and \UEWM\
generally agree within $\sim\!20\%$ suggests that the errors given for the individual $R_0$ measurements are mostly
realistic.  Based on the scatter and \UUEWM\ errors of the various sample means, we adopt here our best estimate for the
distance to the Galactic Center: $R_0\narreq8.2\pm 0.1 \kpc$.  This value is significantly lower than the IAU standard
($R_0\narreq8.5\kpc$).

The weighted sample means for tracers in the different regions of the Galaxy are $R_0\narreq8.24 \pm 0.16\ [0.13, 0.11]
\kpc$ for the GC sample, $R_0\narreq8.17\pm 0.17\ [0.11, 0.09 ] \kpc$ for the B sample, $R_0\narreq8.16 \pm 0.17\ [0.11,
0.13] \kpc$ for the DSN sample, and $R_0\narreq7.95 \pm 0.60\ [0.60, 0.29] \kpc$ for the IH sample.  The fact that
tracers in the Galactic Center, bulge, and disk result in similar estimates for $R_0$ within small errors (also for
\UUEWM\ with largely independent respective methods) suggests that the different Galactic components are well-centered,
and thus that relative sloshing motions between these components must currently be unimportant. Together with the good
alignment of the Milky Way bar and \HI\ disk with the Galactic plane (\S~\ref{s:solaroffset}), this suggests that the
inner Milky Way has settled to a well-determined equilibrium state. Even the inner halo tracers on scales of $10-20\kpc$
are consistent with having the same Galactic Center as the bulge and disk, within $\sim\! 0.6\kpc$ error.

We anticipate significant improvements in many of these distance measurements based on data from the \gaia satellite,
which will provide accurate parallaxes and proper motions for large numbers of Milky Way stars.  These data will greatly
improve our dynamical understanding of the Galactic disk, but also of the bulge and bar outside highly extincted
regions, and lead to much-improved secondary calibrations, e.g., for RR Lyrae and \rcg.  The direct estimate of $R_0$
from stellar orbits around Sgr A$^*$ is expected to improve steadily as the time base line increases and more orbits can
be reliably constrained, and especially with accurate astrometric and spectroscopic monitoring of the next close pericenter
passage (2018 for the star S2).

\begin{marginnote}
\entry{$R_o$}{$8.2\pm0.1\kpc$, Sun's distance from Galactic Center}
\entry{$\zzero$}{$25\pm5\pc$, solar offset from local disk midplane}
\end{marginnote}

\subsection{Solar offset and Galactic plane}
\label{s:solaroffset}

Early estimates of the Sun's vertical position with respect to the Galactic Plane date back to \citet{VanTulder1942}'s
analysis of stellar catalogues, from which was determined $\zzero \narreq 14\pm 2$ pc towards the North Galactic pole.
In support of the early value, \citet{Conti1990} obtained $\zzero \narreq 15\pm 3$ pc using $150$ Wolf-Rayet stars
within 20 kpc of the Sun.  \PIONEER observations of the optical background light in the Galaxy indicated $\zzero \narreq
13\pm 3$ pc \citep{Toller1990}, and modeling the \COBE NIR surface brightness distribution resulted in $\zzero\narreq
14\pm 4$ pc \citep{Binney+97}.

But more expansive studies of the nearby disk show that these are underestimates.  \citet{Chen2001} demonstrated the
importance of correctly treating the larger-scale parameters of the disk population in such studies, using the \sdss
photometric survey.  While the radial scalelength is relatively unimportant, a vertically extended population with a
well established scaleheight is critical.  The following estimates are based on either OB stars, open clusters, or
optical star counts for a range of stellar populations, and are broadly consistent:
$\zzero \narreq 24\pm 3$ pc \citep{Stothers1974};
$\zzero \narreq 28\pm 5$ pc \citep{Pandey1988};
$\zzero \narreq 21\pm 4$ pc \citep{Humphreys1995};
$\zzero \narreq 27\pm 3$ pc \citep{Mendez1998};
$\zzero \narreq 28\pm 6$ pc \citep{Chen1999};
$\zzero \narreq 27\pm 4$ pc \citep{Chen2001};
$\zzero \narreq 24\pm 2$ pc \citep{Maiz-Apellaniz2001}.
While the last estimate from \hipparcos OB stars has the smallest error, the distribution of young stars may be more
sensitive to various perturbations, as illustrated by Gould's belt. Therefore we adopt here the best estimate from the
complete \sdss photometric survey, i.e. $\zzero \narreq 25\pm 5$ pc \citep{Juric2008}, which captures all these values.

\begin{figure}[t]
\begin{center}
\includegraphics[width=10cm]{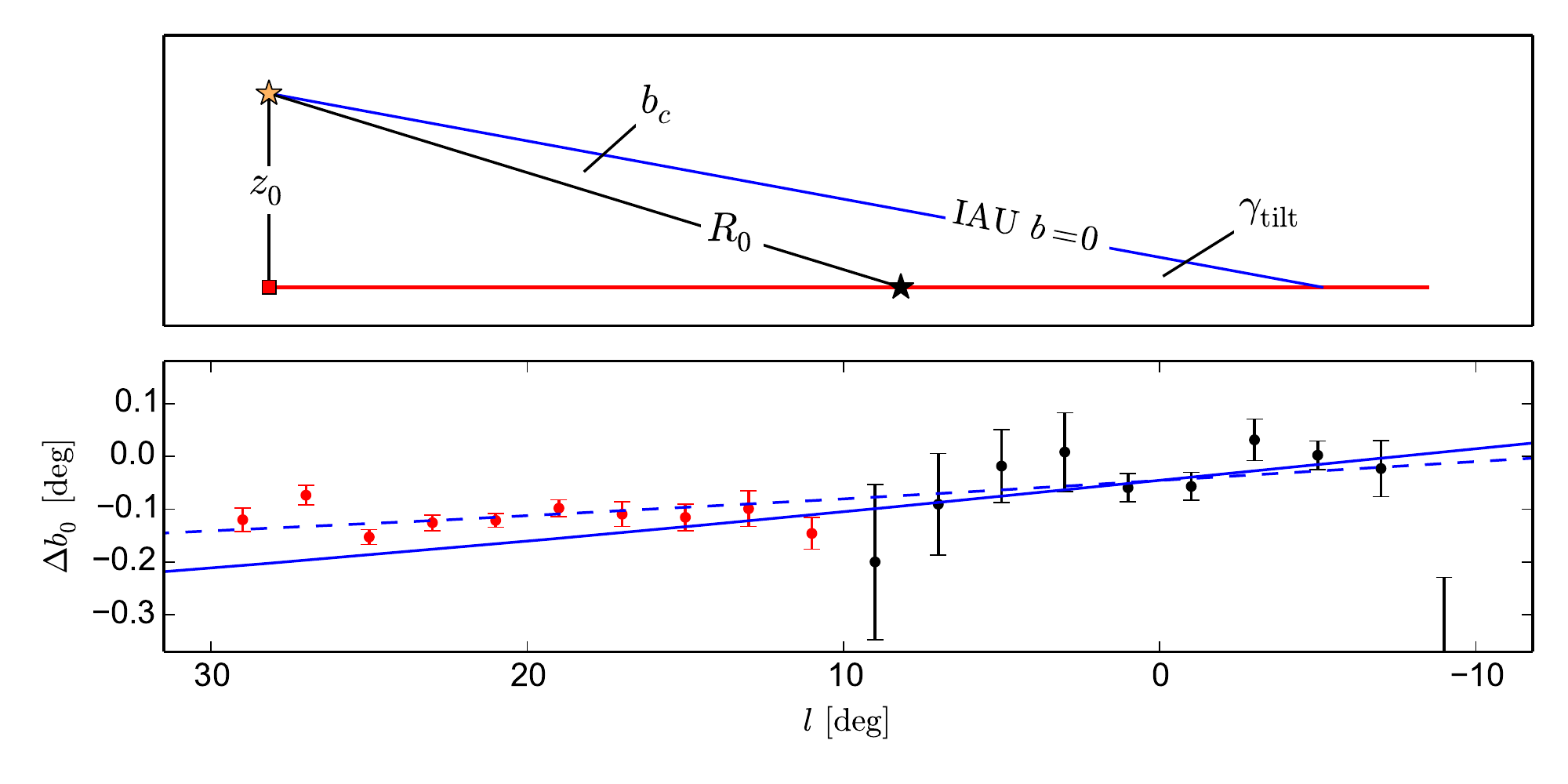}
\end{center}
\caption{Top: illustration of the tilt of the true Galactic disk plane (red) vs.\ the $b\narreq0-$plane (blue), as
  predicted from solar offset $\zzero$ and assuming that the true plane passes through Sgr A$^*$ (black star); after
  \citet{Goodman2014}. Bottom: measured offsets $\Delta b_0$ of bulge (black) and long bar (red) peak starcounts from
  the $b\narreq0-$plane.  The full line shows the predicted offsets for a one-dimensional bar with bar angle $27\dg$
  (\S\ref{s:innergalaxy}), lying precisely in the true Galactic plane defined by $\zzero\narreq25\pc$, $R_0\narreq8.2
  \kpc$, and $b_c \narreq -0.046\dg$, so that $\gamma_{\rm tilt}\narreq0.13\dg$. The dashed line is obtained when
  assuming an additional rotation of the true plane around the Sun-Galactic Center line by $0.14\dg$. Deviations from
  the plane in the long bar region are less than $\sim\!0.05\dg\simeq 5\pc$, of order $0.1\%$ of the length of the long
  bar. Adapted from \citet{Wegg2015}; see also \S\ref{s:longbar}. }
\label{f:galacticplane}
\end{figure}

The Galactic midplane was defined based on the very flat distribution of \HI\ gas in the inner Galaxy, with an estimated
uncertainty in the position of the Galactic pole of $\sim\!0.1\dg$ \citep{Blaauw1960}.  Because the Sun was found to lie
in the \HI\ principal plane within the errors, and the measured offset relative to Population I stars was not considered
reliable, the $b\narreq0$ plane was defined to pass through the Sun and Sgr A (not Sgr A$^*$). Since we know now that
$\zzero\simeq 25\pc$, the true Galactic plane is likely to be slightly inclined relative to the plane $b\narreq0$. If it
is assumed that Sgr A$^*$ lies precisely in the Galactic plane (see \S~\ref{s:blackhole}), the required inclination
angle is $\simeq 0.13\dg$ \citep[see Fig.~\ref{f:galacticplane} and][]{Goodman2014}.  Objects located in the Galactic
plane between the Sun and the Galactic Center then appear at slightly negative latitudes. Near-infrared star counts in
the inner Galaxy have enough signal-to-noise to detect such offsets.  The mean latitude of the peak of \rcg
counts in the Galactic long bar (see \S~\ref{s:longbar}) is indeed found at $b\simeq-0.12\dg$, corresponding to an
offset of $14\pc$ at $\sim\! 6\kpc$ distance \citep{Wegg2015}.  The observed peak latitudes agree with those predicted for
the inclined plane to within $\sim\! 5\pc$ or $\sim\! 0.1\%$ of the half-length of the bar
(Fig.~\ref{f:galacticplane}). The Galactic long bar is thus consistent with lying in a tilted midplane passing through
Sgr A$^*$ and the point $z_0\narreq25\pc$ below the Sun to within $\sim\!0.1\%$. Both stars and \HI\ gas suggest that the
Galactic disk inside the Solar Circle is very nearly flat.

\subsection{Black hole and solar angular velocity}
\label{s:blackhole}

Measurements in the Milky Way have provided the best evidence for a central \smbh to date; see
\citet{Reid2009rev} and \citet{Genzel2010} for recent reviews.  Infrared studies of the motions of gas clouds in the Sgr
A region first indicated a central point mass of several $10^6\Msun$ \citep[e.g.,][]{Lacy1980}. Later, \propm\
measurements of stars in the dense \nsc showed evidence for a Keplerian increase of the stellar velocity dispersion to
several $100\kms$ at $\sim\! 0.01\pc$ of the Galactic Center, corresponding to a central mass of $\sim\! 2-3\times10^6\Msun$
\citep{Eckart1997, Ghez1998}.  Currently, from accurate LOS velocities and astrometric measurements with adaptive optics,
stellar orbits have been determined for some 30 of the so-called S-stars in the central arcsec, including one complete
15.8 year orbit for the star S2. All orbits are well-fitted by a common enclosed mass and elliptical orbit focal
point. From multi-orbit fits, rescaled to $R_0\narreq8.2\kpc$, the total enclosed mass is
$M_\bullet\narreq4.3\times10^6\Msun$ \citep{Ghez2008} and $M_\bullet\narreq4.2\times10^6\Msun$ \citep{Gillessen2009},
and $M_\bullet\narreq(4.20\pm0.36)\times10^6\Msun$ from a joint analysis of the combined VLT and Keck data
\citep{Gillessen2009a}. The error given is combined statistical and systematic and is expected to improve steadily with
increasing time baseline of the measurements; it has an important contribution from a degeneracy between $M_\bullet$ and
$R_0$.

Therefore, external constraints on $R_0$ reduce the range of black hole mass allowed by the orbit
measurements. Combining with their measurement of the \nsc statistical parallax, \citet{Chatzopoulos2015} found
$M_\bullet\narreq(4.23\pm0.14)\times10^6 \Msun$.  Using instead the overall best $R_0\narreq8.2\pm0.1\kpc$ from
\S\ref{s:distance}, the mass of the black hole becomes $M_\bullet\narreq(4.2\pm0.2\times10^6)\Msun$. This corresponds to
a Schwarzschild radius of $\sim\! 0.1 \AU$, and to a dynamical radius of influence $r_{\rm infl}\nsimeq3.8 \pc$, where
the interior mass of the \nsc $M(r\!<\!r_{\rm infl})\narreq 2M_\bullet$ \citep{Chatzopoulos2015}. The inferred mass is
also consistent with the orbital roulette value obtained by \citet{Beloborodov2006} from the stellar motions in the
clockwise disk of young stars around Sgr A$^*$. For $M_\bullet\narreq4.2\times 10^6\Msun$ and a bulge velocity
dispersion of $\sigma\nsimeq113\kms$ (\S\ref{s:bulgekinematics}), the Milky Way falls below the best-fitting $(M_\bullet
- \sigma)-$relation for elliptical galaxies and classical bulges by a factor of $\sim\!5$-$6$
\citep{Kormendy2013,Saglia2016}.

\begin{marginnote}
\entry{$M_\bullet$}{$(4.2\pm0.2\times10^6)\Msun$, mass of Galactic SMBH}
\entry{$\rho_\bullet(<125\AU)$}{$5\times 10^{15}\Msun\pc^{-3}$,  mass density within pericenter of star S2}
\entry{$r_{\rm infl}$}{$3.8\pc$, SMBH's dynamical influence radius}
\end{marginnote}

From the orbit fit, any extended mass distribution within the orbit of S2 can contribute no more than $10\%$ of the
enclosed mass.  The S2 star has approached the central mass within $125\AU$ at pericenter, requiring a minimum interior
mass density of $\rho_\bullet(<125\AU)\narreq5\times 10^{15}\Msun\pc^{-3}$. This is so large that one can rule out any
known form of compact object other than a black hole \citep{Reid2009rev, Genzel2010}. From matching the positions of
SiO maser stars visible both in the NIR and the radio, the position of the compact mass and the radio source Sgr A$^*$
have been shown to coincide within $\nsim 2 {\rm mas}\narreq 16\AU$ \citep{Reid2003, Gillessen2009}. The size of Sgr A$^*$
in $\mm$ radio observations is $<1 \AU$ \citep{Shen2005a,Doeleman2008}. These facts taken together make it highly likely
that Sgr A$^*$ is the radiative counterpart of the black hole at the center of the Galaxy.

The apparent \propm\ of Sgr A$^*$ relative to a distant quasar (J1745-283) has been measured with great precision using
\vlbi \citep{Reid2004,Reid2008}. The \propm\ perpendicular to the Galactic plane is entirely consistent with the reflex
motion of the vertical peculiar velocity of the Sun, with residual $-0.4\pm0.9\kms$, suggesting strongly that the \smbh
is essentially at rest at the Galactic Center. Indeed the Brownian motion of the \smbh due to perturbations from the
stars orbiting inside its gravitational influence radius is expected to be $\sim\!0.2\kms$ \citep{Merritt2007}.  On the
assumption that Sgr A$^*$ is motionless at the Galactic Center, its measured \propm\ in the Galactic plane determines
the total angular velocity of the Sun with high accuracy: $\Omega_{g,\odot} \narreq 30.24 \pm 0.12 \kms \perkpc$. For
$R_0\narreq8.2\pm0.1\kpc$ from \S\ref{s:distance}, the inferred value of the total solar tangential velocity relative
to the Galactic Center is $V_{g,\odot}\narreq248\pm3 \kms$. We will return to these constraints in our discussion of the
Galactic rotation curve in \S\ref{s:rotcur} which brings together many of the major themes of this review.

\begin{marginnote}
\entry{$\Omega_{g,\odot}$}{$30.24\pm0.12$ $\kms\perkpc,\;\;\;$ Sun's total angular velocity relative to Sgr A$^*$}
\entry{$V_{g,\odot}$}{$248\pm3$ $\kms$, Sun's tangential velocity relative to Sgr A$^*$}
\end{marginnote}

\section{Inner Galaxy}  
\label{s:innergalaxy}

\subsection{Nuclear star cluster and stellar disk}
\label{s:NSC}

\citet{Becklin1968} discovered an extended NIR source centered on Sgr A: the MW's nuclear star cluster (\nsc).  The
source had a diameter of $\sim\! 5'\simeq 12\pc$, was elongated along the Galactic plane, and its surface brightness fell
with projected radius on the sky $\propto R_s^{-0.8\pm0.1}$. \nsc are commonly seen in the centers of disk galaxies and many contain
an AGN and thus a \smbh \citep[e.g.][]{Boker2010}.  NIR spectroscopy has shown that most of the luminous stars in the
Galactic \nsc are old ($>5\Gyr$) late-type giant and \rcg stars \citep[the ``old'' \nsc,][]{Pfuhl2011}.  But also a
surprising number of massive early-type stars were found in this volume \citep{Krabbe1995}, including massive young
stars in one and possibly two disks with diameters $1'\mhyphen 2'$ rotating around the \smbh
\citep{Paumard2006,Bartko2009}, and a remarkable concentration of B-stars within $1"$ of the \smbh, the so-called
S-stars \citep{Eckart1995}.  Recent reviews on the \nsc can be found in \citet{Genzel2010} and \citet{Schodel2014}.

The structure and dynamics of the \nsc must be studied in the IR because of the very high extinction towards the GC
\citep[$A_K\nsim2.6$ mag, $A_V\nsim40$ mag,][]{Fritz2011,Nishiyama2008}. Recent analysis of \spitzer /IRAC
$3.6\mum/4.5\mum$ images \citep{Schodel2014a} has shown that the old \nsc is centered on Sgr A$^*$ and point-symmetric,
and it is flattened along to the Galactic plane with minor-to-major projected axis ratio
$q\narreq0.71\pm0.02$. \citet{Chatzopoulos2015} obtain $q\narreq0.73\pm0.04$ from fitting K-band star counts; they also
show that \nsc dynamics requires a flattened star cluster with an axis ratio consistent with this value. The \nsc radial
density profile is discussed in these papers and in \citet{Fritz2016}.  When fitting the $4.5\mum$ data with a Sersic
profile, \citet{Schodel2014a} obtain a total $4.5\mum$ luminosity $L_{4.5,\mathrm{NSC}}\narreq(4.1 \pm0.4)\times 10^7
L_\odot$ and a spherical half-light radius of $r_h\narreq4.2\pm0.4\pc$.

\begin{marginnote}
\entry{NSC}{Nuclear star cluster}
\entry{$L_{4.5,\mathrm{NSC}}$}{ $(4.1 \pm0.4)\times 10^7 L_\odot$, NSC Luminosity}
\entry{$r_{\mathrm{NSC}}$}{$4.2\pm0.4\pc$, half-light radius}
\entry{$c/a$}{$\narreq$ $0.71\pm0.04$, axis ratio}
\entry{$M_{\mathrm{NSC}}$}{$\narreq$ $(1.8\pm0.3)\times10^7\Msun$, NSC mass}
\end{marginnote}

The dynamical mass within $100"\narreq4\pc$ is $(8.9\pm1)\times 10^6\Msun$ \citep{Chatzopoulos2015}; thus
$M/L_{4.5}\narreq0.44\pm0.06 \Msun/L_{4.5,\odot}$ and the total mass of the \nsc for the Sersic model is
$M_{\mathrm{NSC}}\narreq(1.8\pm0.3)\times10^7\Msun$.  An additional error in $M_{\mathrm{NSC}}$ not accounted for in
this estimate comes from the fact that the surface density profile of the \nsc goes below that of the much larger,
surrounding nuclear stellar disk at projected $R_s\gta\!100"$, making its outer density profile uncertain
\citep{Chatzopoulos2015}. The rotation properties and velocity dispersions were measured by \citet{Trippe2008,Fritz2016}
from stellar \propm and \los velocities, and from NIR integrated spectra by \citet{Feldmeier2014}.  The \nsc is
approximately described by an isotropic rotator model, with slightly slower rotation \citep{Chatzopoulos2015}.  There
are indications for a local kinematic misalignment in the \los velocities but not in the \propm \citep{Feldmeier2014,
  Fritz2016} which, if confirmed, might indicate some contribution to the \nsc mass by infalling star clusters
\citep{Antonini2012}; this needs further study.  Another unsolved problem is the apparent core in the \nsc density
profile \citep{Buchholz2009}; this might indicate that the \nsc is not fully relaxed, consistent with the relaxation
time estimated as $\approx\!10\Gyr$ throughout the \nsc \citep{Merritt2013}.

\begin{figure}
\centering
\parbox{6cm}{
\includegraphics[width=5.8cm, bb=300 20 650 260, clip=true]{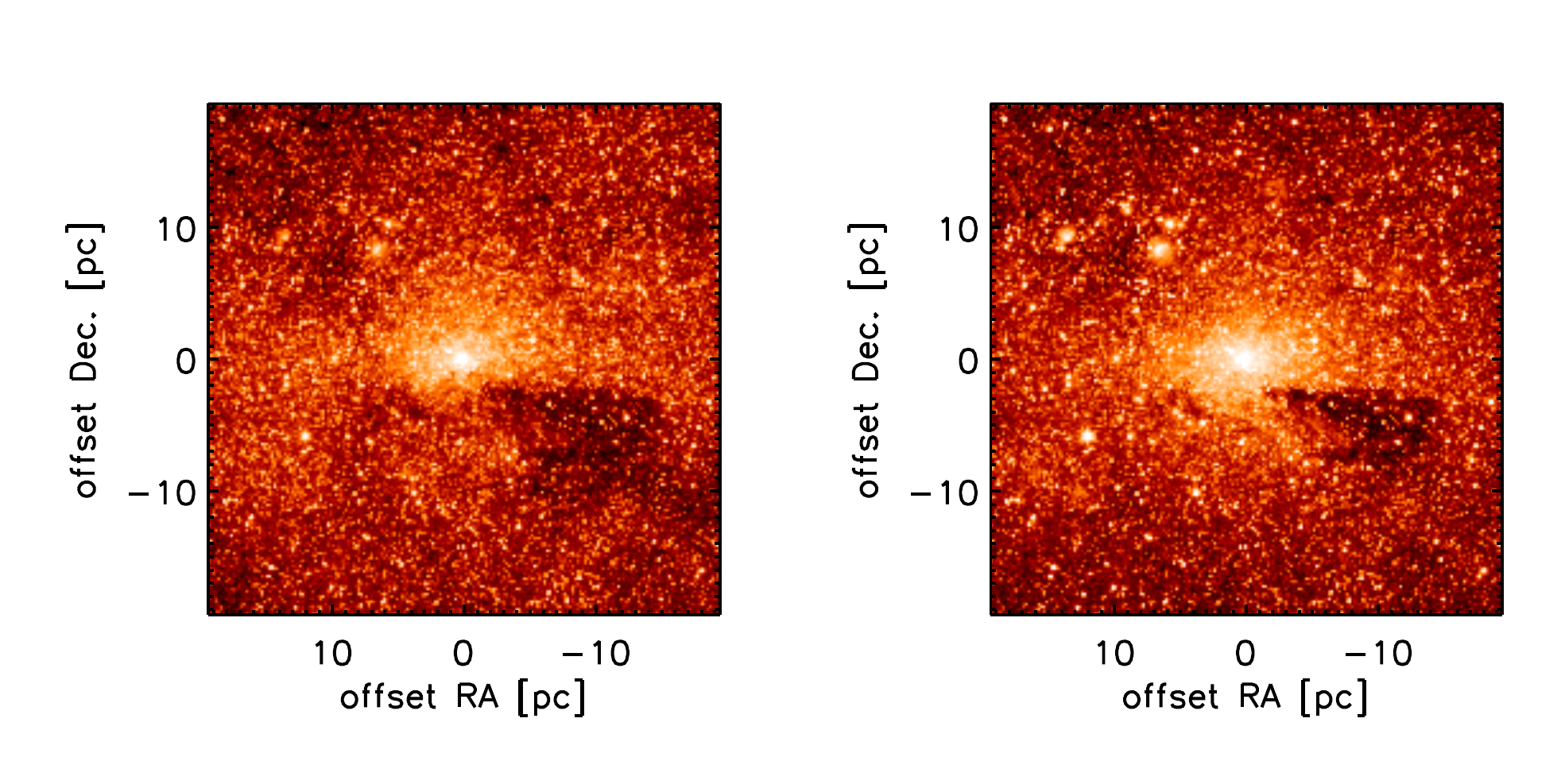}
\includegraphics[width=5.8cm]{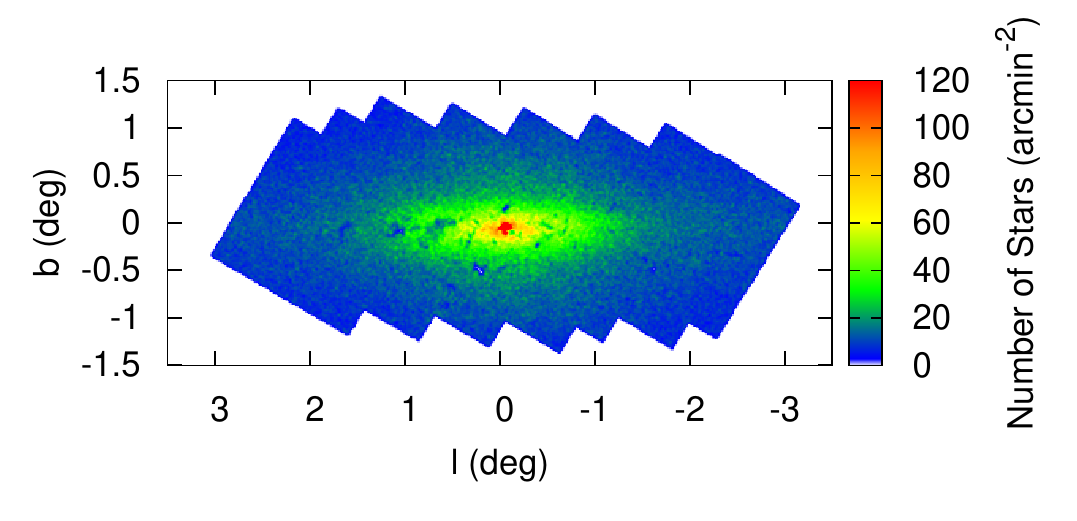}
}
\qquad
\begin{minipage}{6cm}
\includegraphics[width=6cm, bb=20 30 510 530, clip=true]{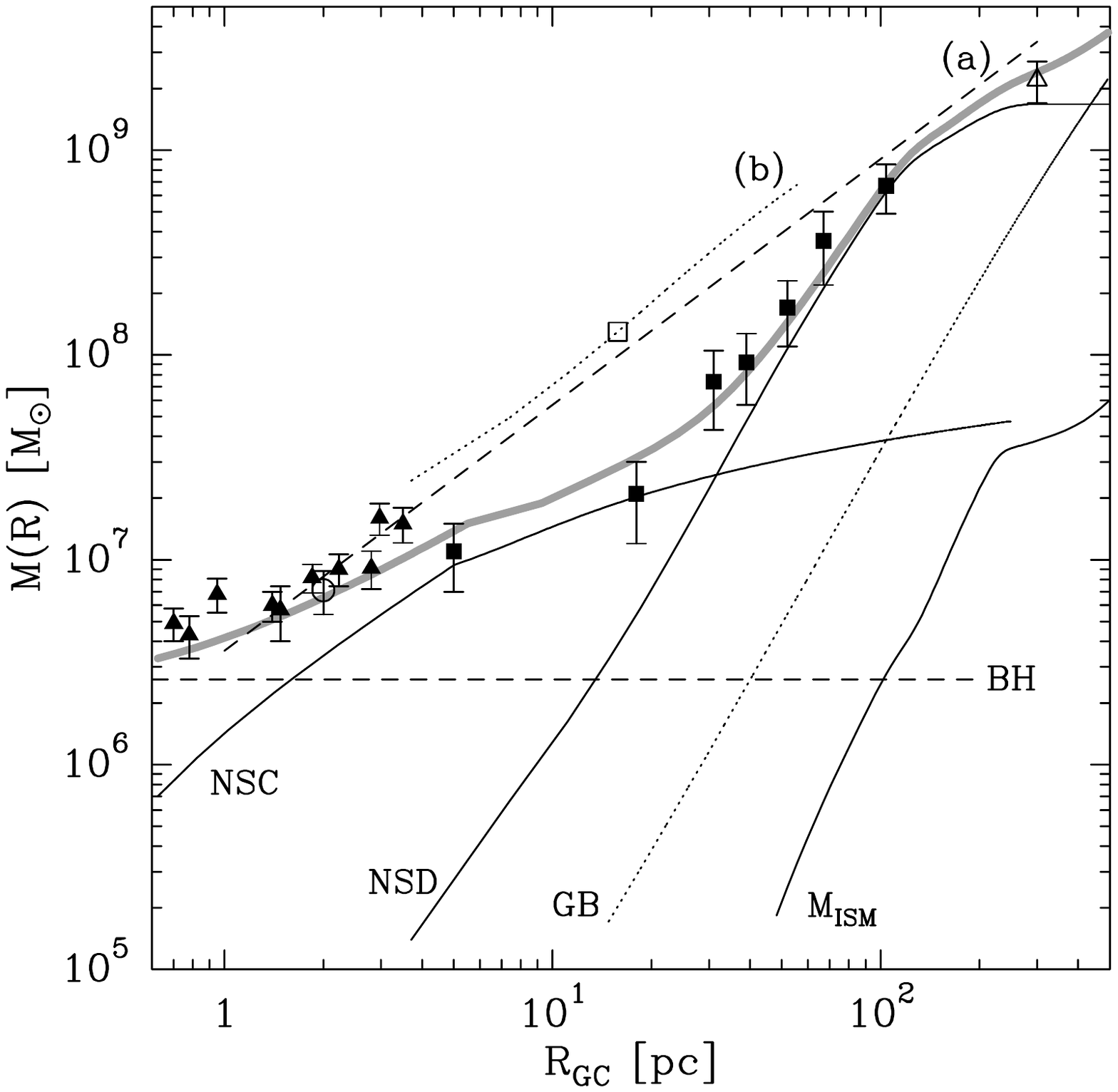}
\end{minipage}
\caption{Top left: extinction-corrected $4.5\mum$ IRAC image of the \nsc from \citet{Schodel2014a}.  Bottom left:
  extinction-corrected star count map of the \nsd from \citet{Nishiyama2013}. The scale is $143\pc/\rm{deg}$ at
  $R_0=8.2\kpc$. The \nsc is seen as the red dot in the center of this image. Right: enclosed photometric mass within
  spherical radius for the \nsc, \nsd, Galactic bulge (\gb), and total (thick line) from \citet{Launhardt2002}, with
  mass measurements at the time overplotted as points with error bars.  }
\label{f:nscnsd}
\end{figure}

The old \nsc is embedded in a nuclear stellar disk (\nsd) which dominates the three-dimensional stellar mass
distribution outside $\nsim30\pc$ \citep{Chatzopoulos2015} and within $\nsim200\!-\!400\pc$ \citep{Launhardt2002}. From
star counts, its vertical density profile is near-exponential with scale-height $h_{\mathrm{NSD}}\simeq 45\pc$
\citep{Nishiyama2013}. This confirms an earlier analysis of \cobe data by \citet{Launhardt2002} which cover a larger
area but with lower resolution. The projected density profile along the major axis ($|l|$) is approximately a power-law
$\propto |l|^{-0.3}$ out to $\sim\!  90\pc$; thereafter it drops steeply towards the \nsd's outer edge at $\sim\!
230\pc$, approximately $\propto |l|^{-2}$. The axis ratio of the \nsd inferred in these papers from the star counts and
NIR data is $\sim\,$3:1 at small radii and $\sim\,$5:1 on the largest scale. The total stellar mass estimated by
\citet{Launhardt2002} is $M_{\mathrm{NSD}} = (1.4\pm0.6)\times 10^9\Msun$, of order 10\% of the mass of the
bulge. The rotation of the NSD has been seen in OH/IR stars and SiO masers \citep{Lindqvist1992,Habing2006}
and with \APOGEE stars \citep{Schonrich2015}, with an observed gradient $\approx\!150\kms/{\rm deg}$
and estimated rotation velocity $\approx\!120\kms$ at $R\nsimeq100\pc$. These data suggest a dynamical mass
on the lower side of the estimated stellar mass range. The NSD is likely related to past star formation in
the zone of $x_2$-orbits near the center of the barred potential \citep[e.g.][]{Molinari2011}.
Clearly, understanding the \nsd better is relevant for the evolution of the Galactic bulge and probably also for
the growth of the \smbh, and further information about its kinematics and stellar population would be important.
Figure~\ref{f:nscnsd} illustrates this still enigmatic Galactic component together with the \nsc.

\begin{marginnote}
\entry{NSD}{Nuclear stellar disk}
\entry{$r_{\mathrm{NSD}}$}{$\nsimeq90\pc$, NSD break radius}
\entry{$h_{\mathrm{NSD}}$}{$45\pc$, NSD vertical scale-height}
\entry{$M_{\mathrm{NSD}}$}{$(1.4\pm0.6) \times10^9\Msun$, stellar mass of NSD}
\end{marginnote}

\subsection{Bulge}  
\label{s:bulge}

For many years, the Galactic bulge was considered as a structure built through mergers early in the formation of the
Galaxy, now called a {\sl classical bulge}. Particularly the old ages of bulge stars inferred from color-magnitude
diagrams supported this view \citep{Ortolani+95, Clarkson+08}. The NIR photometry with the \dirbe instrument on board
the \cobe satellite first established the boxy nature of the bulge \citep{Weiland+94, Binney+97}, later confirmed by the
\twomass star count map \citep{Skrutskie+06}. Recent star count data have unambiguously established that the bulk of the
bulge stars are part of a so-called {\sl box/peanut or b/p-bulge} structure representing the inner, three-dimensional
part of the Galactic bar \citep{McWilliam+Zoccali10, Nataf+10, Wegg+Gerhard13}, consistent with the observed cylindrical
rotation \citep{Kunder+12, Ness+13b}.  This corroborates long-standing evidence for a barred potential in the bulge
region from non-circular motions seen in \HI\ and \CO\ longitude-velocity-($lv$)-diagrams \citep{Binney+91,
  Englmaier+Gerhard99}. The central parts of the Galaxy also contain the dense \nsd and some have argued for a separate,
$200\pc$-scale {\sl nuclear bar} \citep{Alard01, Rodriguez-Fernandez+Combes08}. Finally, the peak in the density of the
{\sl inner stellar halo} is found in this region as well.  Disentangling these various components clearly requires the
best data possible. Results to date and open issues are summarized below.  More extensive reviews of the Galactic bulge
can be found in \citet{Rich2013, Gonzalez2016, Shen2015}.

\subsubsection{The Galactic b/p bulge} 
\label{s:bpbulge}
A large fraction of the bulge stars follows a rotating, barred, box/peanut shaped bulge with exponential density
distribution, similar to the inner three-dimensional part of an evolved N-body bar. The best available structural
information for the dominant bulge population comes from large samples of red clump giant stars (\rcg), for which
individual distances can be determined to $\sim\!10\%$ accuracy. These He-core burning stars have a narrow range of
absolute magnitudes and colors, $\sigma(\mathrm{K}_s)\simeq0.17$ and $\sigma(\mathrm{J-K}_s)\simeq0.05$ and are
predicted to trace the stellar population within $10\%$ for metallicities in the range [0.02,1.5] solar
\citep{Salaris+Girardi02}. In the color-magnitude diagram, \rcg appear spread because of distance, reddening, age
($\sim\!0.03/ \mathrm{Gyr}$ in K$_s$ at age $10\mathrm{Gyr}$), and metallicity (by
$\sigma_{Ks}(\mathrm{Fe/H})\sim\!0.11$ for the measured bulge metallicity distribution).  Among the 25,500 stars of the
\argos survey, \rcg are prominent down to [Fe/H]$\narreq-1.0$, which comprises $\sim\!95\%$ of their sample
\citep{Ness+13a}; i.e., \rcg are representative for most of the bulge stars.

\begin{figure}
\centering
\parbox{6cm}{
\includegraphics[width=6.2cm]{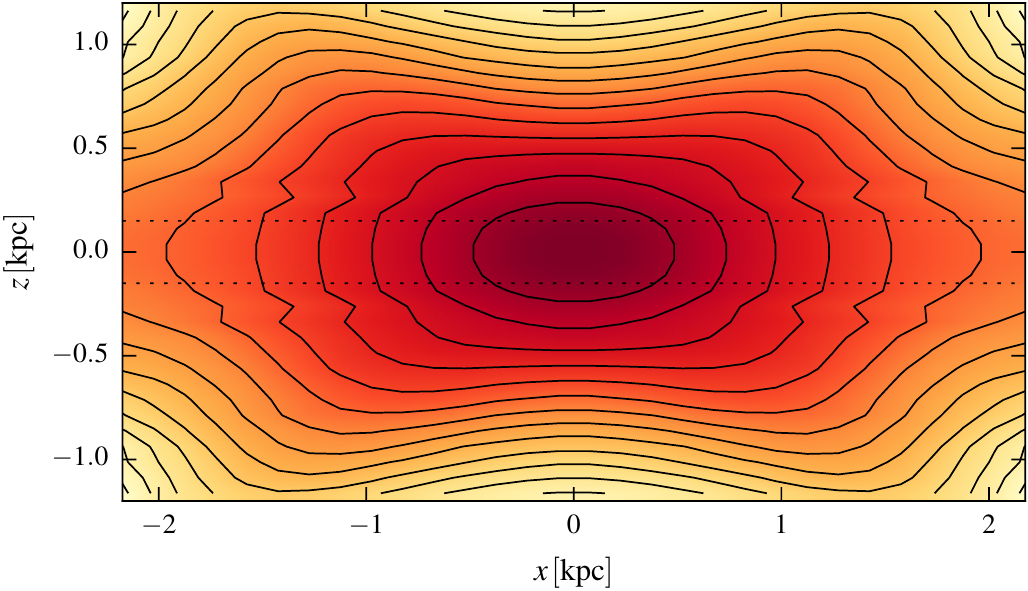}
\includegraphics[width=6cm]{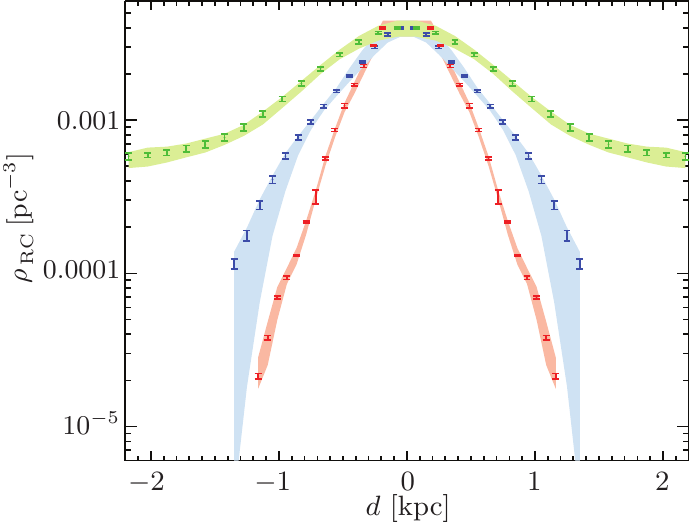}
}
\begin{minipage}{6cm}
\includegraphics[width=5.8cm]{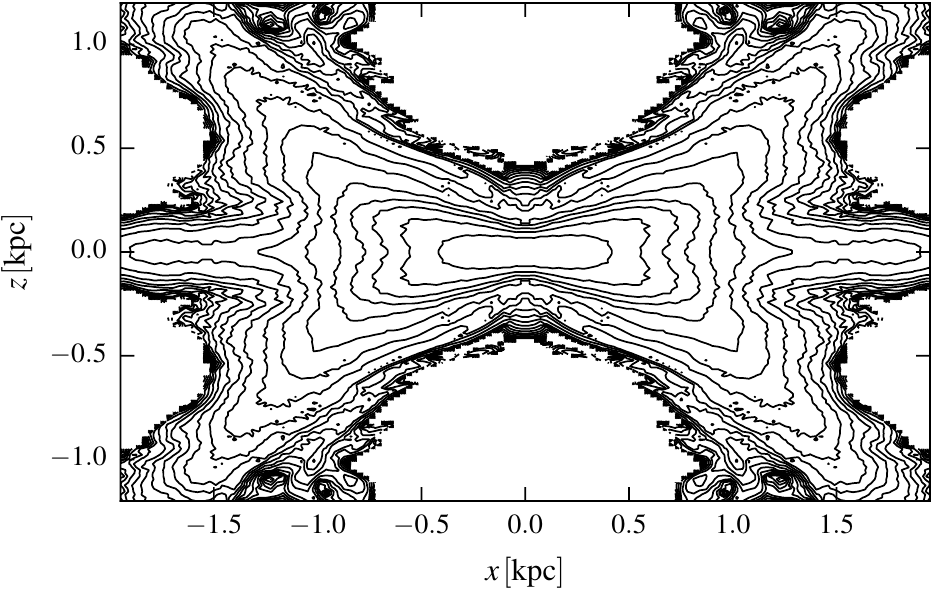}
\includegraphics[width=6cm]{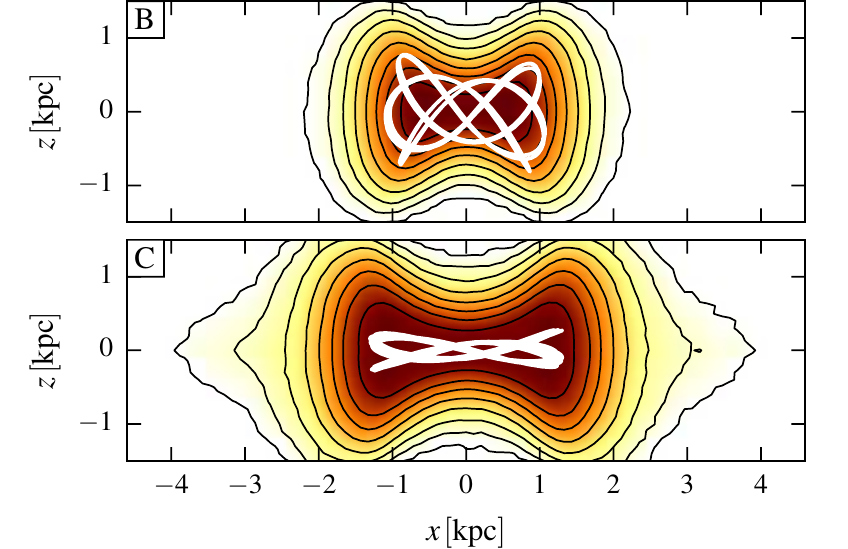}
\end{minipage}
\caption{The Galactic b/p bulge density measured from K-band \rcg star counts \citep{Wegg+Gerhard13}. Top left: side-on
  projection showing the prominent b/p shape. Top right: X-structure in unsharp-masked image \citep{Portail2015}. Bottom
  left: density profiles along the bulge principal axes $(x,y,z)$ in green, blue and red. The error bars show the \rms
  variations between 8-fold symmetric points around the triaxially symmetric 3D-map. Colored regions show estimated
  systematic errors. The density is typically accurate to $\nsim10\%$. Exponential scale-lengths along the
  $(x,y,z)$-axes are $(0.70\ncol 0.44\ncol 0.18)\kpc$ near the center. Bottom right: major orbit classes B,C  supporting the
  b/p-shape in dynamical models of the b/p-bulge \citep{Portail2015a}.}
\label{f:bulgex}
\end{figure}

Using $\sim\!\!8$ million \rcg from the \vvv survey \citep{Minniti2010} over the region $-10\dg\le l\le 10\dg$, $-10\dg\le
b\le 5\dg$, \citet{Wegg+Gerhard13} obtained \rcg line-of-sight density distributions for $\sim\! 300$ sightlines outside
the most crowded region $|b|<1\dg$, and combined these to a 3-dimensional map of the bulge \rcg density assuming 8-fold
triaxial symmetry (Figure~\ref{f:bulgex}). As shown in the figure, \rms variations between 8-fold symmetric points in
the final map are indeed small; there is no evidence for asymmetries in the volume of the \rcg measurement ($\pm 2.2
\times \pm 1.4 \times \pm 1.2\kpc$). The \rcg bulge is strongly barred, with face-on projected axis ratio
$\simeq(1:2.1)$ for isophotes reaching $\nsim2\kpc$ along the major axis; it has a strong b/p-shape viewed side-on, and
a boxy shape as seen from the Sun, consistent with the earlier \cobe and \twomass data. Unsharp masking
\citep{Portail2015} results in a strong off-centered X-shape structure (Fig.~\ref{f:bulgex}), similar to some galaxies
in the sample of \citet{Bureau2006}; see also \citet{Nataf2015}.

The near side of the b/p bulge has its major axis in the first Galactic quadrant ($0\dg<l<90\dg$). The {\sl bar angle}
between major axis and the Sun-Galactic center line found by \citet{Wegg+Gerhard13} is $\BPBphi\narreq27\dg\pm2\dg$,
with most of the error systematic. This is consistent with earlier parametric determinations from \ogle $I$ band \rcg star
counts \citep[$29\dg\pm2\dg$,][]{Cao2013}, \citep[$25\dg\pm2\dg$,][]{Rattenbury2007},
\citep[$20\dg\mhyphen30\dg$,][]{Stanek+97}, from non-parametric inversion of \twomass red giant star counts
\citep[$20\dg\mhyphen35\dg$,][]{Lopez-Corredoira+05}, and from modelling the asymmetry of the \cobe NIR photometry
\citep[$\sim25\dg\pm10\dg$,][]{Dwek+95,Binney+97,Freudenreich1998,Bissantz+Gerhard02}. The global bulge axis ratios
obtained with parametric star count models are typically $(1\ncol 0.4\ncol 0.3)$, similarly to those found from
modelling the \cobe data. However, it is clear from Fig.~\ref{f:bulgex} that a single vertical axis ratio does not
capture the shape of the b/p bulge. The lower left panel shows that the density distributions inside $\nsim 1\kpc$ are
nearly exponential, with scale-lengths $(h_x\ncol h_y\ncol h_z)\narreq(0.70\ncol 0.44\ncol 0.18)\kpc$ and axis ratios
$(10\ncol 6.3 \ncol 2.6)$ \citep{Wegg+Gerhard13}.  Further down the major axis, $h_z/h_x$ increases to $\sim\! 0.5$ at
$x\nsim 1.5\kpc$ where the X-shape is maximal, and then decreases rapidly outwards.

\begin{marginnote}
\entry{Box/Peanut (b/p) bulge}{}
\entry{\BPBphi}{$27\pm2\dg$, b/p bulge bar angle}
\entry{\BPBba}{$0.5\pm0.05$, axis ratio from top}
\entry{\BPBca}{$0.26$, edge-on axis ratio $(x\nsim0)$}
\entry{\BPBhz}{$180\pc$, vertical scale-height $(x\nsim0)$}
\entry{\BPBxX}{$1.5\pm0.2\kpc$, radius of max.~X}
\end{marginnote}

\subsubsection{Inner bulge and disk structure} 
\label{s:innerbulge}
The structure of the inner Galactic disk between the \nsd and $R\nsim2\kpc$ is not well-known due to heavy extinction
and crowding. Observations of maser stars and \vvv Cepheids indicate a barred disk of young stars
\citep{Habing2006,Dekany2015}.  The cold kinematics of young bar stars has likely been seen in \APOGEE \los velocity
histograms \citep{Aumer2015}. The short $h_z=180\pc$ vertical scale height in the bulge is perhaps indicative of a
central disk-like, high-density pseudo-bulge structure, as is seen in many early and late type b/p bulge galaxies
\citep{Bureau2006, KormendyBar2010}. NIR \rcg star counts at $b\narreq\pm1\dg$ have confirmed a structural change in the
\rcg longitude profiles at $|l|\simeq 4\dg$ \citep{Nishiyama2005,Gonzalez+11b, Wegg+Gerhard13}. This has been
interpreted by means of an N-body model in terms of a rounder, more nearly axisymmetric central parts of the b/p bulge
\citep{Gerhard+Martinez-Valpuesta12}. As predicted by the model, the transition at $|l|\simeq 4\dg$ is confined to a few
degrees from the Galactic plane \citep{Gonzalez+12}.

The nuclear bulge within $\nsim200\pc$ is dominated by the \nsd. Based on longitudinal asymmetries in a map of projected
\twomass star counts \citet{Alard01} presented indications for a $200 \pc$ scale nuclear bar separate from the b/p
bulge-bar. However, the large-scale Galactic bar by itself leads to similar inverted asymmetries in the center, just by
projection \citep{Gerhard+Martinez-Valpuesta12}, so the observed asymmetries are not a tell-tale
signature. Unfortunately, the distance resolution of the \rcg is not sufficient to investigate the \los-structure of a
tilted nuclear bar. Thus the most promising test appears to be with models of the nuclear gas flow
\citep{Rodriguez-Fernandez+Combes08}, but this requires understanding the larger-scale properties of the gas flow better
(see \S\ref{s:omega}) which influence the nuclear gas flow. Further studies in the IR, both photometric and
spectroscopic, are clearly needed to shed more light on the inner bulge.

\subsubsection{Does the Milky Way have a classical bulge? Kinematics and metallicities of bulge stars} 
\label{s:bulgekinematics}
Bulges in several disk galaxy formation models have been found to harbour a rapid early starburst component, as well as
a second component which forms later after disk build-up and instabilities, and/or minor mergers \citep{Samland2003,
  Obreja2013}. The former could be associated with a classical bulge even in the absence of a significant early
merger-built bulge.  The Milky Way bulge has a well-established vertical metallicity gradient \citep{Zoccali2008,
  Johnson2011, Gonzalez2013} which has often been taken as the signature of a dissipatively formed classical bulge
\citep[see][]{Pipino2008}.  However, because violent relaxation is inefficient during the bar and buckling
instabilities, preexisting metallicity gradients, such that stars with lower binding energies have lower metallicities,
would survive as outward metallicity gradients in the final b/p bulge \citep{Martinez-Valpuesta2013,
  DiMatteo2014}. Recent spectroscopic surveys have attributed the vertical metallicity gradient to a superposition of
several metallicity components whose relative contributions change with latitude \citep{Babusiaux2010, Ness+13a}. Hence
the signature of a classical bulge must be found with more detailed kinematic and chemical observations.

The mean line-of-sight rotation velocities of bulge stars are nearly independent of latitude, showing {\sl cylindrical
  rotation} as is common in barred bulges.  First found with planetary nebulas \citep{Beaulieu2000}, this was
shown conclusively with the \brava \citep{Kunder+12}, \argos \citep{Ness+13b}, and \gibs \citep{Zoccali2014} surveys.
Rotation velocities reached at $l\nsim10\dg$ are $\nsim 75\kms$.  \los\ velocity dispersions at $l\nsim 0$ are
$\nsim 80\kms$ at $|b|\narreq8\dg$ and increase rapidly towards the Galactic plane, reaching $\nsim 120\kms$ in Baade's
window at $|b|\narreq4\dg$.  Based on the dynamical model of \citet[][see \S\ref{s:bulgemass}]{Portail2015},
mass-weighted velocity dispersions inside the bulge half mass radius are $(\sigma^b_x,
\sigma^b_y, \sigma^b_z) \! \approx \! (135,105,96)\kms$ and the rms is $\sigma^b_{\rm rms} \!\approx\!
113\kms$, to $\approx\!3\kkms$.

The \argos survey mapped the kinematics for different metallicities, showing that higher/lower metallicity stars have
lower/higher velocity dispersions.  \citet{Soto2007} and \citet{Babusiaux2010} found differences between the vertex
deviations of metal-rich and metal-poor bulge stars and argue for the existence of two main bulge stellar populations,
of which only the more metal-rich one follows the bar. \citet{Rojas-Arriagada2014} find two about equally numerous,
metal-rich and metal-poor components in the metallicity distribution of their fields, whereas \citep{Ness+13a, Ness+13b}
find evidence for five populations. The metal-rich components trace the X-shape and hence the barred bulge, but the origin of
the metal-poor stars ([Fe/H]$<\!-0.5$) is currently debated. They could represent an old bulge formed through early
mergers, or a thick disk component participating in the instability together with the inner stellar halo
\citep[e.g.][]{Babusiaux2010, DiMatteo2014}.

Large numbers of RR Lyrae stars found in the \ogle and \vvv bulge surveys have shown that the most metal-poor ([Fe/H]
$\narreq-1.0\pm0.2$), old population does not participate in the b/p-bulge \citep{Dekany2013, Pietrukowicz2015},
consistent with the \argos result that only stars with [Fe/H] $\gta\!-0.5$ participate in the split red clump
\citep{Ness+12}. The RR Lyrae stars show no significant rotation \citep{Kunder2016}. By contrast, the \argos stars with
[Fe/H]$<\!-1$ rotate still fairly rapidly; whether they could be stars from the stellar halo or a low-mass classical
bulge spun up by the b/p-bulge \citep{Saha+12} or whether they could include a component of thick disk stars must still
be checked in detail.

In summary, it is unclear at this time whether the Milky Way contains any classical bulge at all - comparing
N-body-simulated b/p bulge models to the \brava data, \citet{Shen2010} found that the cylindrical rotation in the
Galactic bulge could be matched by their models only if the initial models contained a classical bulge with $\lta8\%$ of
the initial disk mass ($\lta25\%$ of the final bulge mass), and none was needed.  However, there is strong evidence from
structural and kinematic properties that the major part of the Galactic bulge was built from the disk through
evolutionary processes similar to those observed in galaxy evolution simulations, as is also inferred for many external
galaxies \citep[so-called {\sl secular evolution --}][]{Kormendy2013a, Sellwood2014}.

\newcommand\Mten{10^{10}\Msun}

\subsubsection{Mass and mass-to-light ratio in the bulge} 
\label{s:bulgemass}

The stellar mass of the bulge can be estimated from a photometric model combined with a stellar population model.  For
example, \citet{Dwek+95} obtained $1.3 \times \Mten$ from the \cobe NIR luminosity and a Salpeter \imf \citep[$2.0
\times \Mten$ rescaled for Kroupa IMF,][]{Licquia2015}.  \citet{Valenti2015} obtained a projected mass of
$2.0\pm0.3\times \Mten$ from scaling the measured mass function in a small bulge field to the whole bulge using
\rcg. The stellar mass corresponds to the dynamical mass only if the contribution of dark matter in the
bulge region is unimportant.

The dynamical mass in the bulge can be determined either from gas kinematics in the bulge region, or from stellar
kinematics combined with a dynamical model. For a barred bulge, simple rotation curve analysis does not apply, and
analysis of the full gas velocity field requires hydrodynamical models (see \S\ref{s:omega}). Stellar-dynamical models
require a well-determined tracer density, i.e., a NIR luminosity or tracer density distribution. Furthermore, since the
dominant part of the Galactic bulge is the inner b/p part of the Galactic bar, the result depends somewhat on the
spatial region defined as ``the bulge''.

\citet{Zhao1994} built a self-consistent model of the bar/bulge using the Schwarzschild method, and found a total bulge
mass of $2 \times \Mten$.  \citet{Kent1992} modelled the $2.4\mum$ \spacelab emission with an oblate isotropic rotator and
constant mass-to-light ratio, finding a mass of $1.8 \times \Mten$.  \citet{Bissantz+03} determined the circular velocity
at $2.2 \kpc$ to be $190\kms$, modelling gas dynamics in the potential of the deprojected \cobe NIR luminosity
distribution from \citet{Bissantz+Gerhard02}.  Assuming spherical symmetry, this leads to a total bulge mass of about $1.85
\times \Mten$. However, a number of other studies have found lower masses \citep{Licquia2015}.

\begin{figure}
\centering
\parbox{6cm}{
\includegraphics[width=6cm]{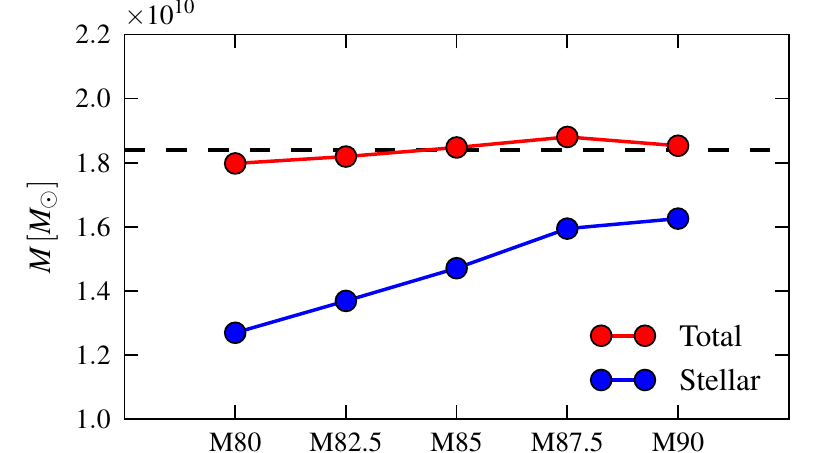}
}
\begin{minipage}{6cm}
\includegraphics[width=6cm]{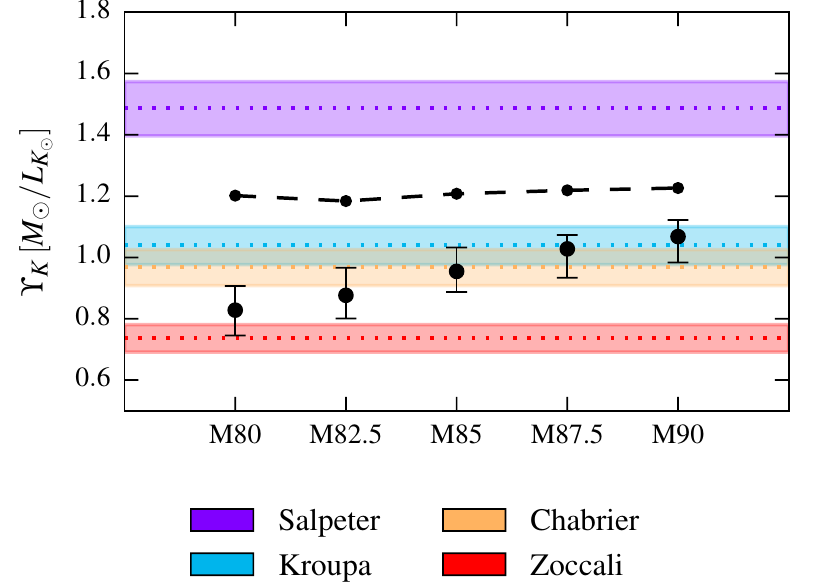}
\end{minipage}
\caption{Left: Mass of the Galactic bulge in the \vvv box for the five dynamical models of \citet{Portail2015} with
  different dark matter halos. The blue curve refers to the stellar mass while the red curve refers to the total mass.
  Right: Stellar mass-to-light ratio in the $K$ band for the same five models. The model errors shown are dominated by
  systematic effects. The different colored lines indicate predictions for different \imf as stated in the legend. The
  most recent measurements \citep{Calamida2015} are close to a Kroupa \imf. The black dashed line is an estimate of the
  highest allowed mass-to-light ratio obtained by turning all dark matter in the \vvv box into stars. This
  rules out a Salpeter \imf for the Galactic bulge with age $10 \Gyr$.  }
\label{f:bulgemass}
\end{figure}

In the most recent study, \citet{Portail2015} find a very well-constrained total dynamical mass of $1.84\pm0.07\times
10^{10}\Msun$ in the \VVV bulge region (the box $\pm 2.2 \times \pm 1.4 \times \pm 1.2\kpc$), by fitting made-to-measure
dynamical models to the combined \vvv \rcg star density and \brava kinematics (Figure~\ref{f:bulgemass}). The data can
be fit well by models with a range of dark-to-stellar mass ratios.  Comparing the implied total surface mass density
with the \cobe surface brightness and stellar population models, a Salpeter \imf for a $10\Gyr$ old population can be
ruled out, predicting significantly more mass than is dynamically allowed. For an \imf between those of
\citet{Kroupa2001, Chabrier2003, Zoccali2000}, $10$-$40\%$ of the mass in the bulge region would required to be in dark
matter.  Recently \citet{Calamida2015} derived the bulge IMF in the \SWEEPS field, removing foreground disk stars, and
found a double-power law form remarkably similar to a Kroupa or Chabrier IMF. The models of \citet{Portail2015} then
predict a total stellar mass in this region of $1.4\mhyphen1.7\times 10^{10}\Msun$, including stars in the inner disk,
and a dark matter fraction of $10$-$25\%$. The estimated total stellar mass in the bulge and disk of the Galaxy is
$M_{\rm tot}^* \approx\!  4.7\mhyphen5.7\times 10^{10}\Msun$ (\S\ref{s:rotcur}), so the ratio of stellar mass in the
bulge region to total is \Mbstar$/M_{\rm tot}^*=0.3\pm0.06$.

\begin{marginnote}
\entry{\Mbdyn}{$1.84\pm0.07\times 10^{10}\Msun$,  dynamical mass in \vvv bulge region}
\entry{\Mbstar}{$(1.4\mhyphen 1.7)\times 10^{10}\Msun$, stellar mass in \vvv bulge region}
\entry{\Mbstar/$M_{\rm tot}^*$}{$0.3\pm0.06$, stellar mass in the bulge region to total}
\entry{$f_{\rm b, DM}$}{$10\%\mhyphen25\%$, dark matter fraction in \vvv region}
\entry{ $(\sigma^b_x, \sigma^b_y, \sigma^b_z, \sigma^b_{\rm rms})$}{$(135,105,96,113)$ 
km/s, mass-weighted velocity dispersions within half-mass radius along $(x,y,z)$ and rms}
\entry{$M_{\rm clb}/$\Mbstar}{$0\mhyphen25\%$, classical bulge (clb) fraction}
\end{marginnote}

The stellar mass involved in the peanut shape is important for constraining the origin of the bulge populations
(\S\ref{s:bulgekinematics}).  \citet{Li2012} applied an unsharp masking technique to the side-on projection of the model
of \citet{Shen2010}, removing an elliptical bulge model from the total. This revealed a centred X-structure accounting
for about $7\%$ of their model bulge. \citet{Portail2015} removed a best-matched ellipsoidal density from the
three-dimensional \rcg bulge density of \citet{Wegg+Gerhard13}, finding that $24\%$ of the bulge stellar mass remained
in the residual X-shape. Most reliable would be a dynamical, orbit-based definition of the mass in the peanut
shape. However, in the bulge models of \citet{Portail2015}, stars in the X-shape do not stream along $x_1v_1$ `banana'
orbits \citep{Pfenniger1991} which follow the arms of the X-shape.  Instead, the peanut shape is supported by `brezel'
orbit families which contribute density everywhere between the arms of the X-structure \citep[][see
Fig.~\ref{f:bulgex}]{Portail2015a}. In these models, the fraction of stellar orbits that {\sl contribute} to the
X-structure account for $40$-$45\%$ of the bulge stellar mass.

\subsection{The ``long bar'' outside the bulge}
\label{s:longbar}

In N-body models for disk galaxy evolution, box/peanut bulges are the inner three-dimensional parts of a longer, planar
bar that formed through buckling out of the galaxy plane and/or orbits in vertical resonance \citep{Combes+90, Raha1991, Athanassoula05}.
There is also evidence that b/p bulges in external galaxies are embedded in longer, thinner bars \citep{Bureau2006}.
Thus also the Milky Way is expected to have a thin bar component extending well outside the b/p bulge. Finding the
Galactic planar bar and characterizing its properties has however proven difficult, because of intervening dust
extinction and the superposition with the star-forming disk at low-latitudes towards the inner Galaxy.

\begin{figure}
\centering
\includegraphics[width=12cm]{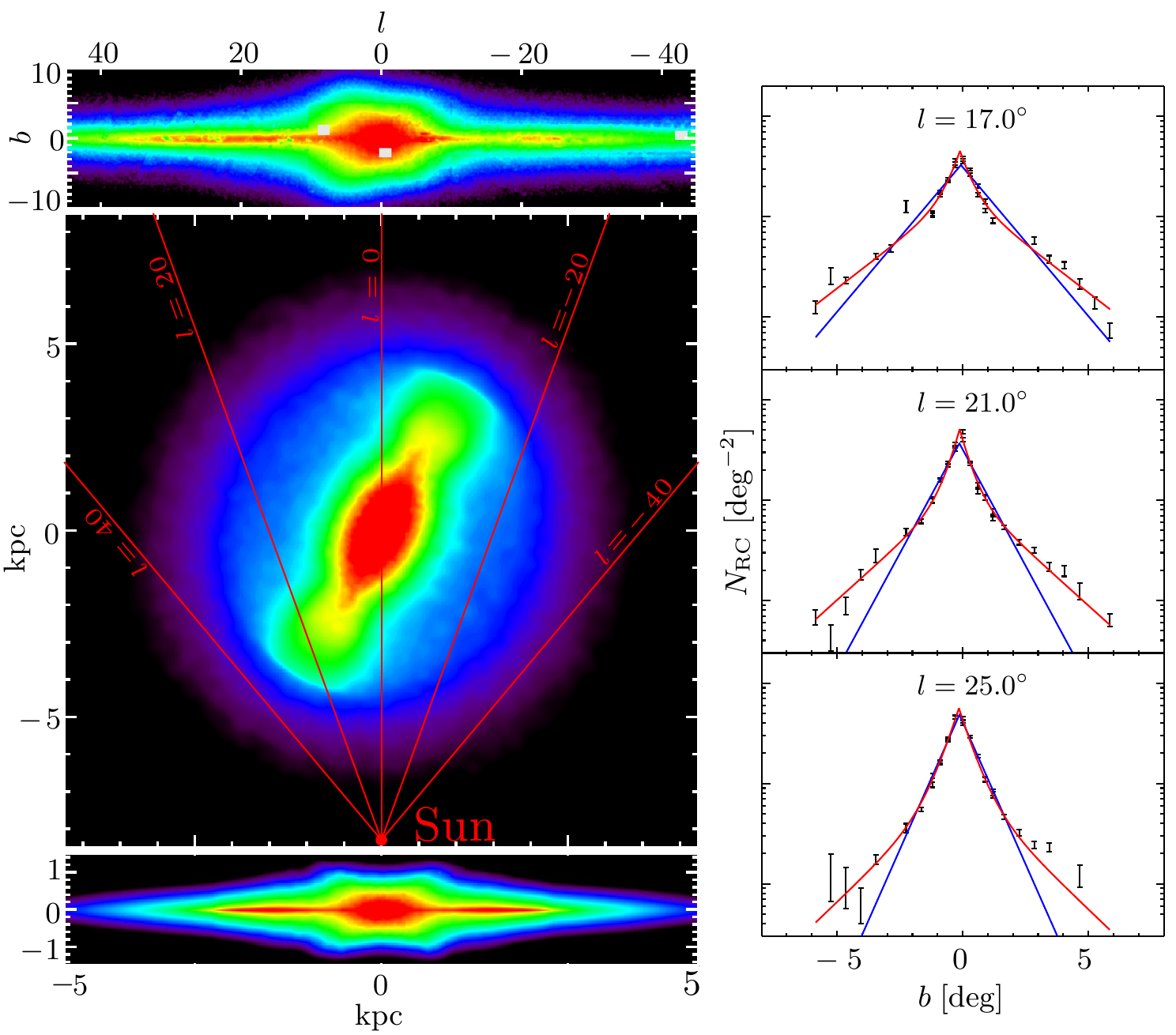}
\caption{Left: projections of the Galactic b/p-bulge and long bar reconstructed from NIR star counts. Top: inner Galaxy
  as seen from the Sun, in bright star counts complete across several NIR surveys. Middle: Projection of best-fitting
  \rcg star count model as seen from the North Galactic Pole. Viewing directions from the Sun are indicated for
  longitudes $|l|=0\dg,20\dg,40\dg$. Bottom left: side-on view showing the transition from the b/p bulge to the long bar
  and disk. Right: Vertical surface density profiles of \rcg stars for several longitude slices in the long bar
  region. Blue lines show single exponential fits. Red lines show the preferred double exponential model consisting of a
  superthin ($h_z=45\pc$) and a thin bar component ($h_z=180\pc$). The fraction of stars in the superthin component
  increases with longitude \citep[adapted from][]{Wegg2015}. }
\label{f:longbar}
\end{figure}

\citet{Hammersley+00} drew attention to an overdensity of stars in the Milky Way disk plane reaching outwards from the
bulge region to $l\simeq 28\dg$.  NIR star count studies with \ukidss and other surveys confirmed this structure
\citep{Cabrera-Lavers+07b, Cabrera-Lavers+08}. Its vertical scale-length was found to be less than $100 \pc$, so this is
clearly a disk feature.  With \spitzer \glimpse mid-infrared star counts, less affected by dust than K-band data,
\citet{Benjamin+05} similarly found a strong bar-like overdensity of sources at positive longitudes.  Because of its
wide longitude extent and the narrow extent along the \los this structure was termed the ``long bar''.

Based on the combined \twomass, \ukidss, \vvv, and \glimpse surveys, \citet{Wegg2015} investigated the long bar in a
wide area in latitude and longitude, $ |b| \le 9\dg$ and $ |l| \le 40\dg$, using \rcg stars and correcting for
extinction star-by-star. They found that the Galactic bar extends to $l \sim\! 25 \dg$ at $|b| \sim\! 5\dg$ from the
Galactic plane, and to $l \sim\! 30 \dg$ at lower latitudes.  Their long bar has an angle to the line-of-sight of $28\dg
- 33\dg$, consistent with the bar angle inferred for the bulge at $|l|<10\dg$. The vertical scale-height of the \rcg
stars decreases continuously from the b/p bulge to the long bar. Thus the central b/p bulge appears to be the vertical
extension of a longer, flatter bar, similar as seen in external galaxies and N-body models.

These recent results are based on a larger and more uniform data base and on a more uniform analysis than the earlier
work on the long bar, using cross-checked star-by-star extinction corrections and a statistical rather than CMD-based
selection of \rcg stars.  This leads to smaller errors in the \rcg magnitude distributions and reduced scatter between
neighbouring fields, particularly near the Galactic plane. These results therefore supercede in particular the earlier
claim that the long bar is an independent bar structure at angle $\sim\! 45\dg$ and misaligned with the b/p bulge.

\begin{figure}
\centering
\parbox{6cm}{
\includegraphics[width=6cm]{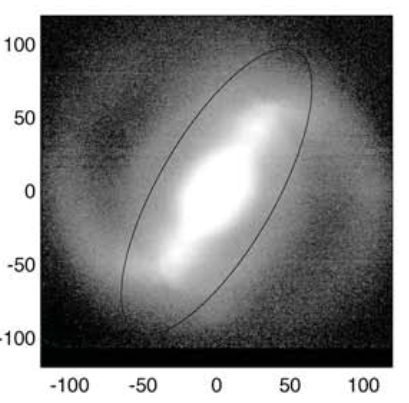}
}
\begin{minipage}{6cm}
\includegraphics[width=5.8cm]{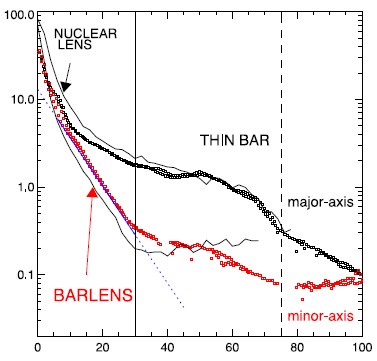}
\end{minipage}
\caption{Left: K-band image of the bar-lens galaxy NGC 4314 from \citet{Laurikainen2011}, with similar morphology as the
  face-on image of the Galactic b/p bulge and long bar in Fig.~\ref{f:longbar}.  Scaled and reflected to the rotation of
  the Milky Way, the Sun would be located at roughly (-110'',-60'') in this image.  Right: major and minor axis surface
  brightness profiles for NGC 4314, from \citet{Laurikainen2014}.}
\label{f:barlens}
\end{figure}

Comparing parametric models for the \rcg magnitude distributions with the data, \citet{Wegg2015} find a total bar (half)
length of $R_{\mathrm{lb}}\narreq 5.0\pm 0.2 \kpc$. Projections of their best model for the combined bulge and long bar
are shown in Figure~\ref{f:longbar}.  The top panel illustrates the asymmetries seen by observers at the Sun, due to the
bar shape and geometry.  The side-on view in Fig.~\ref{f:longbar} clearly shows the Milky Way's central box/peanut bulge
and the decrease of the scale-height in the long-bar region.  In the central face-on view, the projected b/p-bulge
resembles the bar-lens structures described by \citet{Laurikainen2011}, which are considered to be the more face-on
counterparts of b/p-bulges \citep{Laurikainen2014}; see the image of NGC 4314 in Figure~\ref{f:barlens}.

\begin{marginnote}
\entry{Long bar}{}
\entry{$\phi_{\mathrm{lb}}$}{$28\dg-33\dg$, long bar angle}
\entry{$R_{\mathrm{lb}}$}{$5.0\pm 0.2 \kpc$, bar half-length}
\entry{$h_{\mathrm{tlb}}$}{$180\pc$, thin bar scale-height}
\entry{$h_{\mathrm{slb}}$}{$45\pc$, superthin bar scale-height}
\entry{$M_{\mathrm{tlb}}$}{$\sim\!7\pm 1\times10^{9}\Msun$, stellar mass of thin bar}
\entry{$M_{\mathrm{slb}}$}{$\sim\!3\times10^9\Msun$, stellar mass of superthin bar}
\end{marginnote}

In the same analysis, \citet{Wegg2015} find evidence for two vertical scale-heights in the long bar, also illustrated in
Fig.~\ref{f:longbar}. The {\sl thin} bar component has $h_{\mathrm{tlb}}\simeq180\pc$ and its density decreases
outwards roughly exponentially; it is reminiscent of the old thin disk near the Sun. The second {\sl superthin} bar
component has $h_{\mathrm{slb}}\simeq 45\pc$ and its density increases outwards towards the bar end where it dominates
the \rcg counts. The short scale-height is similar to the 60-$80\pc$ superthin disk found in the edge-on spiral galaxy
NGC 891 \citep{Schechtman-Rook2013}. Stars in this component have an estimated vertical velocity dispersion of
$\sigma_z\!\simeq\!20\mhyphen30\kms$ and should be younger than the thin component. However, to have formed \rcg they
must have ages at least $>0.5\Gyr$ but star-forming galaxies have a strong bias towards ages around $\nsim 1 \Gyr$
\citep{Salaris+Girardi02}. Such a younger bar component could arise from star formation towards the bar end or from disk
stars captured by the bar.

The dynamical mass of the long bar component has not yet been determined. The stellar mass was estimated by
\citet{Wegg2015} from the \rcg density using isochrones and a Kroupa IMF. This resulted in a total
non-axisymmetric mass for the thin bar component of $M_{\mathrm{tlb}}\nsimeq 6-8\times10^9\Msun$, assuming a 10 Gyr old,
$\alpha$-enhanced population, and $M_{\mathrm{slb}}\nsimeq 3.3\times10^9\Msun$ for the superthin component, assuming a
constant past star formation rate.  Owing to its $5 \kpc$ half-length and its total mass $\sim\!10^{10} \Msun$, the long
bar may have quite some impact on the dynamics of the Galactic disk inside the solar circle, particularly on the gas
flow and the spiral arms, but perhaps also on surface density and scale-length measurements in the disk (see
Fig.~\ref{f:barlens}).  In Section~\ref{s:omega} below, we summarize constraints on the bar's corotation radius, which
must be larger than $R_{\mathrm{lb}}$.

\subsection{Pattern speed} 
\label{s:omega}

The pattern speed $\Omega_{\rm b}$ of the b/p bulge and bar, or equivalently its corotation radius $R_{\rm CR}$, has great
importance for the dynamics of the bar and surrounding disk. Despite a number of different attacks on measuring
$\Omega_{\rm b}$ its value is currently not accurately known. An upper limit comes from determining the length of the bar and
assuming that, like in external galaxies the Galactic bar is a fast bar, i.e., ${\cal R}=R_{\rm CR}/R_{\mathrm{lb}}=1.2\pm0.2$
\citep{Aguerri2003, Aguerri2015}. Here the $1.0$ lower limit is based on the fact that theoretically, bars cannot extend
beyond their corotation radius because the main $x_1$-orbit family supporting the bar becomes unstable
\citep{Contopoulos1980,Athanassoula1992}. The length of the long bar from starcounts is $R_{\mathrm{lb}}\simeq
5.0\pm0.2\kpc$ and the length of the thin bar component alone is $R_{\mathrm{tlb}}\simeq 4.6\pm0.3\kpc$
\citep{Wegg2015}; thus a strong lower limit is $R_{\rm CR}=4.3\kpc$ and a more likely range is $R_{\rm CR}=5.0-7.0$ kpc,
or $\Omega_{\rm b}\sim\!34-47 \kmskpc$ for $\Vcirc\narreq 238\kms\narreq\mathrm{const.}$ (\S\ref{s:rotcur}).

Early determinations appeared to give rather high values of $\Omega_{\rm b}$. The most direct method applied a modified
version of the Tremaine-Weinberg continuity argument to a complete sample of OH/IR stars in the inner Galaxy
\citep{Debattista2002}, giving $\Omega_{\rm b}=59\pm5\pm10$ (sys) $\kmskpc$ for $(R_0,\Vcirc)\narreq(8\kpc,220\kms)$ but
depending sensitively on the radial motion of the LSR.

More frequently, the pattern speed of the bar has been estimated from hydrodynamic simulations comparing the gas flow
with observed Galactic \CO\ and \HI\ $lv$-diagrams. These simulations are sensitive to the gravitational potential, and
generally reproduce a number of characteristic features in the $lv$-plot, but none reproduces all observed features
equally well. Consequently the derived pattern speeds depend somewhat on the gas features emphasized.
\citet{Englmaier+Gerhard99} and \citet{Bissantz+03} estimated $55-65\kmskpc$ ($R_{\rm CR}=3.4\pm0.3$ kpc) matching the
terminal velocity curve, spiral arm tangents and ridges in the $lv$-plot. \citet{Fux99} obtained $\sim\!50\kmskpc$
($R_{\rm CR}=4-4.5$ kpc) from a comparison to various reference features in the \CO\ $lv$-plot; \citet{Weiner1999}
obtained $42\kmskpc$ ($R_{\rm CR}=5.0$ kpc) from matching the extreme \HI\ velocity contour;
\citet{Rodriguez-Fernandez+Combes08} obtained $30-40\kmskpc$ and $R_{\rm CR}=5-7$ kpc matching to the Galactic spiral
arm pattern. The most recent analysis based on a range of potential parameters is by \citet{Sormani2015}. They
conclude that overall a pattern speed of $\Omega_{\rm b}=40\kmskpc$ corresponding to $R_{\rm CR}=5.6\kpc$ matches best the
combined constraints from the terminal velocity envelope, the central velocity peaks, and the spiral arm traces in the
$lv$-diagram (for $R_0,\Vcirc\narreq8\kpc,220\kms$).

Stellar-dynamical models of the Galactic b/p bulge also depend on $\Omega_{\rm b}$ and give estimated ranges for its
value.  \citet{Shen2010} and \citet{Long2012} find $\Omega_{\rm b}\nsimeq40\kmskpc$ for the same N-body model matched to
the \brava kinematic data.  The recent models of \citet{Portail2015} fitted additionally to the \rcg density from
\citet{Wegg+Gerhard13} give values in the range $\Omega_{\rm b}\sim\!25-30\kmskpc$, placing corotation in the range
$R_{\rm CR}>7.2\kpc$ (for $R_0,\Vcirc\narreq8.3\kpc,220\kms$). These values could depend somewhat on the still uncertain
gravitational potential in the long bar region.

A final method is based on the interpretation of star streams observed in the distribution of stellar velocities in the
solar neighborhood as due to resonant orbit families near the outer Lindblad resonance of the bar
\citep{Kalnajs1991,Dehnen2000}.  \citet{Dehnen2000} estimates $\Omega_{\rm b}=(1.85\pm0.15)\ \Vcirc/R_0$ ($51\pm4\kmskpc$
for $(R_0,\Vcirc)\narreq(8\kpc,220\kms)$. \citet{Minchev2007} find $\Omega_{\rm b}=(1.87\pm0.04)\ \Vcirc/R_0$
($51.5\pm1.5\kmskpc$). \citet{Chakrabarty2007} and others argue that spiral arm perturbations need to be included,
finding $R_0/R_{\rm CR}\simeq 2.1\pm0.1$ and $\Omega_{\rm b}\simeq 57.5\pm 5\kmskpc$. The latest analysis of the
Hercules stream by \citet{Antoja2014} gives $R_0/R_{\rm CR}\simeq (1.83\pm0.02)$, $\Omega_{\rm b}\simeq 53\pm 0.5\kmskpc$,
and $R_{\rm CR}=4.49\pm0.05\kpc$ when rescaled to $(R_0,\Vcirc)\narreq(8.2\kpc,238\kms)$. This is the current most
precise measurement but is model-dependent; it would place corotation just inside the thin long bar and clearly within
the superthin bar. It is just compatible with all the uncertainties; alternatively it may suggest that the Hercules stream
has a different origin than the outer Lindblad resonance.

Considering all these determinations and the systematic uncertainties, we finally adopt a range of $R_{\rm
  CR}\narreq4.5\mhyphen7\kpc$, or $\Omega_{\rm b}\nsimeq 43\pm 9\kmskpc$ for our best estimated
$(R_0,\Vcirc)\narreq(8.2\kpc,238\kms)$. More accurate dynamical modelling of a wider set of stellar-kinematical data, in
particular from \gaia, is expected to narrow down this rather wide range in the coming years \citep{Hunt2014}.

\begin{marginnote}
\entry{$\Omega_{\rm b}$}{$43\pm 9\kmskpc$ Bar pattern speed}
\entry{$R_{\mathrm{CR}}$}{$4.5\mhyphen7.0\kpc$ Bar corotation radius}
\end{marginnote}

\section{STELLAR DISK}
\label{s:disk}
\begin{figure}
\begin{center}
\includegraphics[width=8cm]{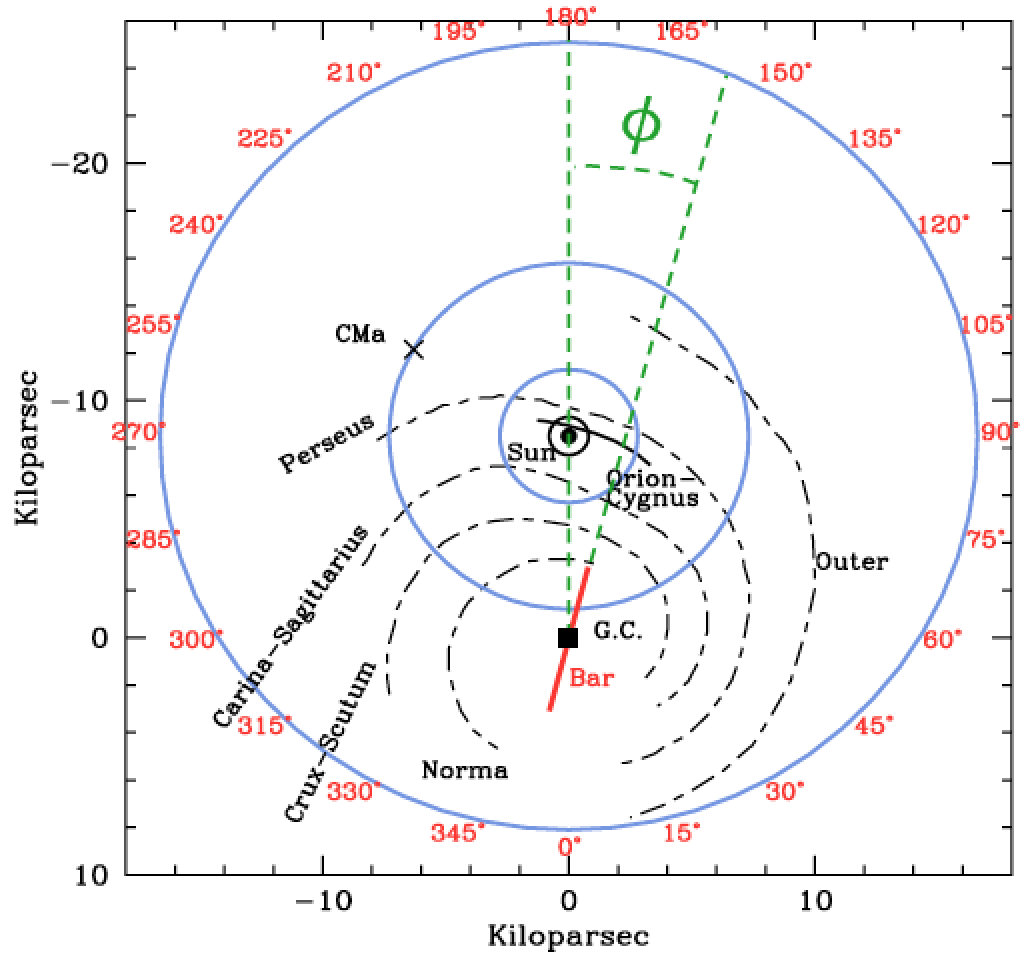}
\end{center}
\caption{Plan elevation of the Galactic disk centred on the Sun's position showing
the orientation and location of the main spiral arms.
The numbered outer circle defines galactic longitude ($\ell$). 
The Canis Major (CMa) overdensity
is in the same general direction as the maximum disk warp 
\citep[Courtesy of][]{Momany2006}.
\label{f:momany3}}
\end{figure}

Our vantage point from within the Galaxy allows us to obtain vast amounts of
unique information about galactic processes but this detail comes at a price.
The Solar System falls between two spiral arms (Fig.~\ref{f:momany3})
at a small vertical distance from
the Galactic Plane (\S 3.3). The thinness of the disk gives a fairly unobstructed view
of the stellar halo and the outer bulge. But deprojecting the extended disk remains
fraught with difficulty because of source confusion and interstellar extinction. Our 
off-centred position at the Solar Radius is a distinct advantage except that it
complicates any attempt to learn about large-scale, non-axisymmetries across the Galaxy.
We now have a better understanding of the structure of the inner Galaxy (\S4)
but the outer disk remains largely mysterious (see \S5.5 below).

Over the past decade, we have learnt much about the Galactic
stellar disk, but few parameters are known with any precision. The local surface density and 
the vertical density profile of the summed disk component (gas, stars, background dark matter) 
are known to high certainty (\S~\ref{s:vertdens}). But while photometric scalelengths are well 
established in external galaxies \citep{Lange2015}, 
the radial scalelengths and vertical scaleheights of the Galactic thin and thick disks are uncertain. 
Unlike for the inner Galactic bar/bulge region (\S4), we are not able to present a fully consistent 
picture for the disk at this time, although the key parameters are discussed and values recommended.
Major reviews of the stellar disk include \citet{Freeman2002,Ivezic2012,Rix2013}; 
the gaseous disk is discussed at length by \citet{Kalberla2008}. 

One of the most interesting developments is the recognition of a Galactic
`thick' disk that is distinct from the dominant thin disk through its unique 
chemistry \cite[e.g.][]{Bensby2014,Masseron2015,Hawkins2015}, 
in addition to its older age and higher elevation. 
Originally recognized by \citet[][]{Gilmore1983}, such disks appear to be 
ubiquitous in the local universe \citep{Yoachim2006}. In external galaxies, deep
stellar photometry reveals that the thick and thin disks have approximately equal 
scalelengths \citep[e.g.][]{Comeron2012}. Whether these old red disks 
have a distinct formation history or have continuous properties with the old thin
disk is unclear. Several authors have presented models where the Galactic thick disk
arises from a combination of stellar migration and/or flaring of the old disk stars,
such that its history is tied to the formation of the thin disk even though its 
mean metallicity may be different \citep[e.g.][]{Schonrich2009,Loebman2011,Minchev2015}.
In view of the distinct chemical signature of the thick disk, it is worthwhile
to quantify its properties separately from the thin disk regardless of its origins.

\subsection{Stellar photometry}
\label{s:diskphot}

Early studies of the Galaxy fitted simple models to the projected star counts in
a given optical or infrared band \citep[e.g.][]{Bahcall1980}. 
Without distance information, these authors found
that a variety of models fit the data 
(including many combinations of two exponential functions) 
due to degeneracies between 
structural parameters. Historically, fits to star counts with vertical distance $z$
have used a variety of functions but, today,
the exponential function is widely used to match both disk photometry \citep{Patterson1940}
and the peaked star counts close to the plane \citep{Wainscoat1989}.

By the end of the decade, we can expect accurate distances for millions
of stars from the \Gaia survey. But for now, we are dependent on
photometric parallaxes for determining stellar distances which
have a long history in their application to Milky Way structure \citep[][]{Gilmore1983,Kuijken1989,Chen2001}. 
The term `parallax' echoes the 
use of nearby bright stars with established trigonometric parallaxes to calibrate the
absolute magnitude-distance relation \citep[e.g.][]{Eggen1951}. (The term
`spectroscopic parallax' refers to the spectral classification used
to determine the absolute magnitude, rather than photometric colours.)

\subsubsection{Scaleheights}

All photometric studies find an exponential scaleheight
$\Zthin = 220-450$ pc for the dominant old thin disk at the Sun's location. The low extreme comes 
from multiband optical studies \citep[e.g.][]{Ojha2001} which include
stellar populations with a wide range of (especially younger) ages. These surveys
suffer from various biases that tend to suppress the scaleheight (see below).
The upper extreme is from early M dwarf studies where confusion with halo M giants
can lead to overestimates of the scaleheight \citep{Reid1993,Gould1996}. 
M dwarfs, which account for roughly half of all stars in the solar neighbourhood (\S~\ref{s:inventory}), trace 
the underlying total stellar mass. 
The much larger M star survey by \citet{Juric2008}, discussed 
below, finds $\Zthin \approx 300$ pc (to within 20\%) after various 
biasses are accounted for. This mid-range value is largely unchanged since 
Schmidt's early determination \citep{Schmidt1963}.

An improvement is to use two or more well calibrated optical bands to compare the 
magnitude counts in different colour bins. This led to Gilmore \& Reid's identification of
the thick disk after constructing the $V$ and $I$ luminosity functions for stars
at or above the main sequence turn-off (MSTO)
at different distances from the plane, and by ruling out biases due to interstellar extinction 
or metallicity gradients perpendicular to the disk. This classic study
observed 12500 stars towards the South Galactic Pole (SGP) brighter than $I=18$ and
provided the first reliable stellar densities vertical to the Galactic Plane $-$ their density
profile continues to compare well with modern derivations (\S~\ref{s:vertdens}). They
estimated $\Zthin \approx 300$ pc in line with modern estimates, and $\Zthick \sim 1450$ pc,
somewhat higher than what is believed today.

After 2000, the quality and angular extent of photometric data from wide-field CCDs 
improved dramatically \citep[e.g.][]{Finlator2000}. The internal accuracies of the multiband 
data led to improved estimates of photometric parallax and metallicity
(robust in the range -2 $<$ [Fe/H] $<$ 0) in wide-field surveys \citep[see][]{Ivezic2012}.
Notably, \citet{Siegel2002} observed select Kapteyn fields to derive photometric
distances for $130,000$ stars. They stressed the need to correct for unresolved
multiplicities (of order 50\%), otherwise stellar distances (and therefore scaleheights) are 
underestimated. The lower scaleheight estimates in the past are likely to have 
been underestimated for this reason. Even for old stars, there is some scaleheight variation
among dwarfs: 280$-$300 pc for early-type dwarfs ($5.8 < M_R < 6.8$) rising to 350 pc for 
late-type dwarfs ($8.8 < M_R < 10.2$). When averaging over old dwarfs, before and after the correction, 
they obtain $\Zthin \approx 290$ pc and $\Zthin \approx 350$ pc respectively; for the thick disk, they find
$\Zthick = 700-1000$ pc and $\Zthick = 900-1200$ pc respectively. (These
are derived from exponential models; sech$^2$ fits lead to 10\% smaller values after
correction for the factor of two difference in scaleheight between the exponential and 
sech$^2$ functions.)

For our subsequent disk analysis, we focus on the \SDSS $ugriz$ northern sky survey with its 
excellent photometric quality ($\sim$0.02 mag). 
With photometric data for 48 millions stars over 6500 deg$^2$, this is the largest 
to date with precise colour$-$magnitude vs. metallicity relations made possible 
using cluster calibrations across the sky \citep{Finlator2000,Chen2001}.
This survey gave rise to three major studies based on photometric parallax
\citep{Juric2008}, photometric metallicities \citep{Ivezic2008} and kinematics
\citep{Bond2010} discussed in the next two sections.

Juric et al (2008) exploit the faint magnitude limit ($r\sim 22$ mag) of the \SDSS\ survey
and target two groups of M dwarfs: a late M group 
with $1.0 < r-i < 1.4$ and an early M/late K group with $0.65 < r-i < 1.0$. The late M group is 
favoured because it is less sensitive to the halo population and to local substructure. 
At the Solar circle, their formal model fits for both
disk components are $\Zthin \approx 245$ pc and $\Zthick \approx 740$ pc before correction
for multiplicity, and closer to $\Zthin \approx 300$ pc and $\Zthick \approx 900$ pc after correction,
both with 20\% uncertainty. These constitute the most reliable values to date because 
of the \SDSS coverage in Galactic longitude and improved photometric distances over
the required physical scales. While the \SDSS and \TWOMASS surveys are widely
used in star count analyses, neither survey is ideal for determining the properties of the thick and 
thin disks {\it simultaneously}. Rather than cross-matching sources common to both catalogues 
\citep[e.g.][]{Finlator2000}, future studies will need to combine both surveys in order to provide better 
input catalogues \citep{Robin2014}.

While the thick disk was originally identified through stellar photometry, 
decompositions based on star counts are subject to degeneracies (\S~\ref{s:norm}).
We include a limited discussion of the photometric estimates because of the historical context 
and because the thin disk values are broadly correct. But we stress that
the thick and thin disk components are better separated through 
their distinct stellar chemistry. Numerous studies (\S~\ref{s:diskchem}) exploit either
stellar abundances based on multiband photometry (large samples, large
measurement errors) or spectroscopy (smaller samples, smaller errors).

\subsubsection{Scalelengths} 
\label{s:scalelength}

While vertical scaleheights are well determined at
optical and IR wavebands due to the low extinction towards the poles, this is not true
of disk radial scalelengths. We have analysed 130 refereed papers on disk parameters, 
with scalelengths ranging from 1.8 to 6.0 kpc.
In order to combat the effects of extinction, for observations that preceded the \SDSS
survey, the infrared point source measurements are the most reliable because they
tell a consistent story. This is particularly true for studies that target a
broad extent in Galactic longitude and observe in the anti-centre direction to ensure
they are less influenced by the presence of the central bar or by substructure. These
include the \spaceshuttle experiment \citep{Kent1991},
\DENIS \citep{Ruphy1996} in the anti-centre direction,
\COBE/FIRAS \citep{Freudenreich1998,Drimmel2001}, \TWOMASS 
\citep{Lopez-Corredoira2002,Cabrera-Lavers2005,Reyle2009}, 
and \GLIMPSE \citep{Benjamin+05}.  A statistical analysis of the main papers (15 in all) on this topic
leads to $\Rthin = 2.6 \pm 0.5$ kpc which includes the highest value ($3.9$ kpc) from 
the \GLIMPSE mid-infrared survey. Our estimate drops to $\Rthin = 2.5 \pm 0.4$ kpc if
we exclude the \GLIMPSE study.

We have already stressed the importance of M stars. \citet{Juric2008} determine 
$\Rthin \approx 2.6$ kpc (20\% uncertainty) for the thin disk. \citet{Gould1996} and 
\citet{Zheng2001}
used the \HST\ to measure a scalelength of $\Rthin = 3.0\pm 0.4$ kpc and $\Rthin =
2.8\pm 0.3$ kpc respectively. {\it All of these values are consistent with the IR measurements.}
A short scalelength finds strong support from dynamical studies of the stellar kinematics in \S5.4. 
From a study of mono-abundance populations (MAP), \citet{Bovy2012b} conclude that different 
populations give a scalelength that is smoothly changing from 2 kpc in the inner disk (older populations) 
to 3 kpc at $R = 12$ kpc (younger populations), or maybe even longer. However, the 
disk is dominated by old populations: a characteristic scale is meaningful when one considers the 
mass density profile of the disk. 
{\it The IR photometric estimate of radial scalelength is 
probably the most useful at the present time, although we recognize that an exponential decline in
the mass distribution is a crude approximation} (Fig. 10).
The thick disk scalelength is discussed in the next section.

For the past thirty years, some have argued for a longer disk scalelength when comparing
the Galaxy, with its high mass and luminosity, to external galaxies 
\citep[q.v.][]{vanderKruit2011}.
The \GAMA survey \citep{Driver2011} includes the largest bulge/disk decomposition survey to date using
the \VIKING K-band imaging survey of 4300 disk galaxies \citep{Lange2015}. This volume-limited survey
has a high level of completeness to a redshift limit of $z<0.06$ ($\Mstar > 10^9$\Msun). Here
only a few percent of galaxy disks exceed the Milky Way's disk mass, and their IR {\it photometric}
scalelengths have a large spread ($4\pm 2$ kpc). The Galaxy's high luminosity and small scalelength may 
not be so unusual.

\begin{figure}
\begin{center}
\includegraphics[width=8cm]{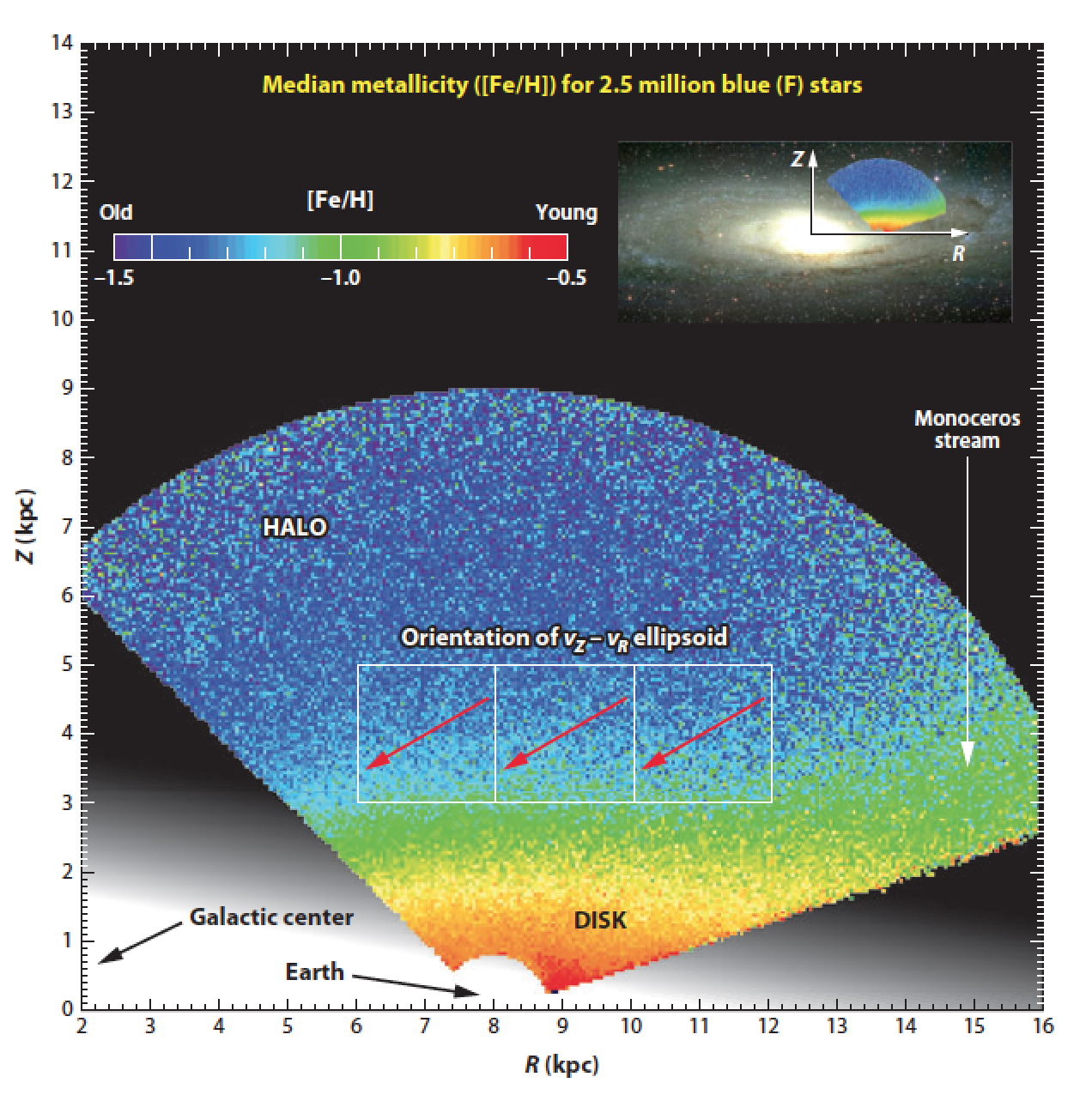}
\end{center}
\caption{A tomographic slice showing the change in stellar metallicity through the Galaxy
perpendicular to the Galactic plane and passing through the Galactic Centre and
the Sun. The underlying stellar density is shown in half tone \citep{Juric2008}.
The dwarf star distances and
metallicities are obtained from photometric parallax and photometric metallicities using the 
\SDSS $ugriz$ data calibrated using globular clusters over a range
of metallicities \citep{Ivezic2008}. The photometric metallicities are robust within the range -2 $<$ [Fe/H]
$<$ 0. This projection highlights the difficulty of fitting simple composite models to the
stellar disk, particularly in the outer reaches where the Monoceros stream and other
substructures become apparent. The direction of the
halo vertex deviation taken from \citet{Bond2010} is also shown 
\citep[Courtesy of][]{Ivezic2012}.}
\label{f:ivezic1}
\end{figure}

\smallskip\noindent{\sl Disk substructure.}
We highlight one spectacular result from the \SDSS survey. The team was able to extract 
tomographic slices through the Galaxy vertical to the plane (e.g. Fig.~\ref{f:ivezic1}). 
Fig.~\ref{f:ivezic1} illustrates a problem with Galaxy model fits and
goes some way to explaining the lack of convergence in disk parameters over
three decades. \citet{Juric2008} explicitly highlight important substructure
across the Galaxy including the `Virgo overdensity' and the `Monoceros ring'
\citep{Newberg2002}.
Substructure is so prevalent that it is not possible to fit a smooth double
exponential disk model in $R$ and $z$ to either the thick or thin disk without taking this
component into account.  This problem is well known for disk-halo fitting 
because the Sgr stream dominates so much of the halo.

\subsubsection{Thick disk normalization} 
\label{s:norm}

Several studies have tried to determine the local density normalization ($\frho = \rho^T/\rho^t$) 
of the thick disk compared to the thin disk with estimates ranging from 1\% to 12\% 
\citep[e.g.][]{Gilmore1983,Siegel2002,Juric2008}. The large uncertainty in $\frho$ is largely 
due to its degeneracy with the derived scalelengths for both components \citep{Siegel2002,Arnadottir2009}: 
higher estimates of \Zthick\ are associated with lower estimates of \frho, and vice versa. 
To aid comparison with most published results, we prefer this form for $\frho$
rather than normalization to the total disk mass \citep[e.g.][]{Piffl2014a}.
We have analysed all results from photometric surveys (25 in all) since the discovery 
paper and arrive at $\frho = 4\%\pm 2\%$.

A detailed analysis of the degeneracy between disk parameters is given 
by \citet{Chen2001} for late-type dwarfs chosen from the \SDSS survey where the 
data are separated into two hemispheres. Our value of \frho\ is in line with their 
likelihood analysis (see their Fig. 9) for a thick disk scale height of $\Zthick = 900\pm 100$ kpc. 
It is also broadly consistent with dynamical fitting to the Solar cylinder 
\citep[e.g. $f_\rho \approx 6\pm 2\%$;][]{Just2010}. 

The error is smaller when we compare the surface density of the thick and thin disks,
i.e. $\fsig = \frho \Zthick/\Zthin$. Here we find $\fsig = 12\%\pm 4\%$. 
Our analysis excludes all papers that do not fit simultaneously for the thin and thick disk.
At this point, no separation is made on the basis of detailed stellar abundance information.
A high thick disk local density ($\frho \approx 6\%$) is also found when spectroscopic 
abundances are used to define the high [$\alpha$/Fe] population \citep{Bovy2015} but this 
depends on how the abundance cut is made. 
A low value for $\frho$ is in conflict with \citet{Juric2008} who determine the thick
disk to be more massive ($\frho \approx 12\%$) at the Solar circle \citep[cf.][]{Fuhrmann2008}. 
Recent claims of a more massive thick 
disk may arise from the survey selection extending into the low [$\alpha$/Fe] population 
\citep[e.g.][]{Snaith2014} or from the use of a tiny survey volume \citep[e.g.][]{Fuhrmann2008}.


\begin{marginnote}
\entry{\Zthin}{$300{\small \pm}50$ pc, thin disk vertical scalelength at \Rzero}
\entry{\Zthick}{$900{\small \pm}180$ pc, thick disk vertical scalelength at \Rzero}
\entry{\frho}{$4\%{\small \pm}2\%$, thick / thin disk local density ratio at \Rzero}
\entry{\fsig}{$12\%{\small\pm}4\%$, thick / thin disk surface density ratio at \Rzero}
\entry{$\Rthin$}{$2.6{\small \pm}0.5$ kpc, thin disk radial scalelength}
\entry{$\Mthin$}{$3.5{\small\pm}1\times 10^{10}\Msun$, thin disk stellar mass}
\end{marginnote}

\begin{figure}
\begin{center}
\includegraphics[width=8cm]{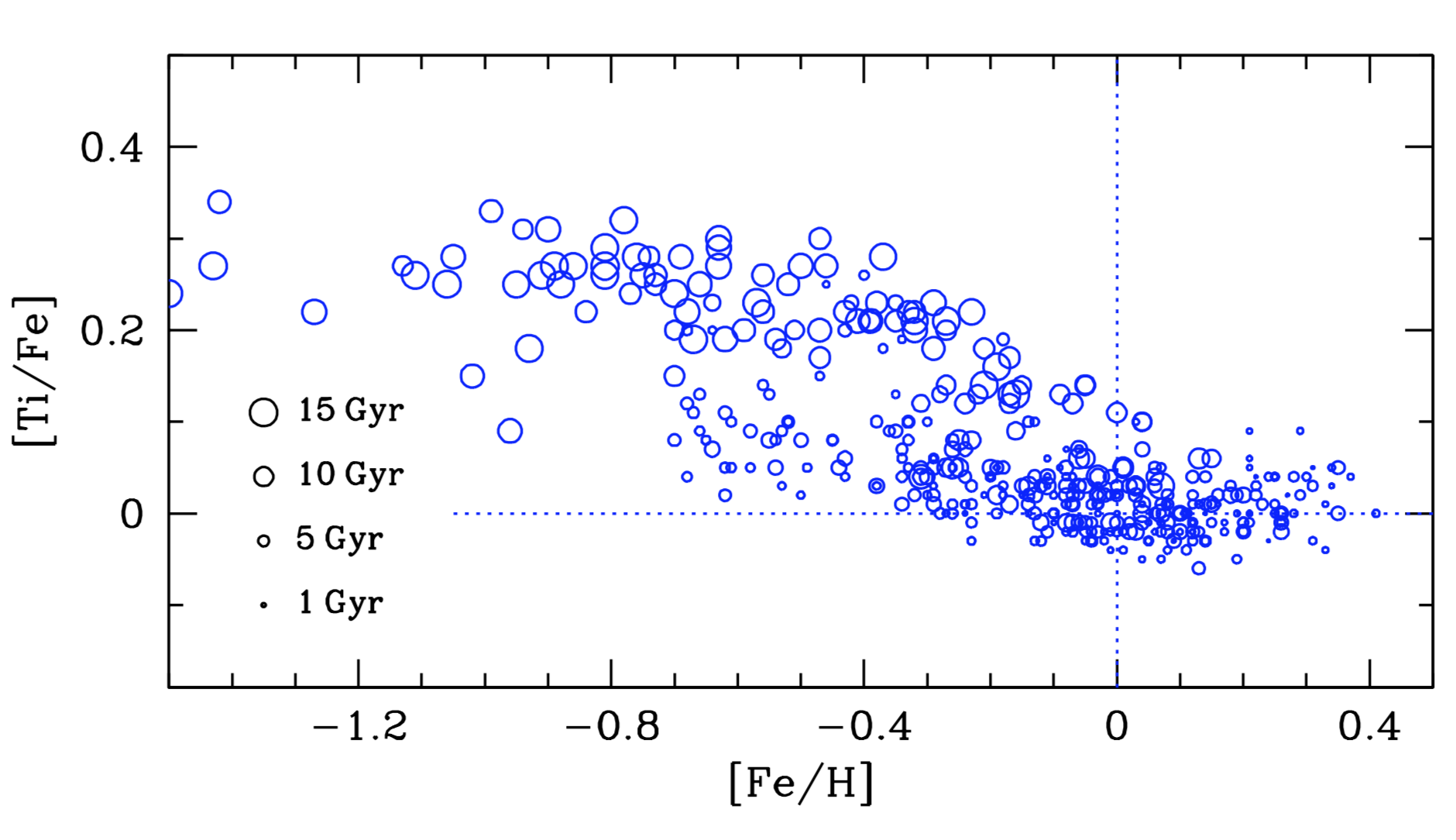}
\end{center}
\caption{[Ti/Fe] vs. [Fe/H] for 700 F, G dwarf stars with age determinations
showing a uniformly old population (thick disk) with
enhanced [$\alpha$/Fe] abundance, and a dominant (thin) disk population with a spread of ages.
The size of the circle scales with age and have uncertainties of order $\sim$Gyr; the metallicity uncertainties are smaller than 0.1 dex. Note
how the two-component disk appears to exist at the solar metallicity extreme
\citep[Courtesy of][]{Bensby2014}.}
\label{f:bensby1}
\end{figure}

\subsection{Stellar chemistry}
\label{s:diskchem}

\subsubsection{Photometric metallicity}

\citet{Ivezic2008} examined the vertical distribution in photometric metallicity of 2 million \SDSS stars
calibrated with \SEGUE spectra of 60,000 F, G dwarfs ($0.2 < g-r < 0.6$).  
An additional refinement was to combine the Palomar Optical Sky Survey (\POSS) and the \SDSS 
data to derive proper motions \citep[e.g.][]{Munn2004}. 
Here the tangential velocity accuracy for stars brighter than $g\sim 19$ is comparable to the radial
velocity accuracy of the \SDSS spectroscopic survey (15\kms\ for a star at 1 kpc;
100\kms\ at 7 kpc). The aim was to look for the thin disk-thick disk transition at $\sim$1 kpc, and the 
thick disk-halo transition at $\sim$2.5 kpc, in metallicity and velocity. 
The \SDSS team confirm earlier trends in declining metallicity and increasing lag at larger
elevations (Fig.~\ref{f:ivezic1}), and find evidence for thick disk stars extending to $z > 5$ kpc. 
They could only discern a gradual transition in photometric metallicity and kinematics across
the thin disk-thick disk divide, in conflict with the traditional two-component fit 
\citep[cf.][]{Bond2010}.
While no kinematic modelling was conducted at this stage (see \S~\ref{s:diskkin}), the \SDSS team 
find the data are more consistent with a gradual continuum from a thin young disk to an
extended older disk. A better understanding of the thick disk vs. thin disk separation 
had to wait for spectroscopic surveys (\S~\ref{s:diskchem}) providing both improved elemental 
abundances and 3D space velocities \citep{Steinmetz2006,AllendePrieto2008}.

\subsubsection{Spectroscopic metallicity}

In the past, some researchers have questioned the existence of a thick disk 
with discrete properties $-$ chemistry, age, kinematics $-$ beyond
the obvious characteristic of physical extent \citep[e.g.][]{Norris1991}.
This is a fundamental question because it hinges on the formation and 
evolution of the major baryonic component of our Galaxy.
In Fig.~\ref{f:bensby1}, the stars of the thick disk 
are mostly older than the thin disk and have a distinct chemistry
\citep{Bensby2003,Schuster2006,Haywood2006,Bensby2014}.
It is recognized today that [$\alpha$/Fe] is enhanced for the thick disk
compared to the thin disk over a wide range in [Fe/H], an effect that is easily seen
in sufficiently high-resolution (${\cal R} \gta 20,000$) spectroscopic data
\citep{Fuhrmann1998,Prochaska2000,Gratton2000,Reddy2003,Soubiran2003,Reddy2006}
even possibly at [Fe/H]$\gta$0 \citep{Haywood2015,Hayden2015,Kordopatis2015}.
Recent studies show that the earlier counter claims likely suffered from underestimating measurement
errors \citep[e.g.][]{Nidever2014} or incorrectly assumed that errors between 
measurements are uncorrelated \citep[e.g.][]{Schonrich2014}. 

The kinematic criteria often used to separate the disks inevitably lead to small stellar samples 
compared to photometric surveys. The Bensby
studies exploit the Geneva-Copenhagen survey of 16000 nearby dwarfs with full 3D space
velocities, and ages and metallicities from Str\"{o}mgren photometry \citep{Nordstrom2004}.
But a clean local separation
is hampered in part by extensive kinematic substructure over the local volume: 
the Hercules stream, for example, comprises both thick and thin disk stars 
\citep{Bensby2007}.
But even without kinematic separation, the chemical signature of 
two distinct populations is evident \citep{Adibekyan2012,Haywood2015}.

A new signature has emerged from the \APOGEE survey that may help to separate the 
disks further. \citet{Masseron2015} show that C/N is enhanced in the thick disk 
compared to the thin disk, presumably due to the effects of dredge-up observed
in old turn-off dwarfs and giants. The associated age (and mass) for the oldest stars is
broadly consistent with Fig.~\ref{f:bensby1}. At the present time, it is not possible to separate
the thick disk from the oldest thin disk stars through age-dating, assuming such a distinction even
exists. The detailed chemistry of the thick disk may provide a better
discriminator than the use of phase space ({\bf x,v}) and be the defining characteristic
of this component. Kinematic criteria will always lead to some overlap. 
\citet{Hawkins2015} emphasize that a more expanded chemical (${\cal C}$) space
may be called for to ensure that the thick disk can be separated from both the thin
disk and the lower halo, i.e. ${\cal C}$([$\alpha$/Fe], [C+N/Fe], [Al/Fe], [Mg/Mn]).
A high quality `chemical tag' defined in this way requires both optical and infrared data at 
high spectroscopic resolution.

The use of chemistry to define thick disk stars now gives a different perspective
on defining the relative density and scales of the two disks.
\citet{Bensby2010} found that the bimodal [$\alpha$/Fe] distribution
continues inside of the Solar Circle, in contrast to the outer disk where the enhanced
[$\alpha$/Fe] population is not detected, which led \citet{Bensby2011} to infer a shorter scalelength 
for the thick disk.
By dividing \SEGUE dwarfs spectroscopically into MAPs, \citet{Bovy2012b} note
the transition between a short scalelength, ``high [$\alpha$/Fe], low
[Fe/H] population," with a scaleheight up to 1000 pc, and a longer scalelength, 
``low [$\alpha$/Fe], high [Fe/H]'' population, with a scaleheight below 400 pc.
But the more extensive \APOGEE survey of 70,000 red giants reveals
that the picture is more complicated \citep{Bovy2015,Hayden2015}. Looking outwards, both tracks 
are evident, except the high [$\alpha$/Fe] sequence disappears beyond 11 kpc
while the low sequence is seen to at least 15 kpc.

Towards the inner disk, the high [$\alpha$/Fe] 
track dominates over the low [$\alpha$/Fe] sequence even though 
an increasing fraction of stars pile up 
towards higher metallicity due to the thin disk abundance gradient. 
Over the innermost disk, the thick disk and bulge are more difficult to 
disentangle because they share kinematic and abundance characteristics, as observed in
red giants  \citep{Alves-Brito2010,Ryde2010,Hill2011,Gonzalez2011}
and microlensed dwarf stars \citep{Bensby2009,Bensby2013b}.

In summary, estimates for the thick disk scalelength range from 1.8 to 4.9 kpc (12 papers)
\citep[e.g.][]{Cheng2012,Larsen2003} but few of these make a distinction based on chemistry. 
There exists now some convergence on \Rthick\ from surveys over very
different volumes and sample sizes. \citet{Bensby2011} estimate an exponential scalelength of 
2 kpc ($\approx$10\% accuracy), in good agreement with the high-[$\alpha$/Fe] MAPs from 
\SEGUE \citep{Bovy2012b}, although \citet{Cheng2012} measure $\Rthick \approx 1.8$ kpc 
using 7000 MSTO dwarfs from the same survey, albeit with larger errors.
More recently, \citet{Bovy2015} find $\Rthick = 2.2 \pm 0.2$ kpc for high-[$\alpha$/Fe] MAPs using 
\APOGEE red clump giants. Overall, we conclude that $\Rthick = 2.0\pm 0.2$ kpc where the 
thick disk is defined in terms of the high-[$\alpha$/Fe] population. When extrapolated to the
Galactic Centre, the ratios $\frho$ and $\fsig$ are 2.6 times higher than the local values. 
Thus, taking into account the shorter scale length of the thick disk,
we estimate $\Mthick = 6\pm 3\times 10^{9}\Msun$, or roughly one fifth of the thin disk mass.

\begin{marginnote}
\entry{$\Rthick$}{$2.0\pm0.2$ kpc, thick disk radial scalelength}
\entry{$\Mthick$}{$6{\small\pm}3\times 10^{9}\Msun$, thick disk stellar mass}
\end{marginnote}

\subsection{Stellar kinematics}
\label{s:diskkin}

The large photometric surveys, with their photometric distance estimates, provide us 
with initial estimates of the structural parameters for each of the major Galactic components. 
Over a decade ago, the community recognized that progress would require kinematic
information for many stars over large swaths of the sky.
Since that time, there has been extensive investment in wide-field stellar kinematic surveys,
some of which are still in progress: \GCS \citep{Nordstrom2004};
\SEGUE \citep{Yanny2009}; \RAVE \citep{Steinmetz2006};  
\APOGEE \citep{AllendePrieto2008}; \LEGUE \citep{Deng2012};
\GES \citep{Gilmore2012}; and \GALAH \citep{DeSilva2015}. 
These surveys look at different parts of the sky and go to different depths. Some
have been rendered dynamically as 3D visualizations at the following website:
https://www.rave-survey.org/project/gallery/movies. The
\GCS survey covers the full sky but is confined to the Solar Neighbourhood;
the \SEGUE, \RAVE and \APOGEE surveys penetrate deeper into the Galaxy than
earlier stellar surveys.
By far the largest of the new surveys, the \Gaia mission will obtain both spectra and
astrometric information for up to $\sim$150 million stars \citep{deBruijne2015}. 
By the end of the decade, we can expect to have radial velocities and stellar 
parameters for millions of stars, for dwarfs out to $\sim$1 kpc, and for giants
out to the halo. 

Here we review the main insights to
emerge from these surveys. We make a distinction between dynamical models 
of the Galaxy \citep[e.g.][\S 6]{Rix2013} and models that fit to the separate
Galactic components without dynamical consistency
 \citep[e.g.][]{Catena2010}, which we refer to as kinematic models.
In a kinematic model, one specifies the stellar motions
independently at each spatial location, and the gravitational
field in which the stars move plays no role. In a dynamical model,
the spatial density distribution of stars and their kinematics are
self-consistently linked by the potential, under the assumption
that the system is in steady state.

When fitting a model, there are important considerations. First, all surveys are
defined by their selection function (e.g. magnitude, velocity, coordinates) and any
biasses must be accomodated by the analysis. 
Secondly, it is important to make a stab at including a plausible star formation history
into the analysis \citep[e.g.][]{Schonrich2009}. Making more stars
in the past places more old stars at higher galactic latitudes today, and therefore higher kinematic
dispersion through the age-velocity dispersion relation \citep[e.g.][]{Aumer2009}. 
But this adds to the complexity because (a) we are introducing 
new variables into an already crowded field; and (b) degeneracy exists between different 
parameters, e.g. the star formation rate and the slope of the initial mass function 
\citep{Haywood1997}.

An increasingly popular approach to fitting is to use Bayesian optimization over
a broad set of free parameters \citep[e.g.][]{Catena2010}. 
As a worked example, in Appendix A, we summarize the \GALAXIA\ framework for fitting up to
20 disk parameters $-$ this code is freely available at http://sourceforge.net/p/galaxia/wiki/Home/. 
The approach is theoretically simple and allows for useful constraints on local kinematic
properties. The framework incorporates a constant star formation rate and a 3D Galactic 
dust model. Uniquely, it is designed to correct for an arbitrary survey selection function and can be 
used to fit data to analytic functions or numerical simulations \citep{Sharma2011}.

\citet{Sharma2014} apply the method in Appendix A to the \RAVE and \GCS surveys by
assuming the structural form of the disk $f(r)$ and then attempting
to fit for $f(v\vert r)$ from the surveys. They use {\it only} sky position
and velocity for each star as these are the most accurate 
measurables. No distance information is supplied other than what is implicit in the model fitting. 
They fit both the Gaussian DF and the dynamically-motivated Shu DF 
\citep{Sharma2013}; the latter performs better because it allows for asymmetric 
velocity distributions (relative to the Sun) 
due to non-circular motions experienced by most stars.

Groups that use Bayesian optimization \citep{Bovy2012,Sharma2014} typically fit
for (i) the age-velocity dispersion relation (see below);  (ii) the radial scalelengths (\RsigthinR, \RsigthickR) of the 
velocity dispersion profile; (iii) the mean stellar motion \Vmean\ with vertical height $z$; and (iv) the solar motion (\Vsolar). 
The \RAVE team find that earlier estimates of the local standard of rest (LSR) are unreliable
if they neglect the vertical dependence of the mean azimuthal motion for a given population. 
Ultimately, even after this correction,
global kinematic measures like $\Theta_0$ are expected to have systematic uncertainties because of the initial
assumption on $f(r)$ and the lack of dynamical consistency. We return to these measures in the next section.

\subsubsection{Age-velocity dispersion relation (AVR)}
\label{s:AVR}

It is well established that the velocity dispersion of a disk stellar population 
$\sigma(R,\phi,z)$ increases with age \citep{Stromberg1925,Wielen1977}. 
Disks heat because a cold thin disk occupies a small fraction of phase space, and fluctuations in the 
gravitational field cause stars to diffuse through phase space to regions of lower phase space density.
These effects are very difficult to model reliably through numerical simulations. For the thin disk,
the AVR is sometimes 
approximated as a power-law in cosmic time \citep[e.g.][]{Aumer2009} such that
\be
\sigma(R,\phi,z) = \sigma_0(R,\phi,z) \left( { {\tau+\tau_{\rm min} }\over{ \tau_{\rm max}+\tau_{\rm min} } }\right)^{\beta_{R,\phi,z}}
\label{e:aumer}
\ee
where ($\tau_{\rm min},\tau_{\rm max}$) are priors.
The power-law indices ($\beta_R, \beta_\phi, \beta_z$) provide important 
information on disk heating parameters \citep{Binney2013,Sellwood2013}; a 
summary of estimates is given in Table~\ref{t:beta}. The widely used value of $\beta = 1/3$ dates 
back to the cloud scattering model of \citet{Spitzer1953}. While useful, this form is not 
universally accepted \citep[e.g.][]{Freeman2001,Quillen2001}; it makes no allowance for the 
thick disk which must be treated separately. From existing data, it is very difficult to distinguish a continually rising
AVR from one that steps or saturates at old ages \citep{Aumer2009,Casagrande2011}.
Some groups attempt to fit for age-metallicity trends in the thin
disk but such fits are less instructive at the present time
\citep[q.v.][]{Freeman2002}, although ultimately this
information will need to be accommodated in a successful model \citep{Sanders2015}.

\begin{table}
\caption{\label{t:beta}
Comparison of measured or quoted $\beta$ indices in the age-velocity dispersion relation from stellar kinematic surveys. The quoted errors are
statistical and do not include systematic errors that are typically larger.
}
\centering
\vspace{0.3cm}
\begin{tabular}{|l|l|lll|l|} 
\hline
Reference & Survey & $\beta_R$ & $\beta_{\phi}$ & $\beta_z$ \\ \hline
Nordstrom et al. (2004) & GCS & $0.31\pm 0.05$ & $0.34\pm 0.05$ & $0.47 \pm 0.05$ \\ 
Seabroke \& Gilmore (2007) & GCS & - & - & $0.48 \pm 0.26$ \\ 
Holmberg et al. (2007) & GCS & 0.38 & 0.38 & 0.54 \\ 
Holmberg et al. (2009) & Hipparcos, GCS & 0.39 & 0.40 & 0.53 \\ 
Aumer \& Binney (2009) & Hipparcos, GCS & 0.31 & 0.43 & 0.45 \\ 
Just \& Jahrei$\beta$ (2010) & Hipparcos & - & - & 0.38 \\  
Sharma et al. (2014) & GCS  & 0.20$\pm$0.02 & 0.27$\pm$0.02 & 0.36$\pm$0.02 \\ 
Sharma et al. (2014) & RAVE & 0.19$\pm$0.02 &  - & 0.3-0.4 \\
Sanders \& Binney (2015) & SEGUE & 0.33 & - & 0.4 \\ \hline
\end{tabular}
\end{table}

\subsubsection{Velocity dispersion profile}
\label{s:VDP}

\citet{Pasetto2012a}  used the \RAVE survey to learn about the variation of velocity dispersion
in the $(R,z)$ plane. They used singular value decomposition to compute the moments of the
velocity distribution. As expected, the thin disk stars follow near circular, co-rotational orbits with a 
low velocity dispersion \citep[e.g.][]{Edvardsson1993,Reddy2003}.
The velocity dispersion falls as a function of distance $R$ from the Galactic Centre, 
consistent with theoretical expectation \citep{Cuddeford1992}.

In an exponential disk, the stellar dispersion declines radially with the disk surface density
$\Sigma^t$ and scale height \Zthin\ as 
$\sigma^t(R) \propto \sqrt{z^t(R)\; \Sigma^t(R)}$. 
\citet{Sharma2014} combined equation~\ref{e:aumer} with an exponential factor in radius
\citep[cf.][]{vanderKruit1986} in order to determine the radial scalelengths (\RsigthinR, \RsigthickR) of the 
dispersion profile \citep[e.g.][]{Lewis1989}. For an isothermal disk, $R_\sigma$ is expected to be
roughly twice the disk density scalelength \citep{Bottema1993}.
The \RAVE study confirms that the dispersion profile declines with radius, yielding
estimates of \RsigthinR\ $\sim$ 14 kpc and \RsigthickR\ $\sim$ 7.5 kpc. 
The thick disk value is in good agreement with a full dynamical analysis which we 
return to below. The thin disk is insufficiently constrained in the \RAVE survey because the elevated
sightlines in latitude do not extend far enough in radius and are susceptible to vertical dispersion 
gradients \citep[cf.][]{Piffl2014a,Sanders2015}.

\citet{Bovy2012a} divided up the \SEGUE survey into MAPs and 
argued that these constituted quasi-isothermal populations. 
(We refer the reader to \citet{Sanders2015} 
for a different perspective on how to treat chemical information in fitting Galactic models.)
Bovy found no break in vertical dispersion between the old thin
and thick disk and suggested that the thick disk is a continuation of the 
thin disk rather than a separate entity \citep[cf.][]{Schonrich2009}.
In contrast, kinematic and dynamic studies $-$ which include a star formation 
history and an age-velocity dispersion relation explicitly for the thin disk $-$ do 
tend to find a break in the vertical stellar dispersion (e.g. Table~\ref{t:jbh1}). 
Future studies that exploit improved stellar ages and chemistry 
will be needed to resolve this issue. 
\subsubsection{Solar motion and LSR}
\label{s:solarmotion}

\citet{Delhaye1965} determined the solar motion \Vsolar\ with respect to the Local Standard of Rest (LSR) defined in the
reference frame of a circular orbit that passes through the Sun's position today.  Formally, for a coordinate system
based at the Sun, where the ${\bf i}$ unit vector points towards the Galactic Centre, ${\bf j}$ in the direction of
rotation, and ${\bf k}$ is upwards from the disk,
$ \Vsolar = \Usol {\bf i}\> + \Vsol {\bf j}\> + \Wsol {\bf k}$.
Delhaye studied different spectral classes and luminosity groups, and arrived at $\Vsolar \approx (9, 12, 7)$\kms\ that
is very respectable by modern standards.

The random stellar motions of a given population is a strong function of their mean age, colour, metallicity and
scaleheight. Care must be taken to account for the asymmetric drift of stellar populations in taking the limit to the
zero-dispersion LSR orbit. Using Str\"omberg's relation, \citet{Dehnen1998} measured ($\Usol,\Vsol,\Wsol$) $=$ (10.0,
5.2, 7.2)\kms\ from the \Hipparcos survey, which for two ordinates are in excellent agreement with Delhaye's early
estimates. $\Vsol$ was later revised upwards close to the original value \citep{Binney2010,Schonrich2010}; the latter
paper showed that Str\"omberg's linear asymmetric drift relation is invalidated by the metallicity gradient in the
disk.  Our values for \Vsolar\ in the margin note are averaged over most estimates since Delhaye's original estimate. We
have removed extreme outliers and ignored early values from any researcher who revised these at a later stage using a
similar method.  Thus, the Sun is moving inwards towards the Galactic Centre, upwards toward the NGP and, given \zzero,
away from the plane.

While ($\Usol, \Wsol$) have converged on Delhaye's original values within the errors (see the margin note), some
uncertainty surrounds $\Vsol$ when considered across the \APOGEE, \RAVE and \LEGUE surveys
\citep{Bovy2012d,Sharma2014,Tian2015}.  One reason for this may arise from local kinematic substructure or any systematic
streaming motion \Vstr\ in the Sun's vicinity \citep{Dehnen2000,Antoja2014,Siebert2011,Williams2013}.  A local spiral
arm density wave, for example, can impose kinematic fluctuations of order 10\kms\ \citep{Siebert2012}. In addition,
because the corotation radius of the Galactic bar may be as close as $\sim\!2\kpc$ inward from the Sun
(\S\ref{s:omega}), systematic streaming velocities may exist in the local Galactic disk due to perturbations from the
bar and adjacent spiral arms.  These could cause deviations of the zero-dispersion LSR orbit from the {\sl average}
circular velocity at $R_0$, defined as the angular rotation velocity of a 'fictitious' circular orbit in the
axisymmetrically averaged gravitational potential, the so-called {\sl rotational standard of rest} \citep[RSR,
see][]{Shuter1982,Bovy2012d}. Analysing two mostly independent samples of stars from \APOGEE, \RAVE and
\GCS, \citet{Bovy2015} modelled the disk velocity field over $\sim\!3$-$4\kpc$ scales and found such effects,
with an implied tangential LSR streaming velocity of $14\pm3\kms$ relative to the RSR. On the other hand,
\citet{Sharma2014} find little difference in $\Vsol$ between the local GCS survey and the \RAVE data which extends to
$\sim\! 2\kpc$, and \citet{Reid2014a} when fitting their maser velocities with a circular orbit velocity field find no
evidence for a deviation of the globally fitted $\Vsol$ from the locally determined value.  Globally determined values
of $\Usol$ \citep{Bovy2012d,Reid2014a} agree well within errors with the locally determined $\Usol=10\pm1\kms$. Here we
adopt $\vert\Vstr\vert \narreq0^{+15}\kms$ because we cannot establish clear agreement on the magnitude of the
streaming motion at the present time.  Future studies are anticipated which compare the impending \gaia data with models
including the central bar and spiral density waves in view of understanding both the random and streaming motions in the
disk.

\begin{marginnote}
\entry{$\Usol$}{$10.0{\small \pm}1$\kms, solar motion in $U$}
\entry{$\Vsol$}{$11.0{\small \pm}2$\kms, solar motion in $V$}
\entry{$\Wsol$}{$7.0{\small \pm}0.5$\kms, solar motion in $W$}
\entry{$\vert \Vsolar \vert$}{$15.5{\small \pm}3$\kms, solar vector motion}
\entry{$\vert \Vstr \vert$}{$0^{+15}$\kms, LSR streaming motion}
\end{marginnote}

\subsubsection{Vertex deviation}
\citet{Binney2014} revisit the \RAVE data but include distance estimates using 
\citet{Burnett2011}. These were not employed by \citet{Sharma2014} because \RAVE distances 
are susceptible to uncertainties in proper motions and stellar parameters, e.g. log-g 
\citep{Zwitter2010,Anguiano2015}.
After excluding young stars, Binney finds that the velocity dispersion for a given stellar population 
increases as one moves vertically in $z$ from the plane \citep[cf.][]{Smith2009}.
Furthermore, at any location in $(R,z)$,
the velocity ellipsoid's long axis (vertex deviation) points towards the Galactic Centre 
(see Fig.~\ref{f:ivezic1}),
indicating that the radial and vertical motions of stars are intimately coupled 
\citep[cf.][]{Siebert2008,Bond2010}.
This important result demonstrates that the stellar motions 
in $R$ and $z$ are entwined through the Galactic potential.

\begin{table}
\caption{\label{t:jbh1}
Parameter estimates from Galactic dynamical models with disk fitting$^a$.}
\begin{center}
\begin{tabular}{llllllllll} 
Parameter                 &    M11                             &     B12$^a$        &   BR13$^b$   &   P14     &   SB15  &  BP15  \\
\hline
Kinematic data		&   \HI\				&   GCS	     &   SEGUE &  RAVE  &  SEGUE  &  RAVE \\
\Rzero\   (kpc)          &   8.3                                &   [8.0-8.4]    &     [8.0]     &    [8.3]      &    [8.0]   &  [8.3]   \\
\Vcirc\   (\kkms)        &   239                                &   [220-241]  &      218        &     [240]       &  [220]  &  [240]  \\
$M_{200}$ (\Msun)        &   1.3$\times 10^{12}$      &      -              &     -    &   1.3$\times 10^{12}$         &    -  &   1.4$\times 10^{12}$ \\
\Mcold (\Msun)$^c$     &   7.1$\times 10^{10}$     &      -              &    6.8$\times 10^{10}$   &   5.6$\times 10^{10}$  &     -   &  6.2$\times 10^{10}$  \\
\fcold (\%)$^d$         &  4.9                                &     -               &   -    &   4.3        &   -    &   4.2   \\
\Rthin\    (kpc)           &    3.0                               &   2.7-3.1     &    2.2        &   2.7=      &    2.3  &  3.7=  \\
\Rthick\   (kpc)          &   3.3                               &   2.1-3.6     &     -        &   2.7=      &    3.5  &  3.7= \\
\hline
\RsigthinR (kpc)$^e$         &    -                                   &   3.6-20      &     -        &   9.0=        &    7.8=   &  2$\times$\Rdisk= \\
\RsigthinZ\ (kpc)         &    -                                   &   -               &     -        &   9.0=        &    7.8=   &  2$\times$\Rdisk= \\
\RsigthickR\ (kpc)        &    -                                  &   -               &     -        &    13       &    6.2=    &  11.6 \\
\RsigthickZ\ (kpc)        &    -                                  &   -               &     -        &    4.2       &    6.2=    &  5.0 \\
\hline
\sigRthin\  (\kkms)    &     -                                  &  40-42        &    -        &     34      &     48     & 35  \\
\sigZthin\   (\kkms)   &    -                                    &  20-27       &     -        &     25      &    31     &  26  \\
\sigRthick\ (\kkms)   &     -                                   &  25-28       &     -        &     51      &    51     &  53  \\ 
\sigZthick\  (\kkms)  &    -                                     &  33-65       &     -        &     49      &    51     &  53  \\
\hline
\end{tabular}
\end{center}
\begin{tabnote}
$^a$The thin and thick disks are treated separately for two distinct potentials and parameter sets (b, c).
$^b$\Mcold\ has been rescaled to \Rzero = 8.3 kpc for ease of comparison with other results. 
$^c$\Mcold\ includes the stellar disk, the bulge and the cold gas disk; the M11 total stellar mass has been corrected
for a gas mass of $0.7\times 10^{10}$\Msun\ in line with the other references, except that BP15 assumes a gas 
mass of $1.7\times 10^{10}$\Msun. $M_{200}$ assumes a spherical halo ($q=1$) with $q<1$ leading to higher values;
\Mcold\ does not include the Galactic corona; thin disk dispersions are evaluated at 10 Gyr. 
$d$\fcold\ is the ratio of \Mcold\ to the total galaxy mass.
$^e$$R_\sigma$ is a parameter that sets the scale of the outward radial decline in velocity dispersion within the disk.
Note: All models (except M11) apply Bayesian fitting of action integrals. Key: [...] indicates a prior; `$=$' indicates values locked in
fitting for \Rdisk. 
References: M11 - McMillan (2011); B12 - Binney (2012); BR13 - Bovy \& Rix (2013); P14 - Piffl et al (2014a); 
SB15 - Sanders \& Binney (2015); BP15 - Binney \& Piffl (2015).
\end{tabnote}
\end{table}
\subsection{Stellar dynamics}
\label{s:diskdyn}

Kinematic models offer greater freedom than physics really allows.
This can lead to systematic errors in parameter estimation which typically 
swamp the statistical errors in the optimization scheme.
The way forward is to consider dynamical models such that the 
spatial density distribution of stars and their kinematics are
linked by the gravitational potential $\Phi$, under the assumption
that the system is in steady state. At present, dynamical models
are used to estimate a subset of parameters 
explored by kinematic methods (see Table~\ref{t:jbh1}). While there has been good
progress in recent years, with the first signs of dynamical self-consistency
beginning to emerge, there is no fully consistent framework at the present
time. We refer the reader to \citet[][\S6]{Rix2013}
for a useful summary of the dynamical methods on offer.

Early methods that operate locally or in annular bins 
\citep[e.g.][]{Bienayme1987,Just2010} have 
given way to holistic treatments over one or more dynamical components
of the Galaxy \citep[e.g.][]{Piffl2014a,Sanders2015}.
Dynamical models assume an equilibrium figure such that
the phase space density of stars $f({\bf x}, {\bf v})$ links only to the 
phase space coordinates through the constants of motion (Jean's
theorem).

\citet{Binney2010,Binney2012}
has argued that action integrals (${\bf J}$) are ideal for building dynamical models
because they are adiabatic invariants. 
The most convenient action integrals are (i) $L_z$,
the approximate symmetry axis of the Galaxy's angular momentum, (ii) $J_z$, to 
describe the action of a star perpendicular to the plane, and (iii) $J_R$, to describe
the radial action of the star. The DF at any point in
the Galaxy has the form $f(L_z, J_z, J_R)$. $\Phi$ is derived through an iterative
process starting with an initial guess $\Phi_i$ to get to the density $\rho_i = \int f\; d^3v$
integrated over the phase-space volume $v$.
Poisson's equation is used to compute an updated $\Phi_{i+1}$ and the process is
repeated until convergence is achieved.
The numerical procedures are non-trivial and computationally expensive, but entirely
feasible for surveys involving thousands of stars \citep{Binney2011}. 

Of the few action integral studies of the disk to date, we highlight the work of \citet{Binney2012} using \GCS;
\citet{Bovy2013} and \citet{Sanders2015} using \SEGUE; \citet{Piffl2014a} and \citet{Binney2015} 
using the \RAVE survey. In Table~\ref{t:jbh1}, these studies attempt to arrive at unbiassed
parameter estimates through fitting the data with a consistent dynamical model.
\citet{Binney2012} introduced important new ideas in model fitting, including the use of the quasi-isothermal DF
to model the disk. But his focus on a very local sample led to the disk dispersions being 
underestimated due to the \GCS bias towards younger stars. As we return to below, this work has been superseded by 
\citet{Piffl2014a} who determined the DF from the \RAVE giants. The \RAVE survey comprises roughly equal 
numbers of dwarfs and giants, most within about 2.5 kpc of the Sun, and is thus more representative of the 
extended disk \citep{Sharma2011}.

\citet{Bovy2013} build on \citet{Binney2012} using 16300 G dwarfs from the \SEGUE survey.
They divide stars in the $(\alpha,{\rm Fe})$ abundance plane into MAPs: 
the more metal-rich MAPs trace the inner disk whereas the metal-poor 
populations trace the disk beyond the Solar Circle. Their goal is to measure the Galactic 
disk's mass profile by identifying a radius for each MAP where the
modelling gives a tight (statistical) constraint on the local surface density. It is unclear
whether (a) most MAPs can be treated as quasi-isothermal populations; and (b) the constraints at 
different radii from the MAPs are mutually consistent, i.e. free of systematic errors
\citep[e.g.][]{Sanders2015}.
The disk scalelength is the most important unknown in 
disentangling the contributions from the disk and the dark halo to the mass distribution near the disk.
In contrast to the photometric radial profile, without dynamical consistency, {\it the mass-weighted radial profile cannot have been reliably measured in earlier kinematic studies.}

\citet{Sanders2015} revisit the \SEGUE analysis and 
instead treat the chemical (and phase) space as a continuous distribution. They introduce
the idea of an `extended distribution function' (EDF) $f({\bf J}, {\bf Z})$ in place of the DF $f({\bf J})$
where {\bf Z} defines the chemical domain. While different abundance groups can have very diverse
kinematics, they all necessarily reside within the same Galactic potential; the
extra information in the EDF allows for a more accurate treatment of the selection function
and associated errors across the survey. 
They determine that the thick disk vertical dispersion \sigZthick\ is
a factor of two larger than the thin disk value (\sigZthin $\approx$ 25\kms), in agreement with 
\citet{Sharma2014} and \citet{Piffl2014a}. 
Unlike either of these studies, they obtain sensible numbers for
both \RsigthinR\ and \RsigthickR\ for the first time, with a disk 
scalelength \Rthin\ ($\approx 2.3$ kpc) 35\% smaller than 
the thick disk \Rthick\ (Table~\ref{t:jbh1}), but in conflict with the \APOGEE survey \citep{Hayden2015}. 
(While $R_\sigma$ defines the scale of the outward decline of the stellar dispersion,
it is not the exponential radial scalelength used in the kinematic studies because DFs
are expressed in terms of integrals of motion, not radii.)
Sanders \& Binney 
state that their analysis is only preliminary because $\Phi$ was kept fixed throughout.
Arguably, this study comes closest to the ideal of chemodynamical self-consistency.
In the margin note, we adopt the \RAVE velocity dispersions \citep{Piffl2014a} as these
are consistent across studies and extend further into the lower latitudes of the disk.
\begin{marginnote}
\entry{\sigRthin}{$35{\small \pm}5\kms$, old thin disk radial velocity dispersion at \Rzero}
\entry{\sigZthin}{$25{\small \pm}5\kms$, old thin disk vertical velocity dispersion at \Rzero}
\entry{\sigRthick}{$50{\small \pm}5\kms$, thick disk radial velocity dispersion at \Rzero}
\entry{\sigZthick}{$50{\small \pm}5\kms$, thick disk vertical velocity dispersion at \Rzero}
\end{marginnote}
\begin{figure}
\centering
\parbox{6cm}{\label{f:piffl1}
\includegraphics[width=6cm]{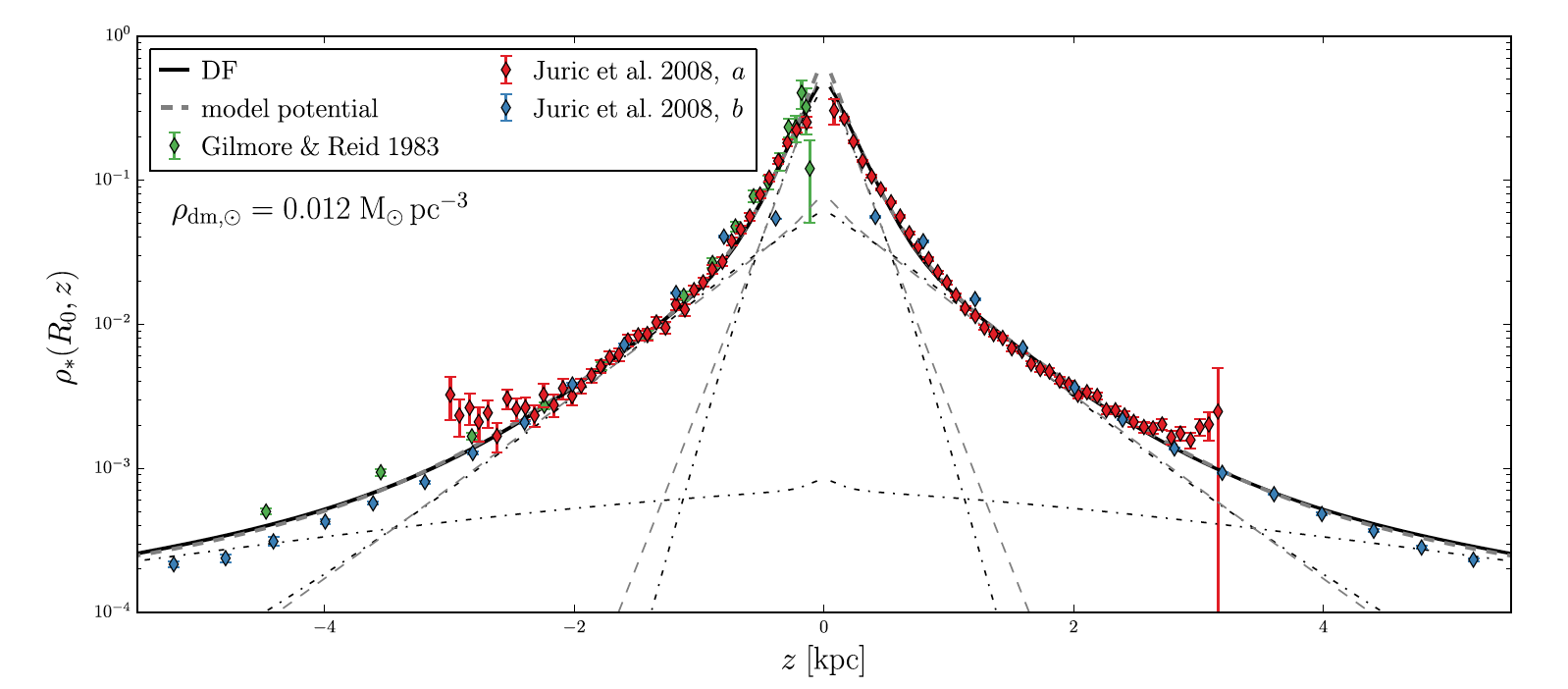}
}
\qquad
\begin{minipage}{6cm}
\includegraphics[width=6cm]{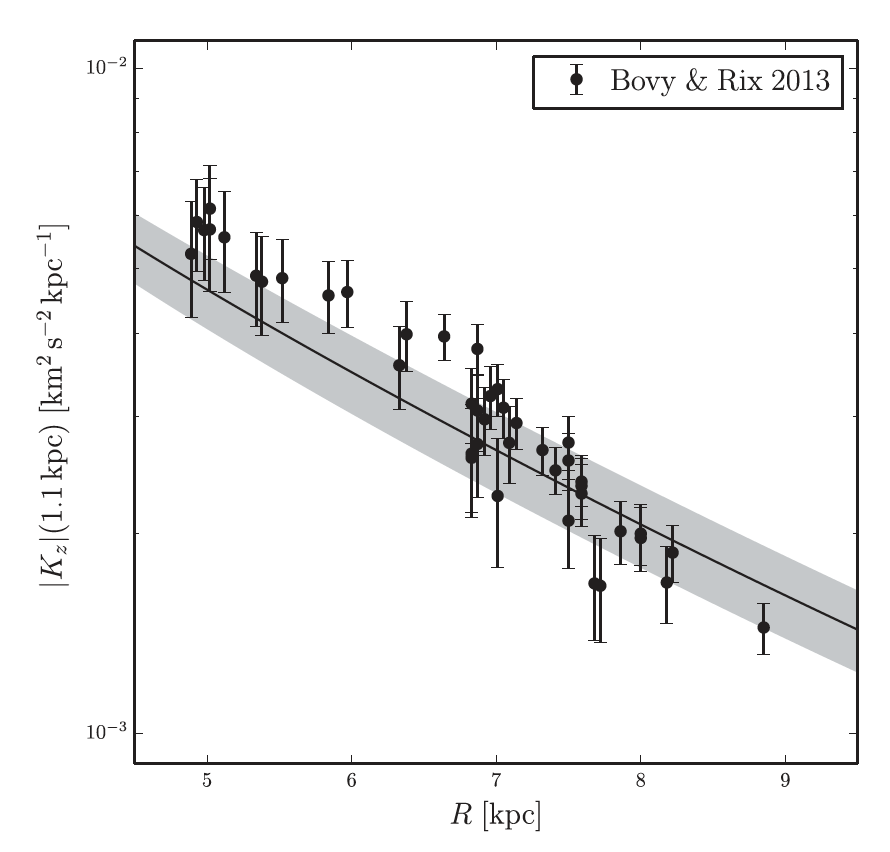}
\end{minipage}
\caption{(Left) The vertical density profile at the position of the Sun where the data points
are from \citet{Gilmore1983} and \citet{Juric2008}. The total density from the potential (sum of disk, bulge and halo), 
shown as a dashed line, follows the predicted total density from the DF using the dynamical
model shown as a solid line. The model fit exhibits a high degree of dynamical consistency over
the local disk.
(Right) Estimates of \Zacc\ over the radial range $5 < R < 9$ kpc at a vertical height of $z = 1.1$ kpc.
The data points are from the \SEGUE dwarf survey \citep{Bovy2013} and the solid line (grey band)
(uncertainty) is a dynamical model fit to the \RAVE survey \citep{Piffl2014a}. While the data
have a slightly shorter scale length compared to the \RAVE model, there is a moderate dynamical
consistency between them \citep[Courtesy of][]{Piffl2014a}.
}
\end{figure}
\subsubsection{Vertical density and acceleration}
\label{s:vertdens}
Measurements of the local baryon and dark matter density have a long tradition in astronomy 
\citep{Kapteyn1922}.
From a survey of K giants towards the SGP,  
\citep[][]{Kuijken1989}
went further and attempted to 
derive the vertical density profile $\rho(z)$ and the gravitational acceleration \Zacc\
induced by the local disk that are related through Poisson's equation $\nabla.\Zacc = -4\pi G\rho$.
This is transformed into a surface density $\Sigma$ such that 
\be
\Sigma = -{{\Zacc}\over{2\pi G}} + \Delta \Sigma
\ee
evaluated over the column $z=\pm 1.1$ kpc or $z = \pm\infty$. As \citet{McKee2015} point out, this must
be done with care. The usual assumption that $\Delta \Sigma = 0$ for a flat rotation curve at the midplane 
leads to an error because, in this instance, the rotation curve cannot be flat at a fixed height off the plane. 
In any event, this is a difficult measurement to get right because
stellar surveys are strongly biassed towards more distant stars \citep[cf.][]{Zheng2001}.

We can be confident that we {\it do} understand the local acceleration of
the disk and the make-up of the local density (\S~\ref{s:inventory}) over a vertical distance of 
$\vert z \vert = 1.1$ kpc \citep{Kuijken1989,Bovy2012}. A dynamical fit to 200,000 giants in the 
\RAVE survey \citep{Piffl2014a} leads to a local determination of \Zacc\ and its local gradient (Fig.~\ref{f:piffl1}) 
which are in generally good agreement with the \SEGUE dwarf analysis.

The vertical density profile $\rho(z)$ determined by 
Gilmore \& Reid for dwarf stars has survived the test of time.
Modern surveys are either too shallow (e.g. \twomass) or too deep (e.g. \SEGUE) to properly represent
the disk although good agreement is found with the \SDSS dwarf photometry \citep{Juric2008} after careful
re-analysis \citep{Piffl2014a}. \citet{Binney2014} note that the kinematics of the cool dwarfs and giants in 
\RAVE are consistent, such that dynamical model fits for giants or dwarfs can reasonably adopt
\citet{Gilmore1983}
as a starting point, as has been done by most dynamical studies.

Given \citet{Gilmore1983}
 or a similar density profile \citep[e.g.][]{Kuijken1989}, the dynamical modelling attempts to find a 
self-consistent
mass model-DF pair. {\it This means that the mass distribution of the stellar disk implied
by the DF in the potential is consistent with the mass distribution of the stellar disk assumed in the mass model.}
The action integral analysis of \citet{Piffl2014a} used $\sim$200,000 giants in the
\RAVE survey. In Fig.~\ref{f:piffl1}, they find remarkable dynamic self-consistency in the local disk 
as defined above.
Moving away from the solar neighbourhood, they infer a declining vertical force with $R$ as expected,
in good agreement with the \citet{Bovy2013} analysis of \SEGUE dwarfs ($5 < R < 9$ kpc).
The latter study derived a mass-weighted scalelength of $\Rdiskmass = 2.15\pm 0.14$ kpc, smaller
than the $\Rdiskmass = 2.68$ kpc (without a quoted error) inferred by the \RAVE analysis. \Rdiskmass\ is
dynamical by nature and is blind to the separate contributions of the ISM and the thin/thick disks.

\subsubsection{Local mass budget}
\label{s:inventory}

In a new study, \citet{McKee2015} revise the local
baryon inventory of \citet{Flynn2006} in light of new observations.
Inter alia, they update the present day stellar mass
function, and the vertical distributions and extents of both
gas and stars. They find the baryon surface density integrated
to infinity is $\Sigma_{\rm bary} \approx 47\pm 3$\ \Msun\ppcsq\ 
comprising brown dwarfs (1.2\ \Msun\ppcsq),
white dwarfs (4.9\ \Msun\ppcsq), ISM gas (13.7\ \Msun\ppcsq),
main sequence and giants (27.0\ \Msun\ppcsq). Interestingly, that least
understood of main sequence stars - the M dwarf - makes up more than half 
of all stars by mass locally.
Over the column $z=\pm 1.1$ kpc, there is general agreement that the 
{\it total} surface density (baryons $+$
dark matter) is $\Sigma_{\rm tot} \approx 70\pm 5$\ \Msun\ppcsq\ 
\citep{Kuijken1989,Catena2010,McMillan2011,Bovy2013,Piffl2014a}.

In terms of {\it local} density, the proportion of each mass constituent is very different 
due to the wide spread in scaleheights. \citet{McKee2015} give the local mass density as 
$\rho_{\rm tot} \approx 0.097\pm 0.013$\ \Msun\ppccb\ ($0.49\pm 0.13$ GeV cm$^{-3}$) 
comprising stars ($0.043\pm 0.04$\ \Msun\ppccb),
gas ($0.041\pm 0.04$\ \Msun\ppccb),
baryons ($0.084\pm 0.04$\ \Msun\ppccb),
and dark matter ($0.013\pm 0.003$\ \Msun\ppccb).

\begin{marginnote}
\entry{\Sigtot}{$70{\small \pm}5$\ \Msun\ppcsq, total mass surface density $\vert z\vert \leq 1.1$ kpc at \Rzero}
\entry{\Rhotot}{$0.097{\small \pm}0.013$\ \Msun\ppccb, local mass density at \Rzero}
\entry{\Etot}{$0.49{\small \pm}0.13$ GeV cm$^{-3}$, local dark matter energy density at \Rzero}
\end{marginnote}

\subsection{Outer disk}
\label{s:outerdisk}

The physical extent and detailed structure of the outer disk is highly uncertain. Over the years,
different authors have claimed evidence for an ``edge'' in the stellar disk in the range $R_{\rm GC}$ $=$ $10-15$ kpc from both optical and infrared surveys 
\citep{Habing1988,Robin1992,Ruphy1996,Minniti2011}.
When looking at external galaxies, what appear to be edges can be 
inflexions in the stellar density, i.e. a break in the exponential density profile. Such ``breaks'' are common \citep{Pohlen2006} but ``continuously exponential'' 
disks are also known and the stars can extend to the observed edge of the HI disk
\citep{Bland-Hawthorn2005,Irwin2005,Ellis2007}.

Many new observations confirm that the outer disk is very complicated. 
The outer disk warps slowly away from the Galactic Plane in both \HI\ \citep{May1993}
and stars \citep{Carney1993}. In addition to the warp, the outer disk flares in
both stars and gas and possesses a fair degree of substructure.
\citet{MoniBidin2012} review the contradicting claims for the flaring stellar
disk, but the comprehensive study of \citet{Momany2006} puts the issue beyond doubt. 
\citet{Carraro2015} reviews the evidence for flaring in the
outer disk in both young (HII regions, open clusters) and old stellar populations (cepheids, pulsars).
The earlier claims of a disk edge did not consider the effect of a warping, flaring disk such that there
is {\it no} strong evidence for a truncation to date in either old or young populations 
\citep{Lopez-Corredoira2002,Sale2009,Carraro2010}.

A complicating factor is the presence of the near-planar Monoceros Ring at $R_{\rm GC} \approx 15-20$ kpc 
(radial width $\approx$ 2 kpc) which appears to corotate at roughly the speed of the outer disk
\citep{Newberg2002,Ibata2003}. The nature of this stream is unknown \citep[cf.][]{Xu2015}:
unlike the Canis Major overdensity which may be related to the warp, the ring appears to have
its own identity after correcting for the warp \citep{Momany2006}, and may even rotate slightly faster than 
the disk \citep{Ivezic2008}.

Before the optical studies, \HI\ observations by \citet{Burton1988} and \citet{Diplas1991} 
identified Galactic gas out to at least $R_{\rm GC}$ $\sim$ $25-30$ kpc 
\citep[see also][]{Hartmann1997}. Molecular gas clouds are seen to 
$R_{\rm GC} \gta 20$ kpc, some with ongoing star formation 
\citep{Yasui2008,Kobayashi2008}.
Further evidence for recent star formation out to these distances is the presence of
young stars and open clusters \citep{Carraro2010,Magrini2015}.
Remarkably, \citet{Kalberla2008} have pushed the \HI\ frontier to $R_{\rm GC} \sim 60$ kpc
which is within range of the orbiting Magellanic Clouds. The stellar/gaseous warp may be highly transient, 
triggered by the passage of the MC or Sgr dwarfs, with bending waves that travel at $\pi G \Sigma/\omega$ $\sim$
20 kpc Gyr$^{-1}$ near the Solar Circle; $\Sigma$ and $\omega$ are the local surface density and
angular velocity of the disk respectively.

\section{Halo} 
\label{s:halo}

In this section, we review the stellar, gaseous, and dark matter halos of the Galaxy, and finally put together a global
rotation curve for the Milky Way. This analysis brings together many of the key themes of the review. The stellar and
dark matter halos share the property that they are three-dimensional structures surrounding the disk and still grow 
by accreting matter. However, they do not necessarily share the same structural properties or formation
histories.  The gaseous halo is an important repository for a part of the Galaxy's baryonic mass, and interacts with
the environment through inflows and outflows.

All three components live in the same gravitational potential, which at intermediate radii ($\sim\!R_0$) is shaped by
the Galactic bulge, disk and dark halo, and at large ($\nsim 100\kpc$) radii is completely dominated by the dark
matter. One important goal of studying the halo and the satellites around the Milky Way is to map out the large-scale
gravitational potential - this is a major science goal of the ongoing \gaia mission.

\subsection{Stellar halo}
\label{s:stellarhalo}

The Milky Way's stellar halo, although containing just $\sim1\%$ of the total stellar mass, is an important component
for unravelling our Galaxy's formation history. It was first identified as a population of old, high-velocity, metal
poor stars near the Sun, similar to the stars in globular clusters. The halo stars showed large random motions, little
if any rotation, and a spheroidal to spherical spatial distribution. Following the influential paper of
\citet{Eggen1962}, the classical view of the halo developed, of a smooth envelope of ancient stars from the time when the
Galaxy first collapsed. Subsequently, \citet{Searle1978} suggested that the halo is built up from independent infalling
fragments, based on their observation that halo globular clusters showed a wide range of metal abundances independent of
Galactocentric distance.

Modern data from large stellar surveys show that the stellar halo has a complex structure with multiple components and
unrelaxed substructures, and continues to accrete matter in the form of smaller galaxies which are then tidally
disrupted in the gravitational field \citep[e.g.,][]{Ibata1997, Belokurov2006, Schlaufman2009}.  This confirms the
predictions of hierarchical galaxy formation models. Because of the long dynamical timescales in the halo, tidal tails,
shells, and other overdensities arising from accreted dwarf galaxies remain observable over Gyrs, thus constituting a
fossil record of the Milky Way's accretion history.  Previous reviews on this subject can be found in \citet{Helmi2008,
  DeLucia2012, Belokurov2013}.

Cosmological simulations reveal that the Galaxy should have accreted $\nsim100$ satellite galaxies which would mostly have
been disrupted by the tidal field, causing the build-up of the stellar halo. Irregular density distributions are
predicted in the outer halo due to shells and tidal streams, with a large variance between different galaxies of the
same dark matter halo mass. In these models, the majority of the halo is often built at early times, $\sim 10\Gyr$ ago,
and most of the stellar halo stars come from the disruption of one or a few massive satellites accreted early-on. The
outer halo is built more recently than the inner halo and the halo properties evolve, reflecting the history of
accretion \citep{Bullock2005, Font2006, DeLucia2008, Cooper2010, Pillepich2014}.

Part of the inner halo may have formed {\sl in situ}, i.e., within the main body of the Galaxy \citep{Abadi2006}. Recent
simulations suggest that a fraction of stars formed in the early Galactic disk could have been ejected into the inner
halo, and further in situ halo stars could have formed from gas stripped from infalling satellites \citep{Font2011,
  McCarthy2012, Tissera2013, Pillepich2015, Cooper2015}, but the quantitative importance of these processes
is not yet fully understood. Observationally, evidence for a {\sl dual halo} has been put forward by
\citet[][]{Carollo2007, Beers2012}, but see \citet{Schonrich2014}.

In this section we review the structural parameters of the Milky Way's stellar halo and put them in the context of these
hierarchical models. Determination of the mass of the dark matter halo based on these data is discussed in
\S\ref{s:darkhalo}.

\subsubsection{Halo flattening and average density profile}
\label{s:halodensity}

The stellar density of the halo is important because it reflects the cumulative past accretion history of the Milky Way.
It has been extensively studied using several tracers for which good distances can be
determined, including RR Lyrae \citep[typical distance accuracy $\nsim 7\%$,][]{Vivas2006}, blue horizontal branch (BHB)
stars \citep[$\nsim 5\%$,][]{Belokurov2013}, red giants \citep[RGB, $\nsim 16\%$,][]{Xue2014}, and near-main sequence
turnoff stars \citep[nMSTO, $\nsim 10\%$ with multiple colors,][]{Ivezic2008}. RR Lyrae and BHB stars,
which trace the old metal-poor populations, are rarer than RGB
and nMSTO stars, but can reach to $r \nsim 100\kpc$ galactocentric radius. RGB trace all halo populations and
currently reach to about $r\nsim 50\kpc$, whereas nMSTO stars are abundant and best for the inner halo, i.e. $r \lta 20\kpc$. 

A significant fraction of these halo tracer stars is found in large substructures (\S\ref{s:halosubstructures}).
Subtracting these leads to an estimate of a {\sl smooth halo} component \citep[e.g.,][]{Deason2011}, which however may
contain smaller, yet unresolved substructures \citep[e.g.,][]{Sesar2013a}.  The spatial distributions of the halo stars
are typically fitted by spherical or axisymmetric density models with single (SPL) or double (DPL) power-law or Sersic
radial profiles, and with one or two flattening parameters for the inner and outer halo; Table~\ref{tab:halo} shows
recent results based on DPL models.  The inner halo flattening is found $q_{\rm in}\narreq0.65\pm0.05$ across various
studies based on data reaching down to $R\nsim5\kpc$ \citep[see also][]{Bell2008,Juric2008}. The inner power-law slope
is encompassed by $\alpha_{\rm in}\narreq-2.5\pm0.3$. (The quoted formal fit errors are often quite small, but
  this could easily depend on the chosen parametric form, the data volume, and on remaining substructures in the
  data.) No evidence is found for an outward gradient in $q$ for the RR Lyrae and BHB samples, but \citet{Xue2015} with
RGB and \citet{Pila-Diez2015} with nMSTO find an increase to $q_{\rm out}\narreq0.8$ by $r\narreq30\kpc$.


\begin{table}
\tabcolsep5pt
\caption{Recent measurements of stellar halo density parameters}
\label{tab:halo}
\begin{center}
\begin{tabular}{@{}l|c|c|c|c|c|c@{}}
\hline
{\bf Reference} & {\bf Tracer} & {\bf $r$ [kpc]} & -$\alpha_{\rm in}$ & {\bf $q_{\rm in}$ } & {\bf $r_{s}$ [kpc]} & -$\alpha_{\rm out}$  \\
\hline
  Xue et al.             2015  &  RGB   &  $10-60$  &  $2.1\pm0.3$   & $0.70\pm0.02$ &  $18\pm1$     & $3.8 \pm 0.1$ \\
  Pila-Diez et al.       2015  &  nMSTO &  $10-60$  &  $2.5\pm0.04$  & $0.79\pm0.02$ &  $19.5\pm0.4$ & $4.85\pm0.04$ \\
  Sesar et al.           2011  &  nMSTO &  $5-35 $  &  $2.62\pm0.04$ & $0.70\pm0.02$ &  $27.8\pm0.8$ & $3.8\pm0.1$   \\
  Deason et al.          2011  &  BHB/S &  $10-45$  &  $2.3\pm0.1$   & $0.59\pm0.03$ &  $27.1\pm1$   & $4.6\pm0.15$  \\
  Faccioli et al.        2014  &  RRL   &  $9-49$   &  $2.8\pm0.4$   & $0.7$ fixed   &  $28.5\pm5.6$ & $4.4\pm0.7$   \\
  Sesar et al.           2013  &  RRL   &  $5-30$   &  $1-2.7$       & $0.63\pm0.05$ &  $16\pm1$     & $2.7\pm0.3$   \\
  Sesar et al.           2010  &  RRL   &  $9-49$   &  $2.8\pm0.2$   & $0.7$ fixed   &  $34.6\pm2.8$ & $5.8\pm0.9$   \\
  Watkins et al.         2009  &  RRL   &  $9-49$   &  $2.1\pm0.3$   & $0.59$ fixed  &  $26.9\pm3.1$ & $4.0\pm0.3$   \\
\hline
\end{tabular}
\end{center}
\begin{tabnote}
  Halo density parameters from recent studies of halo tracer stars with oblate double power-law models.  Columns give
  reference, tracer used, covered Galactic radius range, inner power law slope, inner halo axis ratio, break radius, and
  outer power-law slope (in these models, the outer $q_{\rm out}\narreq q_{\rm in}$). \citet{Watkins2009} and \citet{Sesar2010}
  considered spherical DPL models; the numbers given in these lines are from a reanalyis by \citet{Faccioli2014}.
\end{tabnote}
\end{table}

\begin{marginnote}
\entry{Halo density}{}
\entry{$\alpha_{\rm in}$}{ $-2.5\pm0.3$ Inner density slope}
\entry{$\alpha_{\rm out}$}{$-(3.7\mhyphen5.0)$ Outer density slope}
\entry{$r_s$}{$25\pm10\kpc$ Break radius}
\entry{$q_{\rm in}$}{$0.65\pm0.05$ Inner halo flattening}
\entry{$q_{\rm out}$}{$0.8\pm0.1$ Outer halo flattening}
\end{marginnote}

There is clear evidence that the stellar halo profile steepens with radius; see Table~\ref{tab:halo}. In the context of
DPL models based on data reaching $r\nsim 50\kpc$, the break radius between the inner and outer components scatters in
the range $r_s\narreq25\pm10\kpc$.  For RRL and BHB, the outer power-law slope is in the range
$\alpha_{out}\narreq-4.5\pm0.5$, and \citet{Deason2014} found an even steeper profile beyond $50\kpc$. For RGB the slope
is somewhat shallower, $\alpha_{\rm out}\narreq-3.8\pm0.1$. The overall density profile is similar to an Einasto function
\citep[e.g.,][]{Xue2015}, and qualitatively similar to density profiles predicted by halo formation models
\citep{Bullock2005,Cooper2010}. \citet{Deason2013} use simulations to show that a distinct density break may be related
to the accumulation of stars at their apocenters, following relatively massive accretion events. \citet{Lowing2014}
point out that the measured density parameters depend strongly on the surveyed sightlines and halo accretion history.

\subsubsection{Stellar halo mass and substructure fraction}
\label{s:halosubstructures}

Estimating the stellar halo mass from these density distributions is non-trivial, requiring determination of the mass
normalisation per halo tracer star from stellar population models and calibrations, as well as extrapolating beyond the
survey volume through models. \citet{Bell2008} fitted DPL models to SDSS nMSTO stars and found a best-fit stellar halo
mass within $r\narreq1-40\kpc$ of $\nsim (3.7\pm1.2)\times 10^8 \Msun$.  \citet{Deason2011} gave an estimate for the
ratio of BHB stars per luminosity of $\nsim 10^{-3}$, using data for Galactic globular clusters. With a mass-to-light
ratio $M/L_V\narreq1.4\pm0.5$ for metal-poor Galactic globular clusters \citep{Kimmig2015}, the estimated stellar halo
mass within $10-45\kpc$ becomes $\nsim 3\times 10^8\Msun$.

The Milky Way halo contains numerous substructures which contain a significant fraction of its stellar mass
\citep[see][for a review]{Belokurov2013}. The four largest stellar structures are the Sagittarius Stream, the Galactic
Anti-Center Stellar Structure, the Virgo Overdensity, and the Hercules-Aquila Cloud.  Estimated masses for these
structures are, respectively, $\nsim 0.8-1.5 \times 10^8\Msun$ based on the luminosity from
\citet{Niederste-Ostholt2010} and $M/L_V\narreq1.4\pm0.5$ from \citet{Kimmig2015}; $\nsim 10^8\Msun$
\citep{Yanny2003,Belokurov2013}, $\lta 10^6\Msun$ \citep{Bonaca2012} (iv) $\nsim 10^7 \Msun$ based on
\citet{Belokurov2007}, alltogether summing to $\nsim 2\mhyphen3 \times 10^8\Msun$.  A significant fraction of this
substructure mass is within the volume traced by the \sdss nMSTO stars. We therefore add $\sim\! 50\%$ of this mass to the
result of \citet{Bell2008} to obtain a rough estimate for the total stellar halo mass $M_s\narreq4\mhyphen7 \times
10^8\Msun$. This is somewhat lower than the classical value based on \citet{Morrison1993a}.

\begin{marginnote}
\entry{Halo mass}{}
\entry{$M_{\rm sub}$}{$2\mhyphen3\times10^8\Msun$ Substructure mass}
\entry{$M_s$}{$4\mhyphen7\times10^8\Msun$ Total stellar halo mass}
\end{marginnote}

\citet{Bell2008} also quantified the fraction of mass in substructures from the ratio of the rms deviation of the
density of nMSTO stars to the total density given through a smooth halo model. They find $\sigma/{\rm total} \narreq
40\%$, arguing that much of the halo was accreted from satellite galaxies.  In their study of BHB stars,
\citet{Deason2011} found a lower $\sigma/{\rm total}\nsim 5-20\%$ with some increase for fainter stars (larger
distances); on this basis they argue for a smooth halo with superposed additional substructures.  Reasons for the
discrepancy between both studies could be the less accurate nMSTO distances, leading to blurring of compact
substructures, the lower resolution with the rarer BHB stars, or because the BHB stars trace an older, more mixed
population of stars \citep{Deason2011}.  Resolving this issue requires large samples of stars with accurate distances
and velocities. The problem is that even a fully accreted halo will eventually mix to be smooth above a given
scale, and that mixing times are shortest in the high-density inner regions.

\subsubsection{Halo rotation, velocity dispersion and anisotropy}
\label{s:halokinematics}

\citet{Bond2010} analyzed the largest sample so far of halo star velocities within $\nsim 10\kpc$ from the Sun,
including $\nsim10^5$ SDSS stars with three velocity components. They found (i) a mean rotation of halo stars (their
Fig.~5) of $\vshalo\nsimeq\Vcirc+\Vsol-205\kms \nsimeq 40\kms$ for our adopted $\Vcirc\narreq238\kms$ and
$\Vsol\narreq10.5\kms$ (\S\ref{s:rotcur}); (ii) a velocity ellipsoid whose principal axes align well with spherical
coordinates; and (iii) corresponding halo velocity dispersions $(\sigma_r^{\rm s},\sigma_\theta^{\rm s},
\sigma_\phi^{\rm s})\narreq(141,75,85)\kms$, with a total error in each component of $\sim\!  5\kms$. These values are
in excellent agreement with results from \citet{Smith2009}.  Note that the close alignment of the halo velocity
ellipsoids with spherical coordinates does not imply a spherical potential \citep[see][]{Evans2016}.
 
\begin{figure}
\centering
\includegraphics[width=12.5cm]{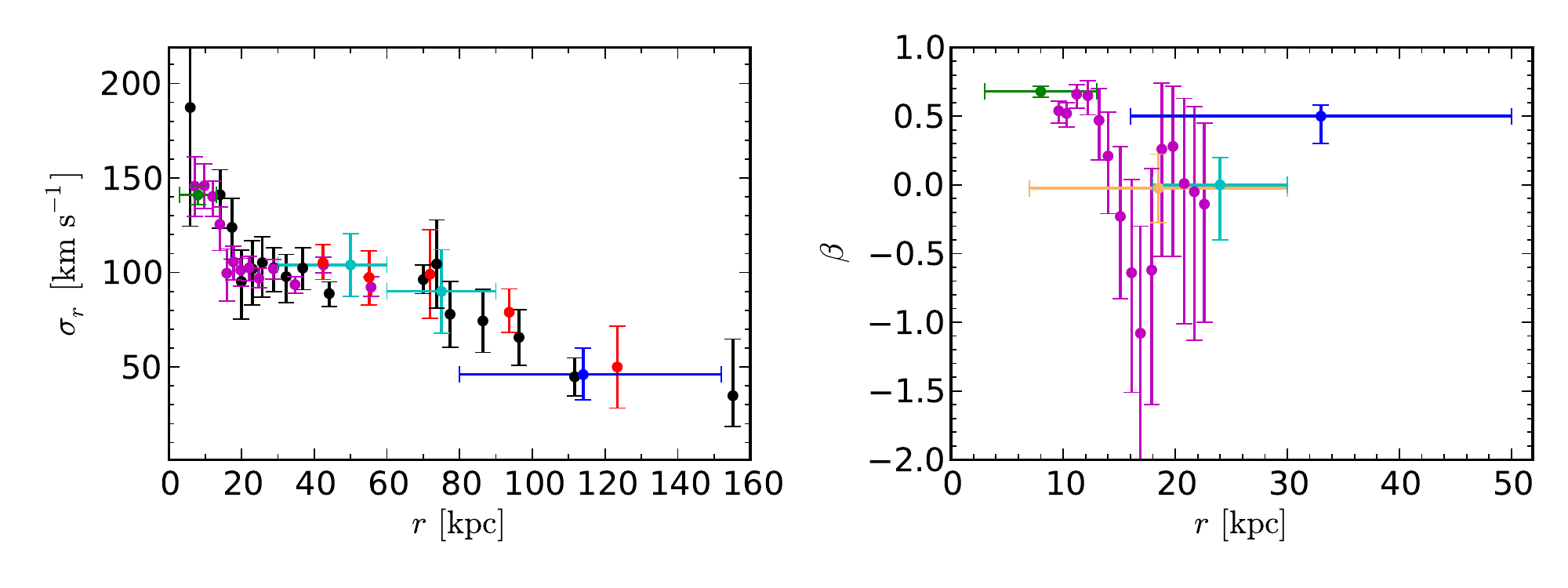}
\caption{Left: radial velocity dispersion profile for several outer halo tracers, including nMSTO stars
  \citep[][green]{Bond2010}, BHB stars \citep[][purple]{Kafle2012} and \citep[][blue]{Deason2013}, blue straggler stars
  \citep[][cyan]{Deason2012}, K-giants \citep[][black]{Kafle2014}, and mixed tracers \citep[][red]{Battaglia2005}.
  Right: measurements of orbital anisotropy in the halo, for nearby \citep[][green]{Bond2010} and distant nMSTO stars
  \citep[][cyan]{Deason2013}, and BHB stars \citep[][purple]{Kafle2012}, \citep[][yellow]{Sirko2004}, and
  \citep[][blue]{Deason2012}.}
\label{fig:halodispersionprofile}
\end{figure}

Fig.~\ref{fig:halodispersionprofile} shows the radial velocity dispersion profile in the outer halo based on several
data sets.  $\sigma_r(r)$ sharply decreases from the local $141\kms$ to $100\kms$ at $r\nsim20\kpc$, then remains
approximately flat until $r \nsim 70\kpc$, and finally decreases to about $35\kms$ at $\nsim 150\kpc$. At the largest
radii, where the stellar density profile is largely unknown (see Table~\ref{tab:halo}), the very low $\sigma_r$ values
are consistent with a tidal truncation of an extrapolated steep $n\propto r^{-4.5}$ power-law \citep{Kafle2014}.

\begin{marginnote}
\entry{Local halo kinematics}{}
\entry{$(\sigma_r^{\rm s},\sigma_\theta^{\rm s},\sigma_\phi^{\rm s})$}{$(141,75,85) \kms$ ($\pm5\kms$) Spherical velocity ellipsoid}
\entry{$\vshalo$}{$\nsim40\kms$ Mean rotation}
\end{marginnote}

Tangential velocities can be estimated from the variation of \los velocities across the sky and from proper motions.
\citet{Fermani2013} used BHB stars reaching to $r\nsim 50\kpc$ and with both kinds of methods find no rotation in either
the inner or outer halo, and no trend with metallicity. Tangential velocity dispersions and the spherical anisotropy
parameter $\beta(r)$ were determined in the outer halo by several authors and are shown in
Fig.~\ref{fig:halodispersionprofile}. The radial anisotropy $\beta\narreq0.5-0.7$ at small and large radii is consistent
with predictions for accreted halos \citep{Abadi2006}. The tangential anisotropy at radii around $r\nsim17\kpc$ was
confirmed by \citet{King2015cc} but was suggested to be transient based on orbit simulations \citep{Bird2015}, perhaps due
to a yet unknown substructure in the halo.

\subsection{Hot halo}
\label{s:hothalo}

The existence of a diffuse hot plasma (or corona) surrounding the Galactic disk has 
been widely discussed since Spitzer's early observation that the ubiquitous HI clouds 
must be confined by an external medium
\citep{Spitzer1956}. But it remains uncertain how much of the gas lies close to the Galactic plane, and
how much of it extends into the halo (Bland-Hawthorn
\& Cohen 2003). Strong evidence for extended coronae have
come from the first reliable detections of hot halos around nearby 
massive disk galaxies \citep[e.g.][]{Anderson2011}. The Galactic corona also 
explains the remarkable observation of gas depletion in all dwarfs within a radius of about 270
kpc, with the exception of the high-mass LMC and SMC dwarfs \citep{Grcevich2009}, 
which is well understood in terms of ram-pressure stripping by a hot
medium \citep{Nichols2011,Gatto2013}. Such an extended hot phase is supported by numerical simulations
\citep{Nuza2014} which focus on the properties and distribution of the multiphase gas in and around two simulated galaxies chosen as an M31-Galaxy analogue. They found the hot ($T \gta 10^5$K) gas has a uniform temperature profile around each of the simulated galaxies, and good agreement between
the observed profile \citep{Miller2015} and the density profile of the simulated Galaxy.

\begin{table}[htbp]

\caption{Estimates of gas density and temperature in the Galactic halo (corona).}
\vspace{0.3cm}
   \begin{tabular}{lccccl}
  	Reference		& 	$d$ [kpc]			& 	$n_p$ [$10^{-5}$cm$^{-3}$]	& 	$T_e$ [$10^6$K]		& \Mhot [$10^{10}$\Msun] & Method \\ \hline
	BR00	&	$\lta 250$		&	$ \gta 2.4^a$			&	$1 - 1.4$			&	1						& ram-pressure stripping\\
	&&&																				&							& of gas in dwarf spheroidals \\ [5pt]
	S+02	&	$15^b$			& 	$100$					&	$1^d$			&	$-$						& pressure equilibrium between \\
				&	$45^b$			&	$30$						&	$1^d$			&							& HVCs (MS) and coronal gas\\ [5pt]
	S+03	&	$\approx 70$			&	$1 - 10$					&	$>1$				&	$-$						& OVI absorption (100 LOS)\\ [5pt]
	BR07	&	$<20$			&	$90^c$					&	$2^d$			&	$0.04^f$					& OVII absorption (25 LOS) \\
	&&&																				&							& uniform spherical halo 		\\ [5pt]
	AB10	&	$50^e$			&	$50^{a,c}$				&	3.5				&	$1.2-1.5$					& LMC pulsar dispersion measures	\\ [5pt]
	G+13	&	$50 -90$			&	$36 - 13^a$				&	$1.8^d$			&	$3.4 - 4.8$				& ram-pressure stripping of gas \\
	&&&																				&							& in Carina and Sextans \\ [5pt]
	MB15	&	$10 - 100$		&	$200 - 4^a$				&	$2^d$			&	$2.7- 4.7$					& OVII / OVIII emission (650 LOS) \\ [5pt]
	S+15	&	$48.2 \pm 5$	&	$11\pm 4.5$			&	$-$				&	$2.6\pm1.4^g$				& ram-pressure stripping \\
	&&&																				&							& of the LMC disk \\ [5pt] \hline
     \end{tabular}
     \begin{tabnote}{}{} Unless otherwise indicated, all coronal masses \Mhot\ are integrated out to \Rvir; the gas
     density is quoted as a total particle density unless otherwise indicated.
     Notation:
     	$^a$Average density out to the given distance;
     	$^b$Distance from the Galactic plane;
     	$^c$Electron density;
    	$^d$Assumed;
    	$^e$Distance from the Sun;
	$^f$Mass enclosed within 20 kpc;
    	$^g$Mass enclosed within $\sim 300$ kpc.
	References: BR00 - \citet{Blitz2000}; S+02 - \citet{Stanimirovich2002}; S+03 - \citet{Sembach2003}; BR07 - \citet{Bregman2007}; AB10 - \citet{Anderson2010};
	G+13 - \citet{Gatto2013};  MB15 - \citet{Miller2015};  S+15 - \citet{Salem2015}.
     \end{tabnote}
   \label{t:halodens}
\end{table}

We tabulate the most recent observations of the Galactic corona in Table~\ref{t:halodens}.
The best evidence for a hot corona comes from bright AGN sight lines with detections of \OVII\ and \OVIII\
in absorption, and in high-resolution x-ray spectra \citep{Paerels2003} of a nearly ubiquitous soft x-ray background (SXRB) with energies $0.1-1.0\keV$ (implying temperatures $\sim 10^{6-7}~\K$), with some contribution of \OVII\ and \OVIII\ in emission \citep{Snowden1997,Henley2012}. The sensitivity of current x-ray spectroscopy limits \OVII\ and \OVIII\ detections to a handful ($\sim 30$) of sight lines. There is an additional difficulty in disentangling the contribution of the Local Bubble, a supernova remnant in which the Solar System is embedded \citep{Snowden1990}, and the contribution from solar wind charge exchange processes, which produce soft x-ray emission throughout the Solar System \citep{Cravens2001}. Consequently, the detailed structure (density, temperature, entropy profile) of the Galactic corona, and hence 
its total mass, is uncertain.

The Galactic hot halo is likely to comprise two main components: one, exponentially decaying, high-metallicity ($Z > 0.3 ~\Zsun$) component with a scale height of a few kpc, which dominates the x-ray observations; and an extended ($\gta 100 ~\kpc$), more diffuse, low-metallicity halo \citep{Yao2007}. Purely exponential gas density profiles (i.e. not characterised by a single temperature) overpredict the coronal temperature and x-ray surface brightness by factors of a few \citep{Fang2013}.

Early attempts to model the corona using x-ray observations assumed (unphysically) that it was isothermal and at a constant density \citep{Bregman2007,Gupta2012}. The studies arrived at wildly different conclusions: the first study derived an electron density $n_e = 9 \times 10^{-4} \pcc$ at $r = 19 ~\kpc$; the second found $n_e \geq 2 \times 10^{-4} \pcc$ at $r \geq 139 ~\kpc$ \citep[cf.][]{Wang2012}. \cite{Fang2013} favoured a picture in which the corona is composed of adiabatic (isentropic) gas in hydrostatic equilibrium with the Galactic potential, in conflict with evidence that the Galactic halo is far from having a constant entropy profile \citep{Miller2015,Crain2010}.
It is now well established that the halo temperature as inferred from x-ray observations is fairly uniform across the sky and with little scatter around $T \approx 2 \times 10^6 ~\K$ \citep{Henley2010} although this
does not demand that the halo is strictly isothermal.

A new all-sky catalog of \OVII\ and \OVIII\ emission lines \citep{Henley2012} has been
studied by \cite{Miller2013} and \cite{Miller2015}. They used this catalogue in combination with x-ray measurements to determine that the halo density is of order $10^{-5} \pcc$ to $10^{-4} \pcc$ at $10 ~\kpc \lta r \lta 100 ~\kpc$. They determine a coronal gas mass of $\sim10^{10} \Msun$ within $r \approx 250 ~\kpc$. These results rely on key assumptions: (i) the density profile of the hot gas is well described by a spherically symmetric $\beta$-model of the form $n(r) \propto r^{-\beta / 2}$, consistent with a truncated King model for the halo potential; (ii) the halo is isothermal with a temperature $T = 2 \times 10^{6} ~\K$; (iii) the gas is in collisional ionisation equilibrium.
Miller \& Bregman's model can be justified if the dark matter halo of the Galaxy is well described by a spherically symmetric isothermal sphere with a core radius $r_c \sim 0.1 ~\kpc$, and if the gas is quasi-isothermal and in approximate hydrostatic equilibrium with the potential. 

\cite{Tepper-Garcia2015} bring all of this work together and search for a physically plausible corona that is consistent 
with the observed stellar halo dynamics and with the UV/x-ray measurements (Table~\ref{t:halodens}).
They normalise the dark matter halo to the density profile and total mass inferred from the
kinematics of halo stars \citep{Kafle2012}. If the dark matter halo is isothermal, the core radius is somewhat
larger than inferred from Miller \& Bregman's model, i.e. $r_c \approx 0.5 ~\kpc$. The halo velocity dispersion implies a gas temperature of $T \sim 10^6 ~\K$, leading to a density profile which is in broad agreement with \citet{Miller2015} and \citet{Nuza2014} with about the same total gas mass. 
The most likely baryonic mass range for
the Galactic corona is \Mhot\ $\sim$ $2.5\pm 1\times 10^{10}\Msun$.

\smallskip\noindent{\sl The Galaxy's baryonic mass fraction.} The dynamical analyses presented in \S~\ref{s:diskdyn}
are relatively consistent in their estimates of the baryonic mass fraction of the Galaxy (stars $+$ cold gas), i.e. 
\Mbary\ $\sim$ $6.3\pm 0.5\times 10^{10}\Msun$ (Table~\ref{t:jbh1}). We now add the likely contribution from the
hot corona \Mhot\ within \rvir\ to arrive at a {\it total} baryonic mass, $8.8\pm 1.2\times 10^{10}\Msun$. This leads to a baryonic
mass fraction out to \rvir\ of \fbary $=$ \Mbarytot/\Mvir\ $\approx$ $6\pm 1\%$ which falls well short of the Universal
value \citep[$\approx$16\%;][]{Hinshaw2003}.

\begin{marginnote}
\entry{\Mhot}{$2.5\pm 1\times 10^{10}\Msun$;
Galactic corona baryonic mass ($r\lta$\Rvir)}
\entry{\Mbary}{$8.5\pm 1.3\times 10^{10}\Msun$;
Galactic total baryonic mass ($r\lta$\Rvir)}
\entry{\fbary}{$7\pm 1\%$; Galactic baryonic mass fraction}
\end{marginnote}

\subsection{Dark halo}
\label{s:darkhalo}

An accurate measurement of the Galaxy's total mass is central to our understanding of how it fits into the Cold Dark
Matter paradigm.  We need to know the mass of the dark matter which has had time to virialize in the Galaxy, the
so-called virial mass \Mvir\ \ defined within the virial radius \rvir. There is some confusion in the literature on the
convention for a galaxy's total mass, i.e. how it should be defined and at what epoch \citep{Shull2014}.  \Mvir\ is usually
expressed as the mass within a region around the centre in which the average density exceeds a multiple of the mean
density of the Universe (either the closure density or the mass density). We follow the definition of 
\citet{Klypin2002} and \citet{Kafle2014}
where $\rho_{\rm vir} = \Delta_{\rm vir} \Omega_{\rm M} \rho_{\rm crit}$ ($\Delta_{\rm  vir} =340$) such that
\be
\rvir = 258 \left({{\Delta_{\rm vir} \Omega_{\rm M}}\over{102}}\right)^{-1/3}
\left({{\Mvir\ }\over{10^{12} \Msun}}\right)^{1/3} {\ \ \rm kpc}
\ee
Since the mass enclosed depends on the product $\Delta_{\rm vir} \Omega_{\rm M}$, our values are within 5\%
of estimators that use $(\Delta_{\rm vir}, \Omega_{\rm M})=(360,0.27)$ \citep[e.g.][]{vanderMarel2012a}.
Note that if the dark matter can be represented by an NFW halo, its scalelength is $r_h \approx 25$ kpc
assuming a concentration parameter $c\approx 10$ \citep[Fig.~\ref{f:CDM};][]{vanderMarel2012a}. Another widely 
used mass estimator is $M_{200}$ where the average density within $r_{200}$ is $\rho_{200} = 200 \rho_{\rm crit}$, 
which is 16\% smaller than \Mvir\ for our adopted parameters \citep{Bryan1998,Klypin2002}.
\Mvir\ is not strictly a total mass because the dark matter profiles are thought to extend (and rapidly
truncate) beyond the virial radius.

\begin{table}
\caption{ \label{t:kafle1} Total mass estimates for the Galaxy.}
\begin{center}
\vspace{0.3cm}
\begin{tabular}{l|l|lll|lll}
Reference &                          Method &     $M_{200}$ &   $M_{100}$  &     $M^r_{\kpc}$  &   $R_{200}$  &    $R_{100}$  &   $r$  \\
\hline
Wilkinson et al. 1999 &     Distribution function based &  1.67 &  2.39 &  0.54 &  243.9 &  346.0 &   50 \\
Sakamoto et al. 2003 &   Mixed halo object kinematics &  1.67 &  2.39 &  0.54 &  243.9 &  346.0 &   50 \\
Dehnen et al. 2006 &           Halo star kinematics &  1.75 &  1.96 &  1.05 &  247.7 &  324.3 &  120  \\
Smith et al. 2007 &                 Escape velocity &  1.03 &  1.43 &  0.39 &  207.7 &  291.7 &   50  \\
Xue et al. 2008 &           Halo star kinematics &  0.87 &  0.91 &  0.40 &  196.1 &  267.0 &   60  \\
Gnedin et al. 2010 &  Hypervelocity star kinematics &  1.33 &  1.74 &  0.69 &  226.4 &  311.3 &   80 \\
Watkins et al. 2010 &   Satellite galaxy kinematics &  2.62 &  3.05 &  2.70 &  283.4 &  375.4 &  300  \\
McMillan 2011 &      Modeling local observables &  1.26 &  1.76 &  0.84 &  222.0 &  312.7 &  100 \\
Kafle et al. 2012 &           Halo star kinematics &  1.21 &  1.99 &  0.21 &  219.1 &  325.5 &   25  \\
Deason et al. 2012 &           Halo star kinematics &  0.87 &  1.03 &  0.75 &  196.0 &  261.9 &  150  \\
Gonzalez et al. 2013 &              Satellite galaxy kinematics &  1.15 &  1.39 &    - &  215.3 &  289.0 &  - \\
Kafle et al. 2014 &           Halo star kinematics &  0.72 &  0.80 &    - &  184.2 &  239.1 &  -  \\
Piffl et al. 2014b &                 Escape velocity &  1.60 &  1.90 &    - &  235.0 &  322.0 &  -  \\
Gibbons et al. 2014 &                 Stream modeling &  0.55 &  0.68 &  0.41 &  168.7 &  227.3 &  100 \\
\hline
\end{tabular}
\end{center}
\vspace{0.2cm}
\begin{tabnote} 
$M_{\Delta}$ is the mass within the radius defined with respect
to the overdensity $\Delta$ (see text);
the radius computed by the author is indicated by $R_\Delta$.
$M^r_{\kpc}$ is the mass within radius $r$ in $\kpc$ as indicated.
All masses are in units of 10$^{12}$\Msun; all radii are in units of kpc.
\end{tabnote}
\end{table}

One constraint for the total mass of the Milky Way comes from the `timing mass' argument (Kahn \& Woltjer 1959): 
the masses of M31 and the Galaxy must be sufficient to overcome universal expansion to explain their present-day 
kinematics
consistent with a head-on collision and a future merger in $\sim$6 Gyr (van der Marel et al 2012b).  Timing mass
estimates in early work (e.g. Li \& White 2008) are now thought to be consistently too high. These have come down
significantly due to improved relative motions for both galaxies and a more accurate estimate of the solar reflex motion
(\S\ref{s:rotcur}). By selecting galaxy pairs in the Millennium simulations (after Li \& White), \citet{vanderMarel2012a}
determine a (virial) timing mass of $4.9\pm 1.6\times 10^{12}$ \Msun\ for the mass within the virial radius of both 
galaxies ($4.1\pm 1.4\times 10^{12}$ \Msun\ for r $\leq$ $r_{200}$). After considering the orbit of M33 about M31, they 
further reduce the total timing mass to $M_{\rm vir,timing}=3.2\pm 0.6\times 10^{12}$ \Msun.
Modern mass estimates for M31 reveal that it is comparable to the Milky Way
\citep{vanderMarel2012a} such that the timing mass provides an upper limit of $\Mvir \lta 1.6\times 10^{12}$ \Msun\
for the Galaxy.

Unlike the timing mass, most mass estimators are limited to the region explored by the available tracer
population, whose spatial distribution and kinematics are used to estimate the enclosed mass.  Estimates of the Milky
Way's mass have been obtained based on the kinematics of halo stars, the kinematics of satellite galaxies and globular
clusters, the evaluation of the local escape velocity, and the modelling of satellite galaxy tidal streams. 
A list of direct mass determinations is compiled in Table~\ref{t:kafle1}. 
Dynamical
analysis of halo star kinematics typically results in relatively low total mass, $M_{200}\lta 10^{12}\Msun$
\citep{Xue2008, Deason2012, Kafle2012, Kafle2014}.  The main uncertainties in such determinations are the lack of
stellar tangential velocities from proper motions, and/or the need to extrapolate from spatially limited samples to the
scale of the virial radius. Such extrapolation is often done using simulated galaxy halos, which then fixes the dark
matter density profile, or by assuming parametric forms for the density distribution and fitting for the best
parameters.  

Mass estimates based on satellite and globular cluster kinematics typically result in higher values,
$M_{200}=1-2\times10^{12} \Msun$ when the Leo I dwarf satellite galaxy is assumed to be bound to the Milky Way; on the
other hand, if Leo I with its large line-of-sight velocity is considered unbound, values more similar to $M_{200}\lta
10^{12}\Msun$ result \citep{Wilkinson1999, Li2008, Watkins2010}. Satellite galaxies reach to larger radii, but here the
main uncertainties come from small numbers and similarly lack of proper motion information.  Determinations of the
escape velocity from radial velocities of stars near the Sun also lead to $M_{200}=1-2\times10^{12} \Msun$
\citep{Smith2007, Piffl2014b}, again with the uncertainties of extrapolating the mass distributions to large
radii. Modelling stellar positions and velocities along the orbit of the Sagittarius dwarf galaxy, or matching the
apocentre radii of its trailing and leading arms, leads to a range of enclosed mass $M^{100}_{\kpc}=0.4-2\times10^{11}
\Msun$ \citep{Gibbons2014}.
The Galaxy's mass can also be estimated by comparing Milky Way properties with various predictions of a CDM 
simulation, such as the number of satellite galaxies larger than a given mass \citep{Cautun2014}.  But these
estimates are not reliable until the models improve. Finally, \citet{McMillan2011} determined a value for the Milky Way
mass from fitting parameterized mass models to a range of observations.

The halo stellar kinematic studies comprise the largest and, arguably, the most reliable data sets. These estimates lead
to a straight average for $M_{200}=1.1\pm0.3 \times 10^{12} \Msun$, or equivalently $\Mvir\ =1.3\pm0.3 \times 10^{12}
\Msun$, consistent with the upper limit from the timing mass. Interestingly, if we derive M31's mass from a simple
scaling of peak rotation, i.e. $(260\kms/\Vcirc)^4 \Mvir$, or a mass that is 40\% higher than the Galaxy, this leads to
a virial timing mass for the Galaxy that is equal to our estimate for \Mvir\ above. The Galaxy's virial mass cannot be
much lower than $10^{12}\Msun$ if it is to provide sufficient angular momentum to the observed baryons over its
lifetime.

\begin{marginnote}
\entry{$\Rvir$}{$282\pm 30 \kpc$, Galactic virial radius scaled to $\Omega_M = 0.3$}
\entry{$M_{200}$}{$1.1\pm0.3 \times 10^{12} \Msun$, Galactic mass within $R_{200}$}
\entry{$\Mvir$}{$1.3\pm0.3 \times 10^{12} \Msun$, Galactic virial mass}
\end{marginnote}

\subsubsection{Halo shape}

Besides the total halo mass, another property of considerable interest is the shape of the dark matter halo. In dark
matter only simulations, halo shapes are strongly flattened, prolate-triaxial, with mean $<c/a> = 0.5\pm 0.1$
\citep[e.g.,][]{Dubinski+Carlberg91, Allgood+06}.  When baryons are included, the halos become more spherical and evolve
towards oblate at all radii, but mostly in their inner parts \citep[e.g.,][]{Kazantzidis+04, Bailin+05,Abadi+10}. In the
Milky Way, constraints on the shape of the dark halo are based on the Sgr orbit, tidal streams, on SDSS kinematics, on
flaring of the HI layer, and on combining rotation curve and vertical acceleration measurements \citep[see review
by][]{Read2014}.

The orbit of the Sgr dwarf, whose leading and trailing arms can each be followed $\sim 180\dg$ across the sky, constrain
the halo shape in the range $R=20 \mhyphen 100\kpc$ \citep{Belokurov2013}. The geometry of the stream on the sky has been shown
to require an oblate near-spherical halo \citep{Ibata2001, Johnston2005}, whereas line-of-sight velocities favour a
prolate shape \citep{Helmi2004}. Thus \citet{Law2010} proposed a triaxial halo model for the MW, in fact nearly oblate
but with short axis in the plane of the disk, with questionable stability.  For this reason, most authors continue to
use spherical models for the outer halo \citep{Ibata2013,Gibbons2014}.

On $20\kpc$ scales, \citet{Koposov2010a} and \citet{Kupper2015} determined the flattening of the dark halo to be
$q_z=0.95\pm0.15$, i.e., essentially spherical, from modelling of the tidal streams of GD-1 and Pal 5.
\citet{Loebman2012} claimed evidence for a strongly oblate shape from SDSS data. \citet{Kalberla2007} found evidence
from HI data for a ring-like distribution of dark matter around $R\sim 15\kpc$.  Near the Sun, the local shape of the
dark matter halo is constrained by the ratio of the local dark matter density to the average enclosed spherical dark
matter density \citep{Garbari2012, Read2014}. Within large error bars the measurements are most consistent with a
spherical or even prolate local halo shape \citep{Garbari2012, Zhang2013, Bovy2013}, i.e., no dark disk
\citep{Read2014}.
On the whole, constraints on the dark matter halo shape in the Milky Way are still weak and no consistent picture has
yet emerged.

\subsection{Rotation curve and baryon fraction profile}
\label{s:rotcur}

Compared to the extended distribution of dark matter, the baryonic mass component in the Milky Way is centrally
concentrated and dominates the mass in the central few kpc.  In this section, we review the total circular velocity at
$R_0$ and the rotation curve of the Milky Way, and then use illustrative dynamical models to estimate the
contribution of stars and gas to the rotation curve, as well as the baryon fraction as a function of radius.

\subsubsection{Solar tangential velocity} 
We recall from \S\ref{s:blackhole} the total angular velocity of the Sun, $\Omgsol\narreq 30.24 \pm 0.12 \kms \perkpc$,
derived from the \vlbi \propm of Sgr~A$^*$ in the Galactic plane and the assumption that Sgr A$^*$ is at rest at the
Galactic Center. For $R_0\narreq8.2\pm0.1\kpc$ (\S~\ref{s:distance}), the corresponding total solar tangential velocity
is $\Vgsol\narreq 248\pm3 \kms$. These values agree within errors with a number of independent recent measurements:
Modelling trigonometric parallaxes and \propm\ of masers in \hmsfr in the Galactic disk, \citet{Reid2014a} derive
$\Omgsol \narreq 30.57 \pm 0.43 \kms \perkpc$, giving $\Vgsol\narreq251\pm5 \kms$.  The analysis of the nearby velocity
field from \SEGUE by \citet{Schonrich2012}, at fixed $R_0\narreq8.2\kpc$, results in $\Vgsol\narreq 248\pm6
\kms$. \citet{Bovy2012d} determine $\Vgsol\narreq242^{+10}_{-3} \kms$ from \APOGEE data, while \citet{Sharma2014} obtain
$\Vgsol\narreq244\kms$ from fitting RAVE data. Here we interpolated to $R_0=8.2\kpc$ and estimate a systematic error
$\sim5\kms$ from their results.  \citet{Kupper2015} find $\Vgsol\narreq254\pm16 \kms$ from modelling the tidal stream of
Pal 5. In what follows we will use $\Vgsol\narreq248\pm3 \kms$ from the \propm\ of Sgr~A$^*$.

\subsubsection{ Circular velocity at $R_0$} In an axisymmetric Galaxy, the circular velocity $\Vcirc$ is simply related
to $\Vgsol=(\Vcirc + \Vsol)$. Here $\Vsol$ is the Sun's peculiar velocity along the direction of rotation with respect
to the LSR, with the LSR defined as the streaming velocity of local stellar populations relative to the Sun in the limit
of vanishing velocity dispersion \citep[e.g.,][]{Schonrich2010}. In \S\ref{s:solarmotion} we estimated
$\Vsol\narreq10.5\pm1.5 \kms$.  In our barred Galaxy, the LSR could itself have an additional streaming velocity $\VLSR$
relative to the circular velocity in the axisymmetrically averaged gravitational potential (the RSR, see
\S\ref{s:solarmotion}), due to perturbations from the bar and spiral arms. As discussed in \S\ref{s:solarmotion}, our
estimate for the total LSR streaming velocity is $\vert\Vstr\vert=0^{+15}\kms$, which could mostly be directed in the
forward direction of rotation. 
If we take $\VLSR\narreq\pm\vert\Vstr\vert\narreq0\pm15\kms$, 
$\Vgsol$ and $\Vcirc$ are
now related by $\Vgsol=(\Vcirc + \VLSR + V_\odot)$; this results in $\Vcirc\narreq238\pm15\kms$ and
$\Omcirc\narreq\Vcirc/R_0\narreq 29.0 \pm1.8 \kms \perkpc$.

\begin{marginnote}
\entry{$\Vgsol\narreq\Vcirc + \VLSR + \Vsol$}{$248\pm3\kkms$, Sun's total tangential velocity relative to Sgr A$^*$}
\entry{$\VLSR$}{$0\pm15\kkms$, tangential velocity of LSR relative to RSR}
\entry{$\Vcirc$}{$238\pm15$ $\kkms$, circular rotation velocity at the Sun}
\entry{$\Omega_o$}{$29.0\pm1.8$ $\kkms\perkpc,\;\;\;$ circular orbit frequency at the Sun}
\end{marginnote}

\begin{figure}
\centering
\includegraphics[width=8cm]{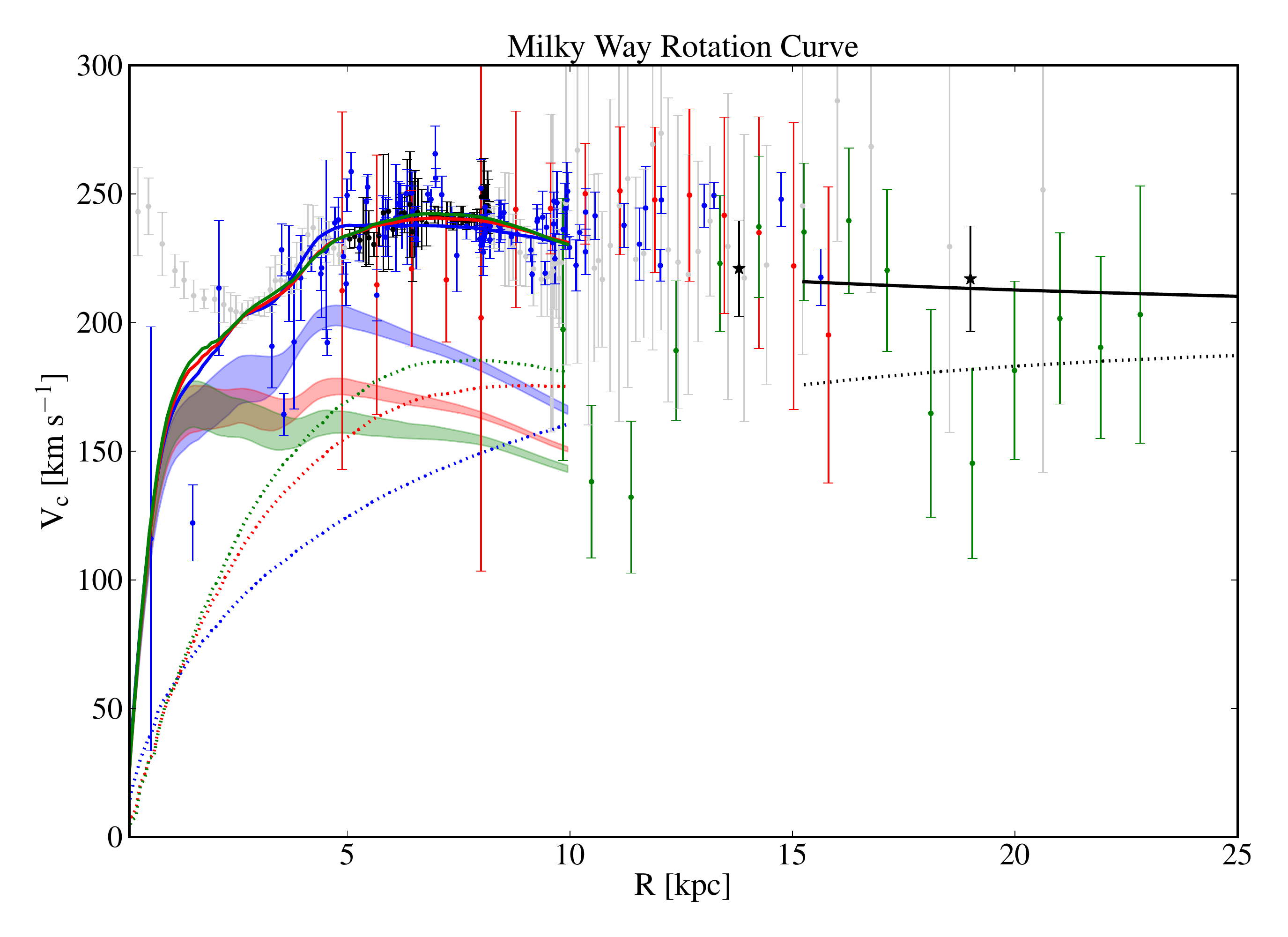}
\caption{Galactic rotation curve. Sources for data points are: maser \propm and \rv\ associated with high mass disk
  stars \citep[][blue]{Reid2014a}, inner Galaxy terminal velocities and outer disk velocities collected by
  \citet[][black ($5\kpc<R<R_0$) and grey (elsewhere)]{Sofue2009}, \propm of disk RCG from
  \citet[][red]{Lopez-Corredoira2014}, Jeans-equation converted \rv\ data for BHB stars \citep[][green]{Kafle2012}, and
  stream modelling for GD-1 \citep{Koposov2010a} and Pal-5 \citep[][black stars]{Kupper2015}.  All data were
  approximately converted to ($R_0\narreq8.2\kpc$, $\Vcirc\narreq238\kms$). The coloured bands show azimuthally averaged
  circular velocities for illustrative dynamical models with bulge, long bar, disk, and dark halo (Portail et al.\
  2016). In the bulge region, these models are based on stellar kinematic data \citep{Portail2015} and thus are more
  reliable than the (misleading) terminal velocities. The bulge and long bar {\sl stellar} mass in these models
  corresponds to a Kroupa IMF $\pm10\%$, whereas the disk has fixed local stellar surface density $38\Msun\pc^{-2}$ and
  scalelength $\RdiskM\narreq (2.15,2.6,3.0)\kpc$ (blue, red, green) and includes a gas disk with surface density
  $13\Msun\pc^{-2}$ and twice the stellar scalelength.  In each case, the lower band shows the rotation curve from the
  baryonic component, the dotted line shows the median dark halo profile, and the upper band shows the total rotation
  curve. In these models, the baryonic component provides ($86\%, 73\%, 65\%)$ of the circular velocity at
  $2.2\RdiskM$. The outer dotted and full lines show the rotation curves for an NFW halo with virial mass
  $\Mvir\narreq1.3\times10^{12}\Msun$ (\S\ref{s:darkhalo}) and concentration $c\narreq16$ which matches with the inner
  halo at $R\nsimeq 12\kpc$, and for this NFW halo combined with the $\RdiskM\narreq 2.6\kpc$ disk.}
\label{f:rotcurv}
\end{figure}

\begin{figure}
\centering
\includegraphics[width=8cm]{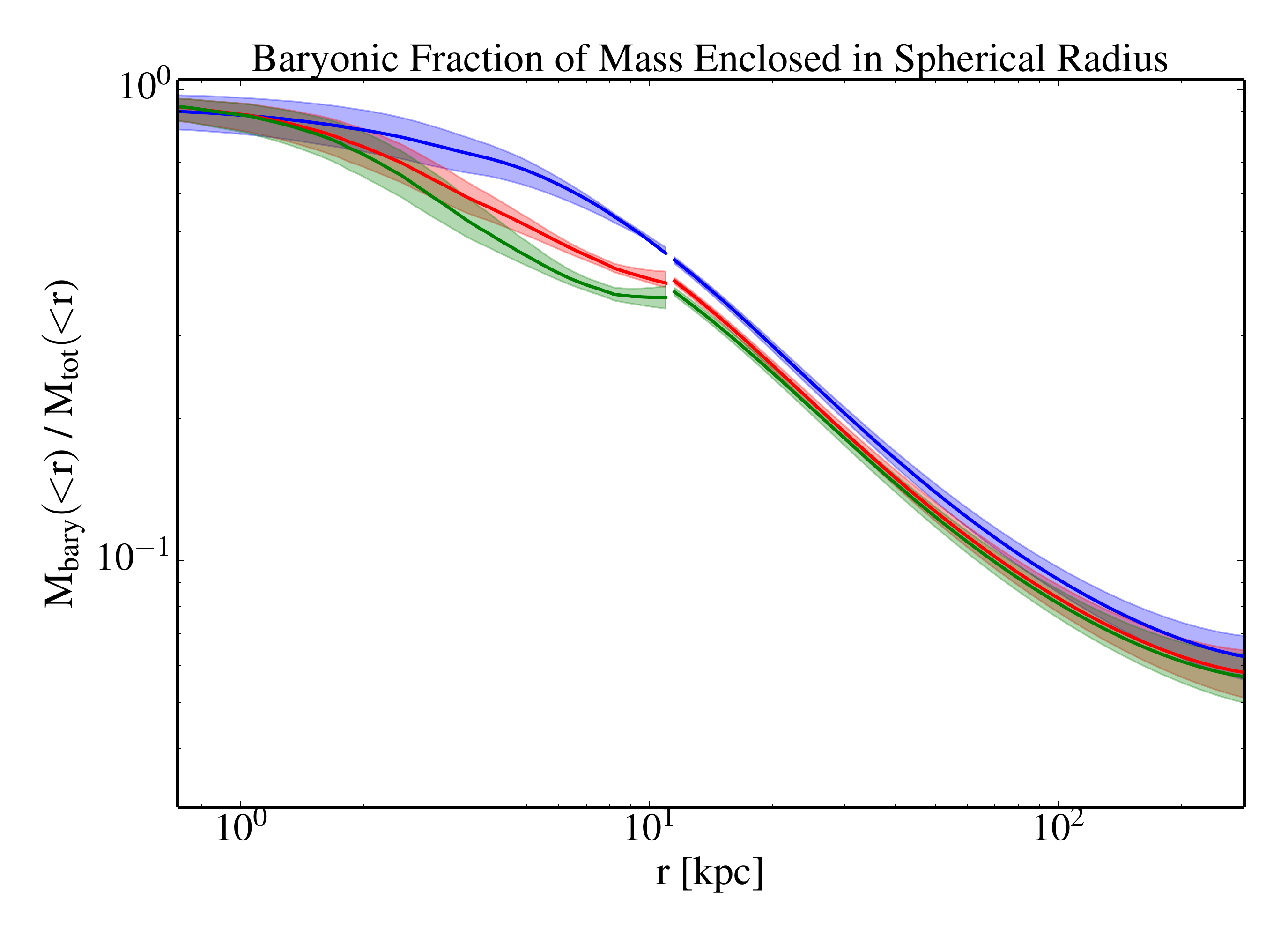}
\caption{Fraction of baryonic mass within radius $r$ including the stellar and cold gas mass from
the dynamical models shown in  Fig.~\ref{f:rotcurv} and the additional mass in hot gas predicted by
\citet{Tepper-Garcia2015} with an assumed uncertainty of $35\%$ (\S\ref{s:hothalo}). 
}
\label{f:barfrac}
\end{figure}

\vspace{5pt}
\noindent {\sl Oort's constants.} Traditionally, Oort's constants $A$ and $B$ were defined for the local disk as a means
to estimate the circular velocity $\Vcirc(R_0)$ and its gradient from radial velocity and \propm\ data for nearby
stellar populations, viz.
\begin{eqnarray}
A-B = {\Vcirc}/R_0; \quad A+B = -(\partial \Theta/\partial R)_{R_0}.
\end{eqnarray}
The quantity $A-B$ has been measured by many authors for different stellar populations \citep{Feast1997, Uemura2000,
  Elias2006} with values in the range $27-32 \kms \perkpc$. For the rotation gradient $-(A+B)$, different authors
find positive, zero and negative values. \citet{Catena2010} argue for $A+B \narreq 0.18 \pm 0.47 \kms \perkpc $ from
an \sdss study of M stars \citep{Fuchs2009}.

\subsubsection{Rotation curve} Figure~\ref{f:rotcurv} assembles rotation velocity measurements from various sources as
explained in the caption. The data indicate a nearly flat rotation curve in the range $R\narreq5-13\kpc$
\citep{Reid2014a} with a slight decrease at larger radii \citep{Kupper2015,Kafle2012}. \propm data from \gaia will
clarify this.  The rotation velocities determined by \citet[][see also original references therein]{Sofue2009} from
terminal velocities and a circular rotation model are unreliable in the region dominated by the Galactic bar
\citep[e.g.,][]{Englmaier+Gerhard99, Fux99}, as is clearly visible in the central $\nsim3\kpc$. Points inside
$R\narreq5\kpc$ (the half-length of the Galactic bar, \S\ref{s:longbar}) are therefore plotted in light shade.

\vspace{5pt}
\noindent{\sl Is the disk maximal?} As shown with the illustrative models in Fig.~\ref{f:rotcurv}, the answer to this
question largely depends on the disk (mass) scalelength. These models include a bulge with dynamical mass fitted to the
\brava kinematic data and stellar mass corresponding to a Kroupa IMF $\pm10\%$ \citep{Portail2015}.  The disk has local
stellar surface density $38\Msun/\pc^2$ \citep{Bovy2013} and scalelengths $\RdiskM\narreq 2.15 \kpc$
\citep{Bissantz+Gerhard02,Bovy2013}, $2.6\kpc$ \citep{Robin+03,Juric2008} and $3.0\kpc$ \citep{Kent1991,Gould1996} and
includes a gas disk with surface density $13\Msun/\pc^2$ and twice the stellar scalelength.  A short scalelength
\RdiskM\ has important implications: the Galaxy's disk (summed over all stellar and gaseous components) is then {\it
  maximal} \citep{Sackett1997} in the sense that the disk and bulge dominate the rotation curve, i.e.\ contribute $85\%$
of the rotational velocity and $\sim\!70\%$ of the rotational support at $2.2\RdiskM$. We included the bulge in the
definition here because it mostly originates from the disk, \S\ref{s:bulge}. For the three scalelengths shown, the
baryonic component provides a median fraction $f_v=$($86\%, 73\%, 65\%$) of the circular velocity and ($74\%, 53\%,
42\%)$ of the radial force at $2.2\RdiskM$. Thus the Milky Way's disk is maximal only for the shortest scalelength
\citep[see also][]{Sackett1997, Gerhard1999, Bovy2013, Piffl2014a}, but even for the longer $\RdiskM$ its
contribution to the rotational velocity is at the upper limit of that inferred for typical spiral galaxies by
\citet[][$f_v\narreq0.4$-0.7]{Martinsson2013}, but see \citet{Aniyan2016}.
 
\subsubsection{Baryonic mass fraction with radius} The dynamical models shown in Fig.~\ref{f:rotcurv} have median total
stellar masses of ($5.7, 5.0, 4.7) \times 10^{10}$\Msun. Of this, the stellar mass of the bulge and the disk embedded in
the bulge region is $\sim 1.5\times 10^{10}$\Msun.
The remaining mass is consistent with the
total disk mass given in Section 5, if we consider the mass of the long bar ($\sim 1\times 10^{10}$\Msun) as part of
the disk mass. Thus our estimate of the total stellar mass in the MilkyWay is $\Mstar = 5\pm 1\times 10^{10}$\Msun.
Adding the mass in cold gas and the
$2.5\times10^{10}$\Msun\ in hot gas resulting from \citet{Tepper-Garcia2015} for the NFW halo shown in
Fig.~\ref{f:rotcurv} ($\Mvir\narreq1.3\times10^{12}$\Msun, \S\ref{s:darkhalo}, $c\narreq16$), the total baryonic mass
fraction of the Galaxy becomes again $0.07\pm0.01$. The resulting total baryonic mass fraction within radius $r$ from stars,
cold gas, and hot gas is shown in Fig.~\ref{f:barfrac}.

\begin{marginnote}
\entry{\Mstar}{$5\pm 1\times 10^{10}$\Msun,
total stellar mass of the Galaxy}
\end{marginnote}
\section{CONCLUDING REMARKS}
The ultimate goal of Galactic research is to understand how the Milky Way has evolved from
cosmological initial conditions to its present state, and how its future evolution will
proceed. Our first task is to describe its current state in some detail; this allows us to
connect to similar galaxies nearby and to studies of galaxies at high redshift. 
Characterizing the Galaxy's dominant components and measuring their main parameters is an
important step in this process.

The last few years have seen significant progress in several areas, driven by the impressive
data from past and on-going surveys such as \SDSS, \VVV, \APOGEE and \RAVE. 
We now have a quantitative description of the box/peanut structure of the Galactic bulge
and, to a lesser extent, of its continuation into the Galactic plane, the long bar. We
know that the thin and thick disks are distinct sequences in abundance ratios and age.
The velocity field in the disk near the Sun has been
charted, and dynamical models have been developed to link these data self-consistently to
the gravitational potential. In the Galactic halo, multiple satellites and substructures
have been discovered and density and velocity measurements have been made to beyond
$>\!50\kpc$.  On the other hand, we still lack good understanding of, e.g., the radial
scale-length of the large-scale Galactic disk, the transition region between the Galactic
bar and the surrounding disk, the properties of the nuclear disk, the mass of the stellar
halo, and the outer rotation curve of the Milky Way. Constructing a complete structural
model for the Milky Way is one of the main challenges for the coming years.

Half of all stars in the Universe formed before a redshift of unity.
Detailed chemistry for millions of stars will provide important new information on the
early formation of the different components. Until recently, most of our understanding
of stars has come from the solar neighbourhood, but high-resolution stellar surveys 
have now begun to reach more representative regions of the Galaxy. We still
do not have a complete chemical inventory for any component over its full physical
extent. In principle, such data for enough stars will enable chemical tagging of 
dynamically distinct sub-systems. If even a few
disrupted stellar systems with different ages can be recovered, 
we can learn about the role of secular 
evolution and stellar migration over cosmic time
from their dispersal across the disk \citep{Bland-Hawthorn2010}. But
accurate elemental abundances are hampered by the difficulty of measuring good 
stellar parameters (e.g. log-g, effective temperature). The differential abundance
technique shows great promise but this works best if the stars have a similar spectral
type, thereby limiting the sample size. But we are encouraged by the revolution that
is now under way to improve the state of the art in achieving consistent stellar
abundances \citep{Jofre2014,Ness2015}.

We are looking forward to a golden age for Galactic research exemplified by \Gaia, the
astrometric space mission that was successfully deployed at the end of 2013.  The ongoing
or upcoming deep all-sky photometric (\DES, \LSST, \JWST, \WFIRST), spectroscopic
(\APOGEETWO, \GALAH, \WEAVE, \PFS, \FOURMOST) and seismological surveys (\COROT, \KTWO,
\TESS, \PLATO) are well placed to advance our understanding of stellar populations. The
large and extremely large telescopes will also play an important role, particularly with
high-resolution spectroscopy (e.g. \MOONS on \VLT, \GCLEF on \GMT).

The \LSST co-added survey will reach 
up to four magnitudes deeper than SDSS: for stars with $0.2<g-r<0.6$ and $g<23.5$, 
\LSST will achieve a metallicity error of 0.1 dex for metal-rich stars and 0.2 dex for
metal-poor stars for about 200 million F/G main sequence stars \citep{Ivezic2008}.
In a single exposure, \LSST will detect metal-poor MSTO stars to $\sim$140 kpc and
horizontal branch stars and RR Lyrae to $\sim$500 kpc, going several times further
in the final co-added data.
\LSST will also provide proper-motion measurements and parallaxes for stars below
$r\sim 20$ where \gaia lacks sensitivity \citep{Ivezic2012}.
This will revolutionize photometric parallax and metallicity
studies such as reviewed in \S 4.
 
\begin{figure*}
\begin{center}
\includegraphics[width=16cm]{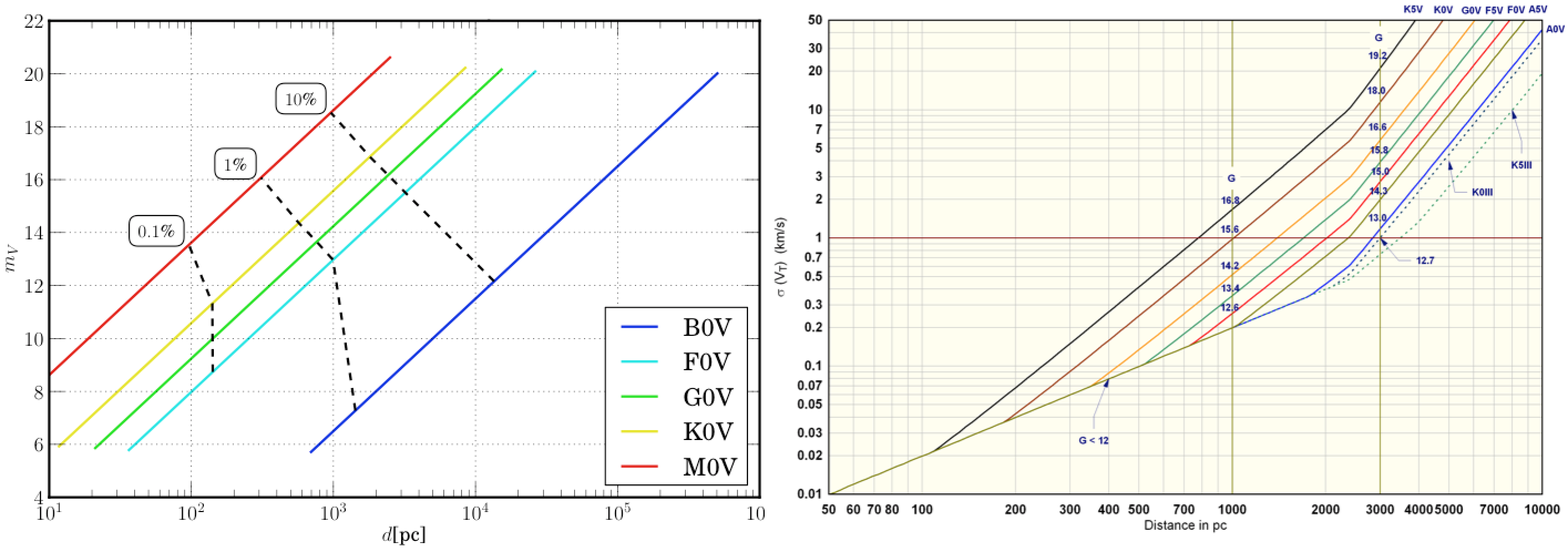}
\end{center}
\caption{(Left) The distance accuracy of \gaia\ for dwarf stars:
the 0.1\%, 1\% and 10\% `accuracy horizons' are shown as dashed lines as
a function of $V$ magnitude. The different dwarf types are indicated by coloured lines.
This figure is made with the {\tt Pygaia} package developed by A.G.M. Brown
\citep[courtesy of][]{Read2014}. (Right) The accuracy in the \gaia\ transverse velocity
for dwarf stars as a function of their distance; the uncertainties arise from errors in 
both the proper motions and parallaxes
\citep[courtesy of][]{Mignard2011}.
}
\label{f:read1}
\end{figure*}

The \gaia astrometric mission will be even more far reaching for Galactic research. 
\gaia\ will have an enormous impact on our understanding of 6D phase space
({\bf x}, {\bf v}) for the stars, particularly in the near field, but with important new 
information for giants extending into the outer halo. 
By the end of the decade, it will have
obtained positions and velocities for up to two billion stars, i.e. phase space information
for a few percent of stars that dominate the visible light in the Galaxy. In Fig.~\ref{f:read1}(left),
many F/G dwarfs to be observed by \LSST will have distances and proper motions 
with 10\% accuracy out to $\sim 3$ kpc from the Sun. Moreover, in Fig.~\ref{f:read1}(right), dwarfs 
brighter than $V\sim 15$ will have transverse velocities at least as good as the measured radial 
velocities. Such precise 6D phase space information will allow us to unravel the 
complex chemistry and dynamics of the local disk components to unprecedented levels.

Putting all the different strands together is not going to be easy. Increasingly
sophisticated dynamical methods will be needed to accommodate the 6D phase space information in a
complex multi-component potential, and to combine this with the chemical information
${\cal C}$([Fe/H],[$\alpha$/Fe], ...), but good progress is already being made 
\citep{Hunt2014,Sanders2015}. Eventually, we will need to consider the
departures from dynamical equilibrium which arise from internal evolution and interactions
with the outside \citep{Binney2013}.

The Galaxy resembles a complex organism which evolves in a self-regulated fashion
according to the laws of gravity, star formation, dynamics, and chemical evolution,
while subject to mass accretion and external influences from its cosmological
neighbourhood. Its low gas content, position in the green valley of the colour-magnitude
diagram, and secularly evolved central parts of old stars indicate that it 
is in its late stages of evolution.
 
N-body and hydrodynamic simulations are becoming increasingly useful to help understand
both formation and evolutionary processes
\citep{Stinson2013,Minchev2014,Feng2016,Scannapieco2015}.
These simulations will need to continue to grow in size to resolve smaller particle masses in both gas and
dark matter, and to ultimately resolve individual star clusters. More thought must be
given to how we ``evaluate'' the goodness of fit of a complex dynamical model, or a
numerical simulation, when comparing to the high-dimensional data space available to
Galactic researchers \citep{Sharma2011}. Through these comparisons Galactic studies
will improve galaxy formation simulations and lead to a understanding of galaxy formation
processes in general.

Ultimately, we may never arrive at a complete understanding of the Galaxy, much
like any complex physical system. Most often, we learn about the physical laws through a 
series of approximations that become progressively more refined. But our search for 
understanding is noble all the same as we seek answers to the many wonders around us.

\def\be{\begin{eqnarray}}   \def\ee{\end{eqnarray}}
\def\ben{\begin{eqnarray*}} \def\een{\end{eqnarray*}}
\def\sec#1{Section~\ref{sec:#1}}
\def\fig#1{Figure~\ref{fig:#1}}
\def\tab#1{Table~\ref{tab:#1}}
\def\equ#1{Eq.~(\ref{e:#1})}
\def\secp#1{Section~\ref{sec:#1}}
\def\figp#1{Figure~\ref{fig:#1}}
\def\tabp#1{Table~\ref{tab:#1}}
\def\equp#1{Eq.~\ref{e:#1}}

\section{ONLINE VERSION: Analytic framework for fitting a parametric model to the Galaxy}

We describe an analytic framework to model the Galaxy that is
defined by a large set of kinematic parameters \citep{Sharma2014}. 
The framework is designed for rapid Bayesian optimization \citep[e.g.][]{Catena2010}
and shares important features with most studies of this kind \citep[e.g.][]{Rix2013}.
The stellar content is modelled as a set of distinct
components: the thin disk, the thick disk, the stellar halo and 
the bulge. The distribution functions, i.e., the number density of stars
as a function of position (${\bf r}$), velocity (${\bf v}$), age ($\tau$),
metallicity ($Z$), and mass ($m$) for each component, is
assumed to be specified a priori as a function $f_j({\bf r,v},\tau,Z,m)$
where $j$ $(=1,2,3,4)$ runs over the four components.
The form of $f_j$ that correctly describes all the properties
of the Galaxy and is self-consistent is still an open question.

For a given Galactic component, we assume stars form at a rate
$\Psi_j(\tau)$ with a mass distribution $\xi(m|\tau)$ (IMF) that is a
function of age $\tau$, where the present day spatial distribution of stars $p({\bf r}|\tau)$ is
conditional on age only. For a velocity distribution $p({\bf v|r},\tau)$ and
metallicity distribution $p(Z|\tau)$, we have
\be
f_j({\bf r,v},\tau,m,Z) = \frac{\Psi(\tau)}{\langle m\rangle}\xi(m|\tau)p({\bf r}|\tau)p({\bf v|r},\tau)p(Z|\tau) .
\label{e:dist_func2}
\ee

The functions take different forms for each Galactic component \citep{Sharma2011}.
The IMF is normalized such that $\int_{m_{\rm min}}^{m_{\rm max}}\xi(m|\tau)dm =1$ and
$\langle m \rangle=\int_{m_{\rm min}}^{m_{\rm max}}m\xi(m|\tau)dm$ is the mean stellar mass.
The metallicity distribution is modeled as a log-normal distribution,
\be
p(Z|\tau)=\frac{1}{\sigma_{\log Z}(\tau)\sqrt{2\pi}}{\rm exp}\left[-\frac{(\log Z-\log\bar{Z}(\tau))}{(2\sigma_{\log
Z}^2(\tau))}\right],\label{e:amr}
\ee
the mean and dispersion of which are given by  age-dependent functions
$\bar{Z}(\tau)$ and $\sigma_{\log Z}(\tau)$.
The $\bar{Z}(\tau)$ is widely referred to as the age-metallicity relation (AMR).
For the thin disk, the well known ``age-scale height'' relation is given by the axis ratio $\epsilon$, viz.
\be
\epsilon(\tau)={\rm Min}\left(0.0791,0.104\left(\frac{\tau/{\rm Gyr}+0.1}{10.1}\right)^{0.5}
  \right).
\ee

\smallskip\noindent{\sl Kinematic model.}
For a useful kinematic model, we want to constrain the velocity distribution $p({\bf v|r},\tau)$.
We assume that everything except for $p({\bf v|r},\tau)$ on the rhs of \equ{dist_func2} is known. 
The functional form for $p({\bf v|r},\tau)$ is restricted because the spatial density distribution and 
the kinematics are linked to each other via the potential. 
The accuracy of a pure kinematic model depends upon our ability to 
supply functional forms of $p({\bf v|r},\tau)$ that are a good approximation to the actual 
velocity distribution of the system.
A proper way to handle this problem would be to use  
dynamically self consistent models, but such models 
are still under development. Here we explore 
kinematic models that provide a reasonable 
approximation to the actual velocity distribution.

\smallskip\noindent
{\sl Gaussian velocity ellipsoid model.}
In this model, the velocity distribution is assumed to be a triaxial
Gaussian,
\begin{eqnarray}
p({\bf v|r},\tau) & = &
\frac{1}{\sigma_{R}\sigma_{\phi}\sigma_{z}(2\pi)^{3/2}}
{\rm  exp}\left[-\frac{v_{R}^2}{2\sigma_{R}^2}\right]
{\rm  exp}\left[-\frac{v_z^2}{2\sigma_z^2}\right] \times{\rm exp}\left[-\frac{(v_{\phi}-\overline{v_{\rm \phi}})^2}{2\sigma^2_{\phi}} \right],
\label{e:veldist}
\end{eqnarray}
where $R,\phi,z$ are cylindrical coordinates.
The $\overline{v_{\phi}}$ is the asymmetric drift and is given by
{\small
\be
\overline{v_{\phi}}^2(\tau,R)  & = &  v_{\rm c}^2(R)+ \sigma_{R}^2 \times\left( \frac{d \ln \rho}{d \ln
R}+ \frac{d \ln \sigma_{R}^2}{d \ln R}+1-\frac{\sigma_{\phi}^2}{\sigma_R^2}+1-\frac{\sigma_z^2}{\sigma_R^2}\right)
\label{e:stromberg}
\ee
}
This follows from Eq. 4.227 in Binney \& Tremaine (2008)
assuming $\overline{v_R\:v_z}=(v_R^2-v_z^2)(z/R)$. This is valid for
the case where the principal axes of velocity ellipsoid are aligned
with the $(r,\theta,\phi)$ spherical coordinate system. If the
velocity ellipsoid is aligned with the cylindrical $(R,\phi,z)$
coordinate system, then $\overline{v_R\:v_z}=0$.
Recent results using the RAVE data suggest that the velocity ellipsoid is aligned
with the spherical coordinates \citep{Siebert2008,Binney2014}.
One can parametrize our ignorance by writing the asymmetric drift as follows:
{\small
\be
\overline{v_{\phi}}^2(\tau,R) & = & v_{\rm c}^2(R)+ \sigma_{R}^2 \left( \frac{d \ln \rho}{d \ln
R}+ \frac{d \ln \sigma_{R}^2}{d \ln
R}+1-k_{\rm ad}^2\right)
\label{e:stromberg_bovy}
\ee
}
This is the form used by \citet{Bovy2012}.

The dispersions of the $R,\phi$ and $z$ components of velocity
increase as a function age due to secular heating in the disk,
and there is a radial dependence such that the dispersion
increases towards the Galactic Center. We model these effects after
\citet{Aumer2009} using the functional form
\begin{eqnarray}
\sigma^{\rm thin}_{R,\phi,z}(R,\tau) & = &\sigma_{R,\phi,z,\odot}^{\rm
thin}\exp\left[-\frac{R-R_{0}}{R_\sigma^{\rm thin}}\right]
\times\left(\frac{\tau+\tau_{\rm
min}}{\tau_{\rm max}+\tau_{\rm
min}}\right)^{\beta_{R,\phi,z}}
\\
\label{e:veldisp1}
\sigma^{\rm thick}_{R,\phi,z}(R) & = &\sigma_{R,\phi,z,\odot}^{\rm thick}
{\rm exp}\left[-\frac{R-R_{0}}{R_\sigma^{\rm thick}}\right].
\label{e:veldisp2}
\end{eqnarray}
The choice of the radial dependence is motivated by the desire to
produce disks in which
the scale height is independent of radius.
For example, under the epicyclic approximation, if $\sigma_z/\sigma_R$
is assumed to be constant, then
the scale height is independent of radius for $R_\sigma=2R_d$
\citep{vanderKruit2011}.
In reality there is also a $z$ dependence of velocity dispersions
which we have chosen to ignore in our present analysis.
This means that for
a given mono age population the asymmetric drift is independent
of $z$. However, the velocity dispersion and asymmetric drift
of the combined population of stars are functions of $z$. This
is because the scale height of stars for a given isothermal
population is an increasing function of
its vertical velocity dispersion.

\smallskip\noindent {\sl Shu distribution function model.}
The Gaussian velocity ellipsoid model has its limitations. In particular,
the distribution of $v_\phi$ is strongly non-Gaussian, being
highly skew to low $v_\phi$. For a two-dimensional disk, a much better approximation to the velocity distribution is provided by the 
Shu (1969) distribution
function.  Moreover, the Shu DF, being dynamical in nature, connects the
radial and azimuthal components of velocity dispersion to each other and to
the mean-streaming velocity, thus lowering the number of free parameters in
the model.

Assuming the potential is separable as $\Phi(R,z)=\Phi_R(R)+\Phi_z(z)$ we can write the
distribution function as
\be
f(E_{R},L_z,E_z)=\frac{F(L)}{\sigma_{R}^2(L_z)}{\rm
exp}\left[-\frac{E_{R}}{\sigma_{R}^2(L_z)}\right] \nonumber
\times \frac{{\rm
exp}\left[-(E_z)/(\sigma^2_z(L_z))\right]}{\sigma_z(L_z)\sqrt{2\pi}},
\ee
where $L=Rv_\phi$ is the angular momentum,
\bea
E_z &=& \frac{v_z^2}{2}+\Phi_z(z) \\
E_R &=& \frac{1}{2}v_R^2+\Phi_{\rm eff}(R,L_z)-\Phi_{\rm eff}(R_g,L_z)
\eea
with
\be
\Phi_{\rm eff}(R,L_z) &= & \frac{L_z^2}{2R^2}+\Phi(R)\simeq
\frac{L_z^2}{2R^2}+v^2_c\ln R \label{e:effpot}
\ee
being the effective potential.
Let $R_g(L_z)=L_z/v_{\rm c}$ be the radius of a circular orbit with
specific angular momentum $L_z$. In \citet{Schonrich2012} \citep[q.v.][]{Sharma2013}, it was shown
that the joint distribution of $R$ and $R_g$ can be written as
{\small
\begin{eqnarray}
P(R,R_g)   & = & \frac{(2\pi)^2\Sigma(R_g)}{g(\frac{1}{2a^2})}{\rm
exp}\left[\frac{2 \ln(R_g/R)+1-R_g^2/R^2}{2a^2}\right],
\label{e:prrl_shu}
\end{eqnarray}
}
where $\Sigma(R)$ is a function that controls the disk's surface density and
\begin{eqnarray}
a & = & \sigma_{R}(R_g)/v_{\rm c} \label{e:a_shu} \\
g(c) & = & \frac{e^c\Gamma(c-1/2)}{2c^{c-1/2}}.
\end{eqnarray}

\noindent
We assume $a$ to be  specified as
\begin{eqnarray}
a & = &a_0(\tau) {\rm exp}\left[-\frac{R_g}{R_\sigma}\right]
= \frac{\sigma_{R,\odot}}{v_{\rm c}}\left(\frac{\tau+\tau_{\rm
min}}{\tau_{\rm max}+\tau_{\rm min}}\right)^{\beta_R}
{\rm exp}\left[-\frac{R_g-R_{0}}{R_\sigma}\right]
\end{eqnarray}
and $\sigma_z$ to be  specified as
{\small
\be
\sigma_{z0}(R_g,\tau) & = &\sigma_{z,\odot}\left(\frac{\tau+\tau_{\rm
min}}{\tau_{\rm max}+\tau_{\rm min}}\right)^{\beta_z}
{\rm exp}\left[-\frac{R_g-R_{0}}{R_\sigma}\right].
\ee
}
Now this leaves us to choose $\Sigma(R_g)$. This should be
done so as to produce disks that satisfy the observational
constraint given by $\Sigma(R)$, i.e., an exponential disk (or disks)
with scale length $R_d$. A simple way to do this is
to let
\be
\Sigma(R_g) & = &  \frac{e^{-R_g/R_d}}{2 \pi R_d^2} .
\ee
However, this matches the target surface density only approximately.
An alternative approach is to use the fast optimization formula proposed in \citet{Sharma2013}
such that
{\footnotesize
\be
\Sigma(R_g) & = & \frac{e^{-R_g/R_d}}{2\pi R_d^2} - \frac{0.00976
a^{2.29}_{0}}{R_d^2}  s\left[\frac{R_g}{(3.74R_d(1+q/0.523)}\right]
\ee
}
where $q=R_d/R_\sigma$ and $s$ is a function of the following form
\be
s(x) & = &k e^{-x/b}((x/a)^2-1),
\ee
with $(k,a,b)=(31.53,0.6719,0.2743)$.

\smallskip\noindent{\sl The gravitational potential $\Phi$.} \label{sec:potmodel}
So far we have described kinematic models in which the
potential is separable in $R$ and $z$. In such cases,
the energy associated with the vertical motion
$E_z$ can be assumed to be the third integral of motion.
In reality, the potential generated by a double exponential
disk is not separable in $R$ and $z$.
For example, the hypothetical circular speed
defined as $\sqrt{R\partial{\Phi(R,z)}/\partial{R}}$ can have
both a radial and a vertical dependence, i.e. 
\be\label{e:defsvc}
v_{\rm c}(R,z) & = & \sqrt{R \frac{\partial \Phi}{\partial R}}
 =  {{\Theta_0+\alpha_R (R-R_{\odot})}\over{1+\alpha_z |z/\kpc|^{1.34}}}.
\ee
The parameters $\alpha_R$ and $\alpha_z$
control the radial and vertical dependencies, respectively. The motivation
for the vertical term comes from the fact that the above formula
with $\alpha_z=0.0374$ provides a good fit to the $v_c(R_0,z)$
profile of Milky Way potential \citep{Dehnen1998, Law2010}.
Both references have bulge, halo and disk components. The former has two double
exponential disks while the later has a Miyamoto-Nagai disk.

To model systems where the potential is not
separable in $R$ and $z$, a simpler
approach is motivated by the fact that,
for realistic Galactic potentials, we expect
the $\overline{v_{\phi}}$ of a single age population to fall with $z$.
\citet{Binney2012} finds that when
vertical motion is present, the effective potential for
radial motion (see Eq. \ref{e:effpot}) needs to be
modified because the vertical motion
also contributes to the centrifugal potential.
Neglecting this effect leads to an overestimation of $\overline{v_{\phi}}$.
As one moves away from the plane, this effect
becomes increasingly important. Furthermore,
as shown by \citet{Schonrich2012}, in a given solar
annulus, stars with smaller $R_g$ will have larger vertical
energy and hence larger scale height. Moreover, stars with
smaller $R_g$ are more likely to be found at higher $z$,
consequently $\overline{v_{\phi}}$ should also decrease
with height.

The fall of $\overline{v_{\phi}}$ with height is also
predicted by the Jeans equation for an axisymmetric system
\be
\overline{v_{\phi}}^2(R,z) & = & \left[R\frac{\partial \Phi}{\partial R}\right] +\sigma_R^2
\left[1-\frac{\sigma_{\phi}^2}{\sigma_R^2}  
 +\frac{\partial{\ln(\rho\sigma_R^2)}}{\partial{\ln
R}}\right] + R\left[ \frac{\partial \overline{v_R
v_z}}{\partial z} +\overline{v_R
v_z}\frac{\partial \ln \rho}{\partial z}\right]. \label{e:jeanaxis}
\ee
The $\overline{v_{\phi}}$ at high
$z$ will be lower both because $R\;\partial\Phi /\partial R$
is lower and because the term in the third square bracket
decreases with $z$, e.g., assuming $\overline{v_R
v_z}=(\sigma_R^2-\sigma_z^2)z/R$.

For the Gaussian model, \citet{Sharma2014} simulate the
overall reduction of  $\overline{v_{\phi}}$ with $z$
by introducing a parametrized form for $v_c(R,z)$ as given
by \equ{defsvc} in \equ{stromberg}. Given this prescription
we expect  $\alpha_z>0.03744$, so as to account for effects other
than that involving the first term in \equ{jeanaxis}.
In reality, the
velocity dispersion tensor ${\bf \sigma}^2$ will have a
more complicated dependence on $R$ and $z$ than
the exponential dependence on $R$.

For the Shu model, \citet{Sharma2014} replace $v_c$ in \equ{a_shu} by
the form in \equ{defsvc}.
However, the prescription breaks the dynamical
self-consistency of the model and turns it into a
fitting formula. In reality, the $\overline{v_{\phi}}$ may
not exactly follow the functional form for the vertical
dependence predicted by our model, but is better than
completely neglecting it.

\section*{DISCLOSURE STATEMENT}
The authors are not aware of any conflicts that might be perceived as affecting the objectivity of this review.

\section*{ACKNOWLEDGMENTS}

The initial idea for a review on the Galaxy's global and structural parameters came from John Kormendy. It has been
challenging to write not least because this is a vibrant and dynamic field of research, and major uncertainties still
exist. But our motto throughout $-$ keep calm and carry on $-$ has sustained us even when no clear picture has emerged
at times.  Inevitably, with imposed page limits, there will be missing topics and references for which we apologize in
advance.  We thank our referee Tim Beers for his oversight of the review, and
Ken Freeman, Rosie Wyse and James Binney for their 
historical perspective, wisdom and insight.
We are indebted to various colleagues for their support and help with figures and tables:
C. Correa, P. Kafle, T. Licquia, F. Mignard, Y. Momany, M. Portail, J. Read, T. Tepper-Garcia, D. Webster, C. Wegg, and for
additional advice, perspectives and comments: M. Arnaboldi, B. Benjamin, T. Bensby, J. Bovy, M. Boylan-Kolchin, R. Drimmel,
S. Gillessen, Z. Ivezic, R. Lange, P. McMillan, D. Nataf, M. Reid,
R. Sch\"odel and R. Sch\"onrich.  
JBH acknowledges the Kavli Institute, UC Santa Barbara for their
hospitality during the early planning of this review.  JBH is supported by an ARC Australian Laureate Fellowship. OG is
grateful for the support of the Max Planck Institute for Extraterrestrial Physics, and acknowledges a visiting fellowship from
the Hunstead Fund for Astrophysics at the University of Sydney and the hospitality of Mount Stromlo Observatory during the final stages of the review.

\bibliographystyle{ar-style2.bst}

\end{document}